\documentstyle[a4,12pt,epsfig,amsmath,psfrag,pstricks,amssymb]{article}

 \def\relic{\Omega_{DM
}}
\def\lsim{\raise0.3ex\hbox{$\;<$\kern-0.75em\raise-1.1ex\hbox{$\sim\;$}}}
\def\gsim{\raise0.3ex\hbox{$\;>$\kern-0.75em\raise-1.1ex\hbox{$\sim\;$}}}

\def\bsg{$b\to s\gamma$}

\def\asusy{a^{\rm SUSY}_\mu}

\DeclareMathAlphabet   {\mathsc}{OT1}{cmr}{m}{sc} 
 
\def\[{\left [} 
\def\]{\right ]} 
\def\({\left (} 
\def\){\right )}

\newcommand{\gappeq}{\mathrel{\rlap {\raise.5ex\hbox{$>$}} 
{\lower.5ex\hbox{$\sim$}}}} 
\newcommand{\lappeq}{\mathrel{\rlap{\raise.5ex\hbox{$<$}} 
{\lower.5ex\hbox{$\sim$}}}}

\newcommand{\bea}{\begin{eqnarray}}
\newcommand{\eea}{\end{eqnarray}}

\newrgbcolor{darkred}{0.7 0 0}
\newrgbcolor{green}{0 1 0}
\newrgbcolor{darkblue}{0 0 0.7}
 \newrgbcolor{purple}{1 0 1}

\begin{document}

\vspace{-1truecm}

\rightline{LPT--Orsay 05/31}
\rightline{DESY 05-097}
\rightline{ULB-TH/05-16}
\rightline{FTUAM 05/9}
\rightline{IFT-UAM/CSIC-05-29}
%\rightline{hep-ph/0506204}
%\rightline{July 2004}   

\vspace{0.cm}

\begin{center}

{\Large {\bf Adiabatic compression 
and indirect detection 

of
    supersymmetric
dark matter }}
\vspace{0.3cm}\\
{\large Y. Mambrini$^{1,2}$, C. Mu\~noz$^{3,4,5}$, E. Nezri$^6$, F. Prada$^7$ 
%$^1$
}
\vspace{0.3cm}\\

$^1$ 
Laboratoire de Physique Th\'eorique, 
Universit\'e Paris-Sud, F-91405 Orsay, France
\vspace{0.3cm}\\

$^2$ 
Deutsches Elektronen-Synchrotron DESY,
Notkestrasse 85, 22607 Hamburg, Germany
\vspace{0.3cm}\\

$^3$ 
Departamento de F\'isica Te\'orica C-XI,
Universidad Aut\'onoma de Madrid,\\
%and Instituto de F\'isica
%Te\'orica C--XVI
%,
%\\
Cantoblanco, 28049 Madrid, Spain
\vspace{0.3cm}\\

$^4$ 
%Departamento de F\'isica Te\'orica C-XI and 
Instituto de F\'isica Te\'orica C--XVI,
Universidad Aut\'onoma de Madrid,\\
Cantoblanco,
28049 Madrid, Spain
\vspace{0.3cm}\\

$^5$ 
Department of Physics, Korea Advanced Institute of Science and
Technology, \\
Daejeon 305-701, Korea
\vspace{0.3cm}\\

$^6$ 
Service de Physique Th\'eorique, CP225, Universit\'e Libre de
Bruxelles,
\\
1050 Bruxelles, Belgium
\vspace{0.3cm}\\

$^7$ 
Instituto de Astrof\'isica de Andaluc\'ia (CSIC), E-18008 Granada, Spain
\vspace{1.cm}\\

\end{center}

\vspace{-0.5cm}

\abstract{Recent developments in the modelling of the dark matter
distribution in our Galaxy
point out the necessity to consider  some physical processes to satisfy
observational data. In particular,
models with adiabatic compression, which 
include the effect of the baryonic gas in the halo, 
increase significantly the dark matter density in the central 
region of the Milky Way. On the other hand, the non-universality 
in scalar
and gaugino sectors of supergravity models can also increase significantly 
the neutralino annihilation cross section. We show that 
the combination of both effects 
gives rise to a gamma-ray flux arising from the galactic center 
largely reachable by future experiments like GLAST. 
We also analyse in this framework the EGRET excess data above 
1 GeV, as well as the recent data from CANGAROO and HESS.
The analysis has been carried out imposing the most recent experimental 
constraints, such as the lower bound on the Higgs mass,
the \bsg\ branching ratio, and the muon $g-2$.
In addition, 
the recently improved upper 
bound on $B(B_s \to \mu^+ \mu^-)$ has also been taken into
account.
The astrophysical (WMAP) bounds on the dark matter density have also been
imposed on the theoretical computation of the relic neutralino
density through thermal production.
}

\newpage

\vspace{3cm}

\newpage

\tableofcontents

\vspace{3cm}

\newpage

\pagestyle{plain}

\section{Introduction}

It is now well established that luminous matter makes up only a small
fraction of the mass observed in Universe. 
A weakly interacting massive particle (WIMP) is one of the leading
candidates
for the ``dark'' component of the Universe.
Indeed, WIMPs may be present in the right amount to explain
the matter density observed in the analysis 
of galactic rotation curves \cite{Persic}, cluster of
galaxies,
and large scale flows \cite{Freedman},
implying
$0.1\lsim \relic h^2\lsim 0.3$.
Actually,
the recent data obtained by the WMAP satellite \cite{wmap03-1}
confirm that dark matter is present, and they lead to the value
$0.094\lsim\relic h^2\lsim 0.129$.

One of the most promising methods for the
indirect detection of WIMPs consists of detecting the gamma rays produced by
their annihilations in the galactic 
halo \cite{nuevas}-\cite{Khlopov}.
The amount of gamma-ray fluxes observed will depend of course on the nature
of WIMPs through their annihilation cross sections, but also on their
density.
This is the reason why the inner center of our galaxy 
($\sim 100$ pc from the center),
where the dark matter density is large, 
is the main theater of dark matter searches
through gamma-ray signatures. 
Atmospheric Cherenkov telescopes or space-based
detectors are used for this search. 
For example,
one of the space-based experiments, EGRET, 
detected a signal \cite{EGRET} that apparently is difficult 
to explain with the 
usual gamma-ray background.
%\footnote{Some attempts with more exotic, but
%consistent triaxial isothermal halo \cite{modi2} or with modifications
%of the ``conventional model" based on the locally observed electron spectrum
%\cite{modi} seems to reproduce EGRET data over much of the sky too.}.
Also, the atmospheric telescope CANGAROO-II claims a significant
detection of gamma rays from the galactic center region
\cite{Cangaroo1}.
In addition, the atmospheric telescope HESS has also observed 
gamma-rays \cite{HESS} in 
Sagittarius A$^*$ direction.
% (although they would correspond to a 
%neutralino heavier than 12 TeV).
Projected experiments might
clarify the situation. This is the case of the
space telescope GLAST \cite{GLAST}, which is scheduled
for launch in 2006, and will be able to provide 
a larger sensitivity. 

Concerning the nature of WIMPs, 
the best motivated candidate is the lightest
neutralino, a particle predicted by the 
supersymmetric (SUSY) extension of the standard model \cite{mireview}. 
In particular, in most of the parameter space of the
minimal supersymmetric standard model (MSSM)
the lightest supersymmetric particle (LSP) is the lightest neutralino.
Thus it is absolutely stable and therefore a candidate for dark
matter. In addition, the neutralino is an electrically neutral
particle, and 
this is welcome since otherwise it  
would bind to nuclei and would be excluded as a candidate
for dark matter from unsuccessful searches for exotic heavy 
isotopes.

Recently, SUSY dark matter candidates have been studied
in the context of realistic halo models including
baryonic matter \cite{Prada:2004pi}.
Indeed, since the total mass of the inner galaxy is dominated by baryons,
the dark matter distribution is likely to have been influenced by 
the baryonic potential. In particular, its density is increased, and 
as a consequence typical halo profiles such as
Navarro, Frenk and White (NFW) \cite{Navarro:1996he} 
and
Moore et al. \cite{Moore:1999gc} 
have
a more singular behaviour near the galactic center.
The conclusion of the work in Ref.~\cite{Prada:2004pi} 
is that the gamma-ray flux produced by the annihilation of neutralinos
in the galactic center is increased significantly,
and is within the sensitivity of projected experiments, 
when density profiles with baryonic compression are taken into account.

Nevertheless, 
this result has been obtained assuming particular values for the
neutralino mass and annihilation cross sections $\sigma_i$, 
and therefore a more detailed analysis computing explicitly 
$\sigma_i$ in the framework of SUSY must be carried out.
It has also been discussed recently
that another way of increasing the
gamma-ray flux produced by the annihilation of neutralinos
is to consider the possibility of non-universality in the scalar
and gaugino sectors of the 
MSSM \cite{hooper,Bottino,Mambrini:2004ke, Mambrini:2004kv}.
In particular, in the context of supergravity (SUGRA) 
%the annihilation
%cross sections 
%$\sigma_i$, and therefore 
the flux
can be increased significantly.
This is so 
when departures from the minimal supergravity (mSUGRA)
scenario, where the soft terms of the MSSM are assumed to be
universal at the unification scale and radiative electroweak
symmetry breaking is imposed, are taken into account.
Using non-universal soft scalar and gaugino masses 
the annihilation cross section of neutralinos can be increased, producing
a larger gamma-ray flux \cite{Mambrini:2004ke, Mambrini:2004kv}.

The aim of this work is to discuss in detail both effects, 
halo models with baryonic compression
and non-universal soft terms in SUGRA, on the gamma-ray flux.
We will see that 
a significant enhancement is obtained, and regions of the
SUGRA parameter space are compatible with the sensitivity
of present and projected experiments. 
%in particular
%GLAST, EGRET or HESS datas. 

Let us remark that in addition to the astrophysical bounds
on the dark matter density discussed above,
the most recent experimental constraints are also considered in the
analysis.
In particular, we implement the lower bounds on the masses
of SUSY particles and Higgs boson, as well as the experimental
bounds on the branching ratio of the \bsg\ process and on 
$\asusy$, for which the more stringent constraint from $e^+e^-$ 
disfavors 
important regions of the SUGRA parameter space 
(see e.g. Ref. \cite{Bertone:2004ag,Mambrini:2004ke}). 
In addition,
we have also taken into account  the last data concerning the 
$B_s \to \mu^+ \mu^-$ branching ratio.
It is now known that the upper bound on this process constrains 
strongly the parameter space
of non-universal SUGRA models, and has important implications
for direct searches of dark matter \cite{ko,ko2}. 
Thus an analysis in the case
of indirect detection is necessary and has been carried out here.
%Summarizing, BLABBLABAL.

The paper is organized as follows.
In Section~2 we will discuss in general the
gamma-ray detection from dark matter annihilation, 
paying special attention to its halo model dependence.
In particular, the modifications produced in the different profiles
due to the effects of baryonic compression will be studied.
The particle physics dependence will be reviewed in
Section~3 for the case of neutralinos in SUGRA.
We will study the general case
when
scalar and gaugino non-universalities are present.
In addition, the most recent experimental and astrophysical
constraints will be taken into account in the discussion.  
In Section~4 both effects, halo models with baryonic compression
and non-universal soft terms, will be considered for the
study of the gamma-ray flux. 
The theoretical predictions will be compared with 
the sensitivity of present and projected experiments, 
such as EGRET, CANGAROO, HESS, and GLAST.
Finally, the conclusions are left for Section~5.

\section{Astrophysical inputs and adiabatic compression}

As discussed in the Introduction, we are interested in the
annihilation of dark matter particles 
in the galactic 
center \cite{center}.
There are two possible types of gamma rays that can be produced
by the annihilation. First, gamma-ray lines
from processes $\chi \chi \to \gamma \gamma$ 
\cite{Ullio1} and $\chi \chi \to \gamma Z$ \cite{Ullio2}.
This signal would be very clear since the photons are
basically mono-energetic. Unfortunately,
the neutralino does not couple directly to the photon,
the Feynman diagrams are loop suppressed, and therefore
the flux would be small \cite{Berezinsky}.
For more recent analyses of this possibility see 
Refs.~\cite{Ullio3,Wang,Mambrini:2004ke}.

On the other hand, continuum gamma rays produced
by the decay of neutral pions generated in the cascading
of annihilation products
will give rise to
larger fluxes \cite{Salati}.
We will concentrate on this possibility in the following.

%Although 
%we will concentrate on this possibility in the following, 
%we have seen than adiabatic compression would be able to
%give mono--energetic signals which reach the sensitivity of GLAST.

%$10^{-13}$ cm$^{-2}$\ s$^{-1}$, and therefore much below the sensitivity of GLAST.

Let us then discuss the theoretical formulas for the computation of
the continuum gamma rays, and in particular
the astrophysical inputs one has to use.
As we will see below, the gamma-ray flux 
can be significantly enhanced for specific
dark matter density profiles.

\subsection{Gamma-ray flux}
\label{gammaray}

For the continuum of gamma rays, the
observed differential flux at the Earth  
coming from a direction forming an angle
$\psi$ with respect to the galactic center is
\begin{equation}
%\frac{d \Phi_{\gamma}}{d \Omega d E}
\Phi_{\gamma}(E_{\gamma}, \psi)
=\sum_i 
%\frac{1}{2}
\frac{dN_{\gamma}^i}{dE_{\gamma}}
 \langle\sigma_i v\rangle \frac{1}{8 \pi m_{\chi}^2}\int_{line\ of\ sight} \rho^2
% (r(l,\psi))
\ dl\ ,
\label{Eq:flux}
\end{equation}
\noindent 
where the discrete sum is over all dark matter annihilation
channels,
$dN_{\gamma}^i/dE_{\gamma}$ is the differential gamma-ray yield,
$\langle\sigma_i v\rangle$ is the annihilation cross section averaged over its
velocity
distribution, $m_{\chi}$ is the mass of the dark matter particle,
and $\rho$ is the dark matter density. 
Assuming a spherical halo, one has
$\rho=\rho(r)$ with the galactocentric distance
$r^2=l^2+R_0^2-2lR_0 \cos \psi$,
where $R_0$ is the solar distance to the galactic center
($\simeq$ 8 kpc).
It is customary 
to separate in the above equation
the particle physics part from the halo model dependence introducing the
(dimensionless) quantity
\begin{equation}
J(\psi)=\frac{1}{8.5 ~\mathrm{kpc}}
\left(
\frac{1}{0.3 ~\mathrm{GeV/cm^3}}
\right)^2
\int_{line\ of\ sight}\rho^2(r(l,\psi))\ dl\ .
\end{equation}
Thus one can write
% gamma-ray flux can now be expressed as 
\begin{eqnarray}
\Phi_{\gamma}(E_{\gamma}, \psi)
%\Phi_{\gamma}(E_{thr})
& = & 
0.94\times 10^{-13}\ \mathrm{cm^{-2}\ s^{-1}\ GeV^{-1}\ sr^{-1}} 
\nonumber \\
&\times & \mbox{}
%\nonumber
%\\
%\~A  
%\frac{1}{2}
\sum_i
%\int_{E_{thr}}^{m_{\chi}}dE_{\gamma}
\frac{dN_{\gamma}^i}{dE_{\gamma}}
\left(
\frac{\langle\sigma_i 
v\rangle}{10^{-29} {\mathrm{cm^3 s^{-1}}}}
\right)
\left(
\frac{100 ~\mathrm{GeV}}{m_{\chi}}
\right)^2
{J}(\psi)\ .
\label{Eq:totflux}
\end{eqnarray}
Actually, when comparing to experimental data
one must consider the integral of $J(\psi)$
over the spherical region of solid angle 
$\Delta \Omega$ given by the angular acceptance of the detector
which is pointing towards the galactic center, i.e. the quantity
$\bar{J}(\Delta \Omega)$ with 
\begin{equation}
\bar{J}(\Delta \Omega)\equiv \frac{1}{\Delta \Omega}\int _{\Delta \Omega}
J(\psi)\ d\Omega\ ,
\label{Jbarr}
\end{equation}
must be used in the above equation.
For example, for EGRET $\Delta \Omega$  is about
$10^{-3}$ sr whereas for GLAST, CANGAROO, and HESS it is $10^{-5}$ sr.

The gamma-ray flux can now be expressed as 
\begin{eqnarray}
\Phi_{\gamma}(E_{thr})
& = & 
0.94\times 10^{-13}\ \mathrm{cm^{-2}\ s^{-1}} 
\nonumber \\
&\times & \mbox{}
%\nonumber
%\\
%\~A  
%\frac{1}{2}
\sum_i
\int_{E_{thr}}^{m_{\chi}}dE_{\gamma}\frac{dN_{\gamma}^i}{dE_{\gamma}}
\left(
\frac{\langle\sigma_i 
v\rangle}{10^{-29} {\mathrm{cm^3 s^{-1}}}}
\right)
\left(
\frac{100 ~\mathrm{GeV}}{m_{\chi}}
\right)^2
\bar{J}(\Delta \Omega) \Delta \Omega\ ,
\label{Eq:totflux}
\end{eqnarray}
where $E_{thr}$ is the lower threshold energy of the detector,
Concerning the upper limit of the integral, note that
neutralinos move at galactic velocity and therefore
their annihilation occurs at rest.

\subsection{Halo models}
\label{crucial}

A crucial ingredient for the calculation of 
$\bar{J}$ in (\ref{Jbarr}), and therefore for the calculation of the flux of gamma rays,
is the dark matter density profile of our galaxy. 
The different profiles that have been proposed in the literature
can be parameterised as \cite{model}
\begin{equation}
\rho(r)= \frac{\rho_0  [1+(R_0/a)^{\alpha}]^{(\beta-\gamma)/\alpha}  }{(r/R_0)^{\gamma} 
[1+(r/a)^{\alpha}]^{(\beta-\gamma)/\alpha}}\ ,
\label{profile} 
\end{equation}
where $\rho_0$ is the local (solar neighborhood) 
halo density 
% ($\simeq$ 0.3 GeV/cm$^3$), 
and 
$a$ is a characteristic length.
Although we will use $\rho_0=0.3$ GeV/cm$^3$ throughout the paper,
since this is just a scaling factor in the analysis,
modifications to its value can be straightforwardly taken into account
in the results.
Highly cusped profiles are deduced
from $N$-body simulations\footnote{For analytical derivations see e.g. 
the recent work \cite{Hansen}, and references therein.}.
In particular, NFW \cite{Navarro:1996he} 
obtained 
a profile with a behaviour $\rho(r)\propto r^{-1}$ at small
distances.
A more singular behaviour,  $\rho(r)\propto r^{-1.5}$, was obtained by
Moore et al. \cite{Moore:1999gc}. 
However, these predictions are valid only for halos 
without baryons. One can improve simulations in a more realistic way by taking into
account the effect of the normal gas (baryons). This loses its 
energy through radiative
processes falling to the central region of forming galaxy.
As a consequence of this redistribution of mass,
the resulting gravitational potential is deeper, and the dark matter
must move closer to the center increasing its density. 

This increase in the dark matter density is often treated using
adiabatic invariants. The present form of the adiabatic compression model
was numerically and analytically studied by Blumental et
al. \cite{Blumenthal:1985qy}. 
This model assumes spherical symmetry, circular orbit for
the particles, and conservation of the angular momentum $M(r) r = $ const., where
$M(r)$ is the total mass enclosed within radius $r$. The mass distributions in the initial 
and final configurations are therefore related by 
$M_i(r_i) r_i = [M_b(r_f) + M_{DM}(r_f)] r_f$,
where $M_i(r)$, $M_b(r)$ and $M_{DM}(r)$ are the mass profile
of the galactic halo before the cooling of the baryons (obtained through $N$-body simulations),
the baryonic composition of the Milky Way observed now, and the to be determined
dark matter component of the halo today, respectively. 
This approximation was tested in numerical simulations \cite{Jesseit,Gnedin}.
Nevertheless, a more precise approximation can be obtained including
the possibility of 
elongated orbits \cite{Prada:2004pi}. In this case, the mass inside the
orbit, $M(r)$, is smaller than the real mass, the one the particle `feels' during 
its revolution around the galactic center. As a consequence, the
modified compression model 
is based on the conservation of the product $M(\overline r) r$, where $\overline{r}$
is the averaged radius of the orbit. The time average radii is given by 
$\overline{x} \sim 1.72 x^{0.82}/(1 + 5 x)^{0.085}$, with $x \equiv r/r_s$, and
$r_s$ the characteristic radius of the assumed approximation. 

\begin{figure}
\begin{center}
\begin{tabular}{cc}
\psfrag{rho}[c][r][1][90]{\scriptsize $\log{[\rho_{DM}(r)]}$}
 \psfrag{logRkpc}[c][c]{\scriptsize $\log{[R({\rm kpc})]}$}
 \psfrag{Fit}[c][c]{\scriptsize {\darkblue Fit ($a=20,
 \alpha=1,\beta=3,\gamma=1$)}}
\psfrag{Interpolation}[c][c]{\scriptsize {\green Interpolation}}
\psfrag{DATAS}[l][r]{\scriptsize {\darkblue DATA}}
\psfrag{NFW}[l][r]{\scriptsize {\bf a)}  NFW}
\psfrag{uncompressed}[c][c]{\scriptsize }
\includegraphics[width=0.5\textwidth]{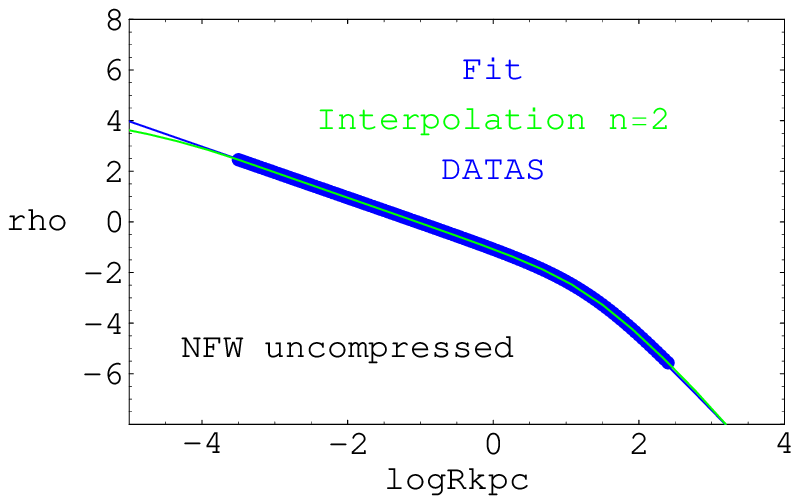}&
\psfrag{rhoDM}[c][r][1][90]{\scriptsize $\log{[\rho_{DM}(r)]}$}
 \psfrag{logRkpc}[c][c]{\scriptsize $\log{[R({\rm kpc})]}$}
\psfrag{Fit}[c][c]{\scriptsize {\darkblue Fit ($a=28, \alpha=1.5,\beta=3,\gamma=1.5$)}}
\psfrag{Interpolation}[c][c]{\scriptsize {\green Interpolation}}
\psfrag{DATAS}[c][c]{\scriptsize {\darkblue DATA}}
\psfrag{Moore}[l][r]{\scriptsize {\bf b)}  Moore et al}
\psfrag{uncompressed}[c][c]{\scriptsize }
\includegraphics[width=0.5\textwidth]{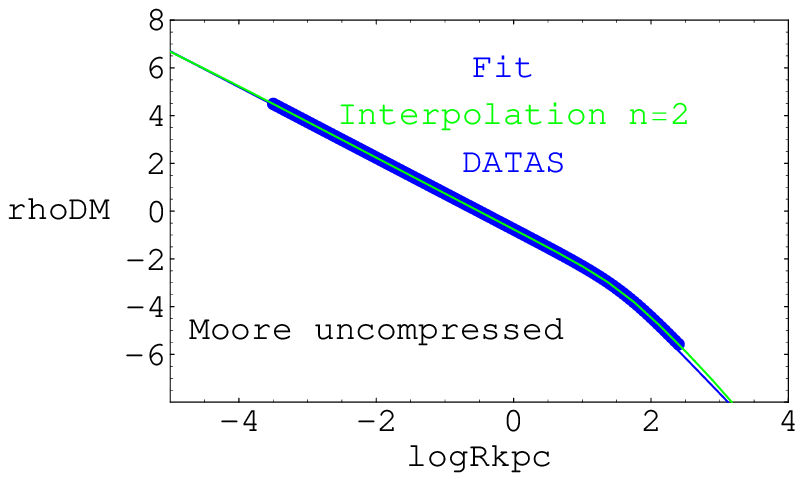}\\
%a) & b)\\
\psfrag{rho}[c][r][1][90]{\scriptsize $\log{[\rho_{DM}(r)]}$}
 \psfrag{logRkpc}[c][c]{\scriptsize $\log{[R({\rm kpc})]}$}
\psfrag{Fit}[c][c]{\scriptsize {\darkred Fit ($a=20, \alpha=0.8,\beta=2.7,\gamma=1.45$)}}
\psfrag{Interpolation}[c][c]{\scriptsize {\green Interpolation}}
\psfrag{DATAS}[c][c]{\scriptsize {\darkred DATA}}
\psfrag{NFW}[l][r]{\scriptsize {\bf c)}  ${\rm NFW_c}$}
\psfrag{compressed}[c][c]{\scriptsize}
 \includegraphics[width=0.5\textwidth]{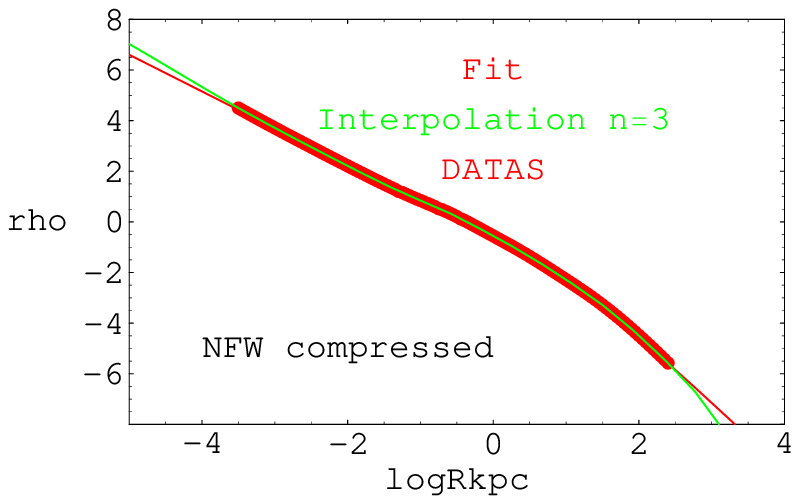}&
\psfrag{rhoDM}[c][r][1][90]{\scriptsize $\log{[\rho_{DM}(r)]}$}
 \psfrag{logRkpc}[c][c]{\scriptsize $\log{[R({\rm kpc})]}$}
\psfrag{Fit}[c][c]{\scriptsize {\darkred Fit ($a=28, \alpha=0.8,\beta=2.7,\gamma=1.65$)}}
\psfrag{Interpolation}[c][c]{\scriptsize {\green Interpolation}}
\psfrag{DATAS}[c][c]{\scriptsize {\darkred DATA}}
\psfrag{Moore}[l][r]{\scriptsize {\bf d)}  ${\rm Moore_c}$}
\psfrag{compressed}[r][r]{\scriptsize }
\includegraphics[width=0.5\textwidth]{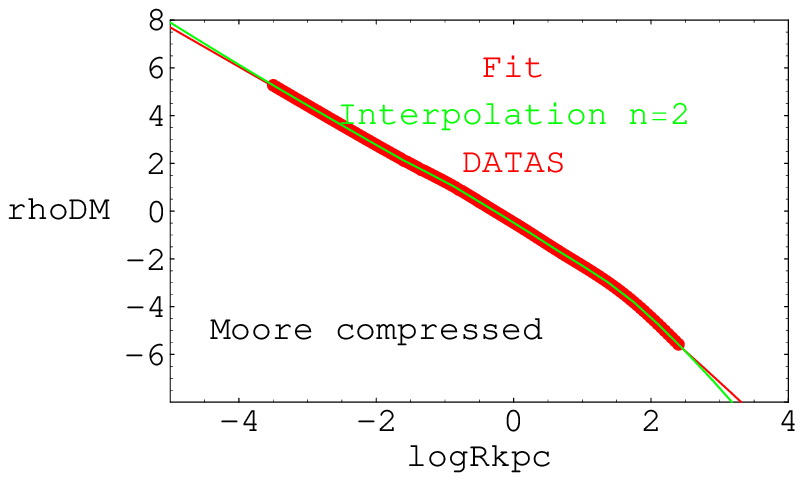}\\
%c) & d)
\end{tabular}
\caption{\small Fit, interpolation and data for: a)  NFW,  b)  Moore
 et al., c)  NFW compressed,  and d)  Moore et al. compressed .}
\label{fig:fitdatainter}
\end{center}
\end{figure}

The models and constraints
that we used in this work for the Milky Way can be found in Table I 
of Ref.~\cite{Prada:2004pi}.
We have fitted the resulting data of that work 
with the power-law parameterisation of eq.~(\ref{profile}).
The results are shown in Fig.~\ref{fig:fitdatainter},
and summarized in Table~\ref{tab}.
There we label the resulting NFW and Moore et al. profiles with adiabatic
compression by $\rm{NFW}_c$ and $\rm{Moore}_c$, respectively.
As one can see, at small $r$ the dark matter density profile following the adiabatic cooling
of the baryonic fraction is a steep power law $\rho \propto r^{-\gamma_c}$
with $\gamma_c \approx 1.45 (1.65)$ for a $\rm{NFW_c}$($\rm{Moore_c}$)
compressed 
model\footnote{It is worth noticing that we obtain 
the same order of magnitude as the authors of Ref.~\cite{gfandmerrit}.}.
We observe for example that the effect of the adiabatic compression on
a NFW profile is basically 
to transform it into a Moore et al. one (see Table \ref{tab}).

Let us remark that
NFW and Moore et al. profiles can be considered as a lower and upper
limit,
respectively. For example, the one recently proposed by Diemand, Moore
and Stadel in Ref.~\cite{Diemand:2004wh}, where $\rho(r)\propto
r^{-1.16}$,
is between both `standard' profiles.
It will be interesting for the analysis in 
Sect.~\ref{comparison} to use
an average value of  $\bar J$ for compressed scenarios.
In particular, we will consider a value $\bar J'=5\times  \bar J_{\rm{NFW_c}}$,
and we will refer to this scenario as $\rm{NFW'_c}$ for simplicity. 
It is worth noticing in this sense than some mechanisms can be advocated to reduce
the effect of the 
compression like angular momentum transfer 
to dark matter or formation of the central black hole by spiraling and merging of two
black holes. The study of such complex mechanisms is far beyond the scope of this work.

Of course, these results have important implications 
for the computation of the gamma-ray fluxes from the galatic 
center. 
In particular, in Table~\ref{tab}
we see that for 
$\Delta \Omega=10^{-3}$ sr 
one has
$\bar{J}_{\rm{NFW_c}}/ \bar{J}_{\rm{NFW}} \simeq 145$ and
$\bar{J}_{\rm{Moore_c}}/ \bar{J}_{\rm{Moore\ et\ al.}}\simeq 77$.
Thus the effect of the adiabatic compression is very strong,
increasing the gamma-ray fluxes about two orders of magnitude.
Similarly, 
for $\Delta \Omega=10^{-5}$ sr one obtains
$\bar{J}_{\rm{NFW_c}}/ \bar{J}_{\rm{NFW}} \simeq 953$ and
$\bar{J}_{\rm{Moore_c}}/ \bar{J}_{\rm{Moore\ et\ al.}}\simeq 95$.

Let us finally remark that the  $\bar{J}$
calculation has been regulated by assuming a constant density for
$r<10^{-5}$ kpc.
This procedure has consequences essentially for
divergent $\bar{J}$
when $\gamma \geq$ 1.5.
To check our results we have also calculated $\bar{J}$ in
the following way.
We used a cut off in density, $\rho(r<R_{s})=0$,  $R_{s}$ being the
Schwarchild radius of a $3 \times 10^6 M_{\odot}$ black hole. In addition we
also considered dark matter annihilation and estimate its effect by requiring an upper
value for dark matter density,
$\rho(r<r_{max})=\rho(r_{max})=m_{\chi}/(\langle \sigma
v \rangle.t_{BH})$ ( $t_{BH}\sim10^{10}$ yr, being the age of the black
hole). We also applied the same procedure with a more realistic
  cut-off value
$\rho(r<10^{-6}\ {\rm pc})=0$ suggested by dark matter particle
scatterings on stars \cite{merrit04}. 
This kind of effects can be significant only for
the Moore$_c$
halo model we consider in this paper and increase our
$\bar{J}$ 
value by
less than an order of magnitude for all the $\langle \sigma
v \rangle$ values corresponding to our wide SUSY parameter range.
In this sense our analysis is conservative.
More
refine treatments of dark matter annihilation effects in the 
innermost region of the 
galaxy can be found in \cite{gfandmerrit,Athanassoula:2005dw}.

\begin{center}
\begin{table}
\centering
\begin{tabular}{|c|cccccc|}
\hline 
&a (kpc)&$\alpha$&$\beta$&$\gamma$&$\bar{J}(10^{-3} {\rm sr}) $&$\bar{J}(10^{-5} {\rm sr})$  \\
\hline 
NFW & 20 & 1 & 3 & 1 & $1.214 \times 10^3$ & $ 1.264 \times  10^4$\\
$\rm{NFW_c}$ & 20 & 0.8 & 2.7 & 1.45 & $1.755  \times 10^5$ & $ 1.205  \times 10^7$\\
Moore et al. & 28 & 1.5 & 3 & 1.5 & $1.603  \times 10^5$ & $ 5.531  \times 10^6$\\
$\rm{Moore_c}$ & 28 & 0.8 & 2.7 & 1.65 & $1.242  \times 10^7$ & $ 5.262  \times 10^8$\\
\hline 
\end{tabular}
\caption{NFW and Moore et al. 
density profiles without 
and with
adiabatic compression ($\rm{NFW}_c$ and $\rm{Moore_c}$ respectively)
with the corresponding parameters, and values of $\bar{J}(\Delta\Omega)$.}
%for $\Delta\Omega=10^{-3}, 10^{-5}\ {\rm sr}$.}
\label{tab}
\end{table}
\end{center}

\section{Neutralino dark matter}

For the computation of the gamma-ray flux 
discussed in the previous section, in addition to the halo profile
it is also crucial the value of the annihilation cross section
$\sigma_i$.
Of course, for determining this value the theoretical framework must be established.

As mentioned in the Introduction, we will work in the context of the
MSSM. Let us recall that in this framework
there are four neutralinos, $\tilde{\chi}^0_i~(i=1,2,3,4)$, 
since they are the physical superpositions of the fermionic partners of the neutral
electroweak gauge bosons, called bino ($\tilde{B}^0$) and wino ($\tilde{W}_3^0$), 
and of the fermionic partners of the  
neutral Higgs bosons, called Higgsinos ($\tilde{H}^0_u$, $\tilde{H}_d^0$). 
Thus one can express the lightest neutralino as
\begin{equation}
\tilde{\chi}^0_1 = {Z_{11}} \tilde{B}^0 + {Z_{12}} \tilde{W}_3^0 +
{Z_{13}} \tilde{H}^0_d + {Z_{14}} \tilde{H}^0_u\ .
\label{lneu}
\end{equation}
It is commonly defined that $\tilde{\chi}^0_1$  is mostly gaugino-like 
if $P\equiv \vert {Z_{11}} \vert^2 + \vert {Z_{12}}  \vert^2 > 0.9$, 
Higgsino-like
if $P<0.1$, and mixed otherwise.

\begin{figure}
%    \begin{center}
\centerline{
%      \hspace{-0.5cm} 
\epsfig{file=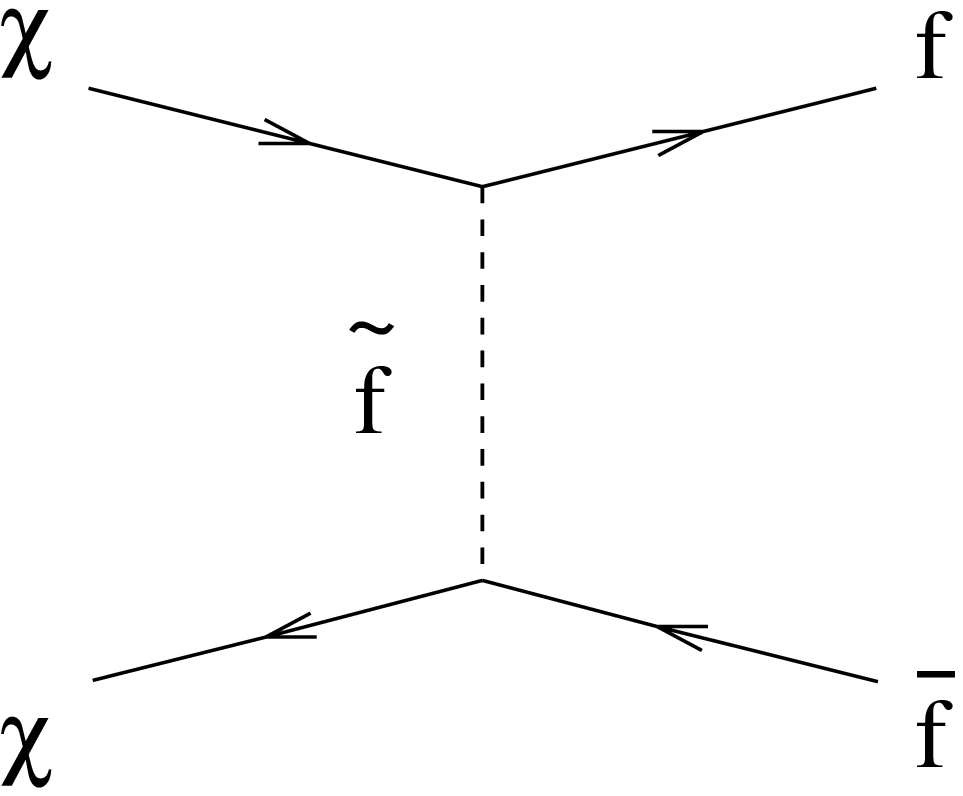,width=0.15\textwidth}\hskip 1cm
    \hspace{1cm}     \epsfig{file=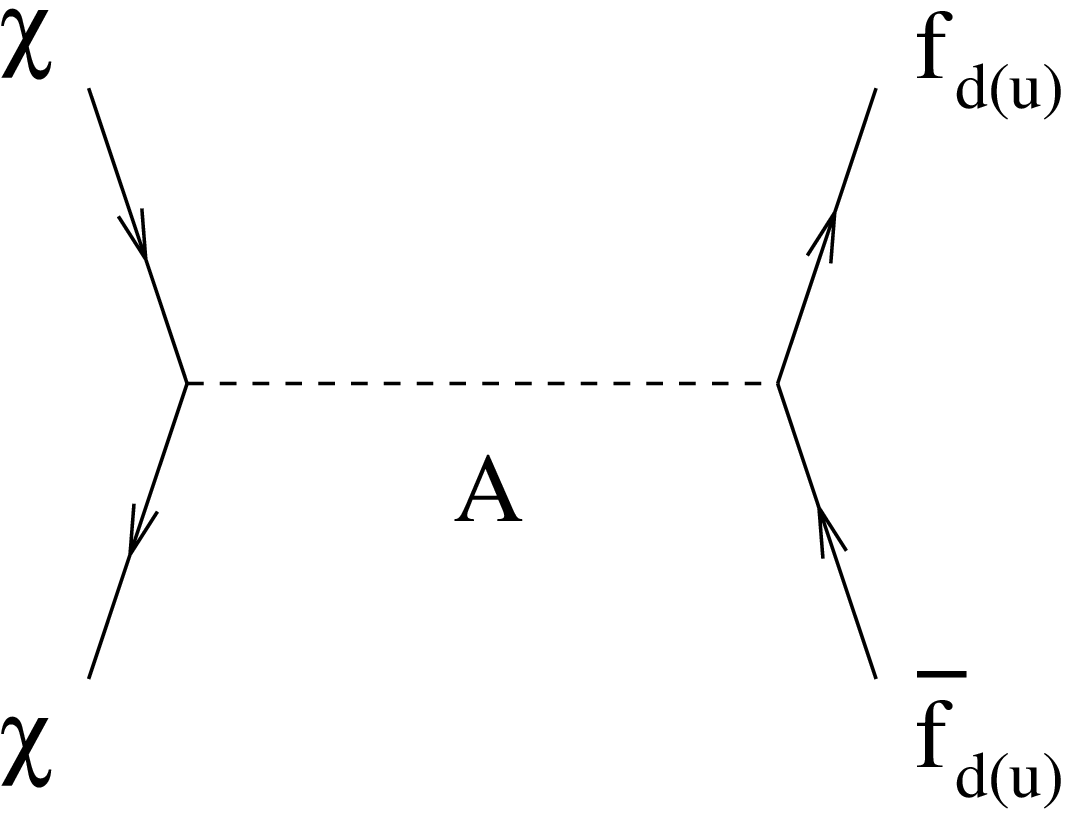,width=0.15\textwidth}\hskip 1cm
       \epsfig{file=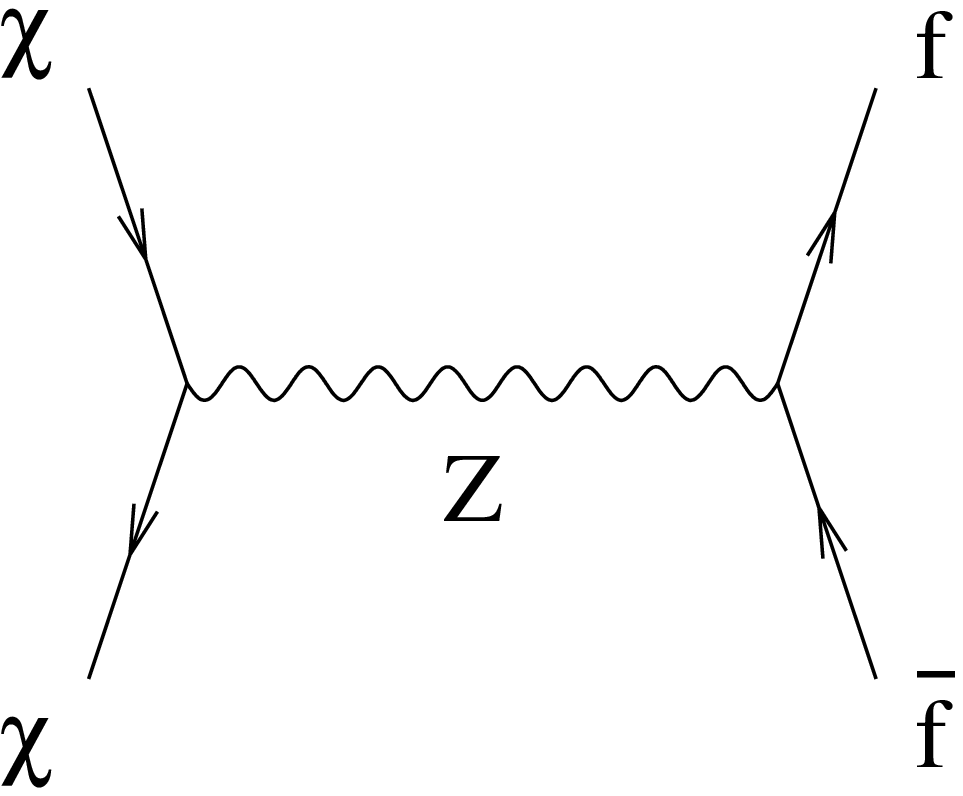,width=0.15\textwidth}\hskip 1cm
      \hspace{1.5cm} 
\epsfig{file=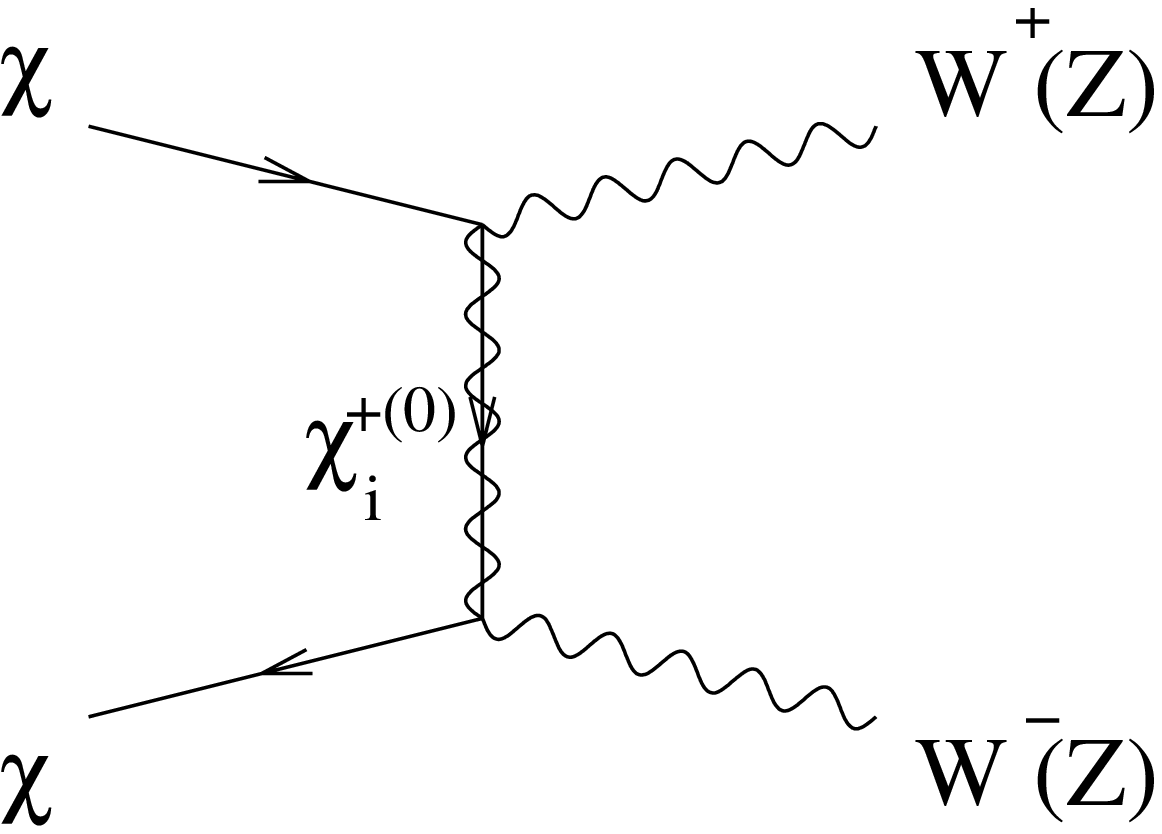,width=0.15\textwidth}\hskip 1cm
        }
 $\propto \frac{m_{\chi} m_f}{m_{\tilde f}^2}Z_{11}^2~~~$
%\hskip cm
$\propto \frac{m_{\chi}^2}{m_A^2}
\frac{Z_{11}Z_{13,14}}{m_W}  m_{f_d} \tan \beta (\frac{m_{f_u} }{\tan\beta})~ 
~~$
%\hskip 1cm
$\propto \frac{m_f m_{\chi}}{m_Z^2} Z_{13,14}^2~$
%\hskip 1cm 
$\propto \frac{[-Z_{14} V_{21}^* + \sqrt{2} Z_{12} V_{11}^*]^2 (-Z_{13} Z_{31}^* +  
Z_{14} Z_{41}^*)^2}{1+m_{\chi_i^{+(0)}}^2 /
m_{\chi}^2 - m_{W(Z)}^2 / m_{\chi}^2}$
          \caption{{\footnotesize Dominant neutralino annihilation
diagrams. Relevant parts of the amplitudes are shown explicitly
Terms between parenthesis correspond to $f_u$ and $Z$ final states
in second and fourth diagrams. V and Z are chargino and neutralino
mixing
matrices.
}}
        \label{fig:feynmandetail}
%    \end{center}
\end{figure}

In figure~\ref{fig:feynmandetail} we show the relevant 
Feynman diagrams contributing to neutralino annihilation.
As was 
remarked e.g. 
in Ref.~\cite{Mambrini:2004ke,manunonU,Birkedal-Hansen:2002am,Bottino},
the annihilation cross section
can be significantly enhanced depending on the
SUSY model under consideration.
We will concentrate here on the 
SUGRA scenario, where the soft terms are
determined at the unification 
scale, $M_{GUT}\approx 2\times  10^{16}$ GeV, after SUSY breaking,
and radiative electroweak symmetry breaking is imposed.

%================= FIGURE 3ab : New Feynman Diagrams ===================

%=========================================================================

\subsection{Supergravity models}

Let us discuss first the mSUGRA scenario,
where the soft terms of
the MSSM are assumed to be universal at  $M_{GUT}$. 
Recall that in mSUGRA
one has only four free parameters:
the soft scalar mass $m$, the soft gaugino mass $M$, 
the soft trilinear coupling $A$, and 
the ratio of the Higgs vacuum
expectation values, 
$\tan\beta= \langle H_u^0\rangle/\langle H_d^0\rangle$.
%$\tan\beta$.
In addition, the sign of the Higgsino mass parameter, $\mu$,
remains also undetermined by the 
minimization of the Higgs potential.

Since in this scenario the lightest neutralino  $\tilde{\chi}^0_1$ is
mainly bino,
only  ${Z_{11}}$ is large and then the contribution of 
diagrams in Fig.~\ref{fig:feynmandetail} will be generically small,
the first being suppressed by $\tilde{f}$ mass.
As a consequence,
the predicted annihilation cross section $\sigma_i$ is small,
and therefore the flux is generically below the present accessible 
experimental regions (unless $\tan\beta\gsim 50$ 
combined with a Moore et al. profile are considered) 
%{\bf manu : this is
%  not really true for HP/focus point region ... }. 

However, as discussed in detail in 
Ref.~\cite{Mambrini:2004ke} in the context of indirect detection,
$\sigma_i$ can be increased 
in different
ways when the structure of mSUGRA for the soft terms is abandoned. 
In particular, it is possible to enhance the annihilation
channels involving exchange of the CP-odd Higgs, $A$, by reducing the
Higgs mass. In addition, it is also possible to  increase the Higgsino components of
the lightest neutralino ${Z_{13,14}}$. 
Thus annihilation channels through Higgs exchange become more important
than in mSUGRA.
This is also the case for $Z^-$, $\chi_1^\pm$, and  $\tilde{\chi}_1^0$-exchange
channels.
As a consequence, the gamma-ray flux will be increased.

In particular, the most important effects are produced
by the non-universality of Higgs and gaugino masses.
These can be parameterised, at $M_{GUT}$, as follows
\begin{equation}
  m_{H_{d}}^2=m^{2}(1+\delta_{1})\ , \quad m_{H_{u}}^{2}=m^{2}
  (1+ \delta_{2})\ ,
  \label{Higgsespara}
\end{equation}
and 
\begin{eqnarray}
  M_1=M\ , \quad M_2=M(1+ \delta'_{2})\ ,
  \quad M_3=M(1+ \delta'_{3})
  \ ,
  \label{gauginospara}
\end{eqnarray}
%respectively.
We will concentrate in our analysis on the following representative cases:
\begin{eqnarray}
a)\,\, \delta_{1}&=&0\ \,\,\,\,\,\,\,\,,\,\,\,\, \delta_2\ =\ 0
\,\,\,\,\,\,\,\,\,,\,\,\,\, 
\delta'_{2,3}\ =\ 0\ , 
%(mSUGRA)
\nonumber\\
b)\,\, \delta_{1}&=&0\ \,\,\,\,\,\,\,\,,\,\,\,\, \delta_2\ =\ 1\
\,\,\,\,\,\,\,\,,\,\,\,\, 
\delta'_{2,3}\ =\ 0\ , 
\nonumber\\
c)\,\, \delta_{1}&=&-1\ \,\,\,\,,\,\,\,\, \delta_2\ =\ 0\
\,\,\,\,\,\,\,\,,\,\,\,\, 
\delta'_{2,3}\ =\ 0\ , 
\nonumber\\
d)\,\, \delta_{1}&=&-1\ \,\,\,\,,\,\,\,\, \delta_2\ =\ 1\
\,\,\,\,\,\,\,\,,\,\,\,\, 
\delta'_{2,3}\ =\ 0\ \,\, , 
\nonumber\\
e)\,\, \delta_{1,2}&=&0\ \,\,\,\,\,\,\,\,,\,\,\,\, \delta'_{2}\ =\ 0
\,\,\,\,\,\,\,\,\,,\,\,\,\, 
\delta'_{3}\ =\ -0.5\ ,
\nonumber\\
f)\,\, \delta_{1,2}&=&0\ \,\,\,\,\,\,\,\,,\,\,\,\, \delta'_{2}\ =\ -0.5
\,\,\,\,\,\,\,\,\,,\,\,\,\, 
\delta'_{3}\ =\ 0\ .
\label{3cases}
\end{eqnarray}
\noindent Case {\it a)} corresponds to mSUGRA with universal soft terms,
cases {\it b)}, {\it c)} and  {\it d)} 
correspond to  non-universal Higgs masses, 
and finally cases {\it e)} and {\it f)} to non-universal gaugino masses.
The cases {\it b)},  {\it c)},  {\it d)}, and {\it e)} were discussed in Ref.~\cite
{Mambrini:2004ke},
and are known to produce gamma-ray fluxes larger than in mSUGRA, whereas case
{\it f)} will be of interest when discussing heavy WIMP signals predictions 
in the perspective of atmospheric Cherenkov telescopes like e.g. CANGAROO.

We will discuss all these cases taking into account in the 
computation of the gamma-ray flux the effect
of the adiabatic compression.
Except for the calculation of the ${\bar J}$ factor, 
for the evaluation of the fluxes we used
the last DarkSusy released version \cite{darksusynew}.
To solve the renormalization group equations for the soft
SUSY-breaking terms between $M_{GUT}$ and the electroweak scale,
we used the Fortran package SUSPECT \cite{Suspect}.

Depending on the non-universal case, the main parameters that enter in
the flux computation, namely the neutralino mass $m_{\chi}$ and 
the annihilation rate 
$N_{\gamma}\langle \sigma v \rangle \equiv 
\sum_{i}
\int_{E_{thr}}^{m_{\chi}}dE_{\gamma}\frac{dN_{\gamma}^i}{dE_{\gamma}}
\langle\sigma_i 
v\rangle$ 
can vary significantly 
within the parameter space. Running a scan on the gaugino mass parameter 
$M$ between 0 and 2 TeV, we expect a neutralino mass lying between 200 GeV and 
1 TeV, the lower bound being mainly restricted by accelerator constraints on
the chargino or lighter higgs mass. 
We can however reach 3 to 4 TeV neutralinos respecting all 
the experimental and cosmological constraints if we extend the range 
of the scanning on $M$ to 10 TeV. 
On the other hand, the WMAP relic density constraint restricts the 
annihilation cross section $N_{\gamma}\langle \sigma v \rangle$ to lie from
$\sim 1 \times 10^{-28} {\mathrm{cm^3 s^{-1}}}$ (when the
coannihilation process with the lighter stau dominates, cases {\it a)} 
or {\it c)} with low tan$\beta$) to 
$\sim 2 \times 10^{-26} {\mathrm{cm^3 s^{-1}}}$ (when the $A$-pole
region dominates, case {\it c)}, {\it d)} and {\it e)} with tan$\beta=35$)
\footnote{In some fine tuned regions of parameter space with especially 
large values of $m$ (Focus Point region) it is possible to reach 
$\langle \sigma v \rangle \sim 10^{-25} {\mathrm{cm^3 s^{-1}}}$
in the cases {\it b)}, {\it d)} and {\it e)}}.

Before carrying out the analysis in detail, it is worth remarking that in these
studies it is crucial to reproduce the correct phenomenology.
Thus, in the next section, the most recent experimental and
astrophysical
constraints which can affect the computation will be discussed.

\subsection{Experimental and astrophysical constraints}

We have taken into account in the computation 
the most recent experimental and astrophysical
bounds. These may produce important constraints 
on the parameter space of SUGRA models, restricting therefore the regions
where the gamma-ray flux has to be analyzed.

In particular, concerning the astrophysical bounds, the effect
of $0.1\lsim \relic h^2\lsim 0.3$,
on the  relic neutralino density computation has been considered.
Due to its relevance, the WMAP narrow range
$0.094\lsim\relic h^2\lsim 0.129$ has also been analyzed in detail.
Actually, we have also considered the possibility that not all the dark matter is made
of
neutralinos, allowing
$0.03<\Omega_{\tilde{\chi}^0_1} h^2<0.094$. In this case we have 
to rescale the density of 
neutralinos in the galaxy $\rho(r)$ in Eq.~(\ref{profile}) by a factor
$\Omega_{\tilde{\chi}^0_1}h^2/ 0.094$.

\begin{figure}[!]
    \begin{center}
\centerline{
       \epsfig{file=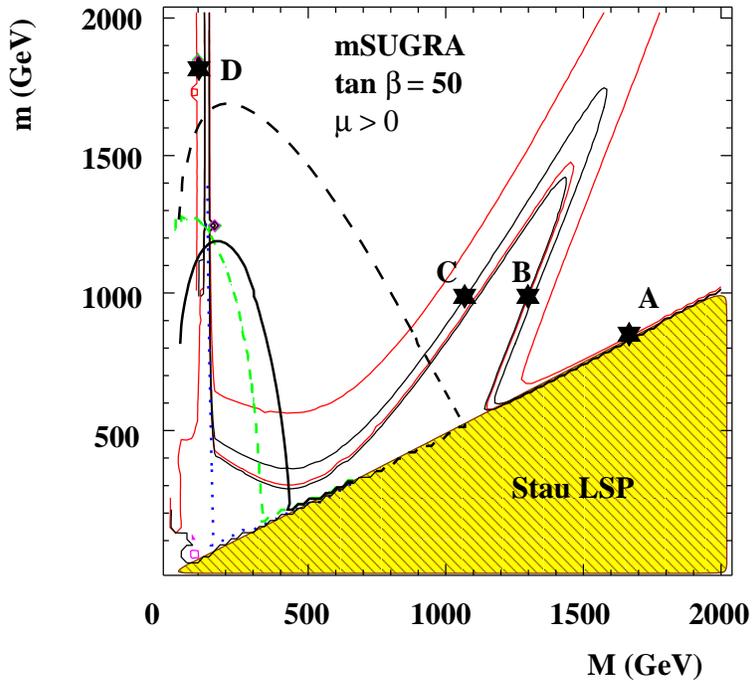,width=0.6\textwidth}
       }
          \caption{{\footnotesize 
Experimental and astrophysical bounds
in the parameter space of the mSUGRA 
scenario ($m$, $M$) for $\tan \beta=50$, $A=0$
and $\mu > 0$.
The region to the left of the 
light grey (green) dashed line
is excluded by the lower bound 
on the Higgs mass.
The region to the left of the dotted line
is excluded by the lower bound on the chargino mass
$m_{\tilde\chi_1^{\pm}}>103$ GeV.
The region to the right of the 
black dashed line
corresponds to
$\asusy< 7.1\times 10^{-10}$, and would be excluded by
$e^+e^-$ data.
The region to the left of the solid line is excluded by $b\to s\gamma$.
The region at the bottom (Stau LSP) is excluded because the lightest stau
is the LSP.
The region between  grey (red) contours fulfils
$0.1\leq \Omega_{\tilde{\chi}_1^0}h^2\leq 0.3$, whereas that between black contours
indicates the WMAP range $0.094<\Omega_{\tilde{\chi}_1^0}h^2<0.129$.
}}
        \label{fig:mSUGRAtb50scan}
    \end{center}
\end{figure}

We illustrate this effect in Fig.~\ref{fig:mSUGRAtb50scan}
for the case of mSUGRA, i.e. case {\it a)} of (\ref{3cases}).
There, the cosmologically allowed regions are shown in the parameter
space ($m$, $M$) for $\tan \beta=50$, $A=0$
and $\mu > 0$.
As discussed above, in most of the parameter space the lightest 
neutralino is mainly bino, and as a consequence the annihilation cross
section is small producing a too large relic abundance.
Nevertheless, let us recall that there are basically four corridors where the
above bounds can be satisfied.
There is
the narrow coannihilation branch of the parameter space, i.e. 
the region where the stau is the next to the LSP producing efficient
coannihilations (for\footnote{Recall that we can deduce these values in the
plot from the value of $M$, since in mSUGRA 
$m_{\tilde{\chi}^0_1}\simeq {M_1}\simeq 0.4\ M$.} $500\lsim m_{\tilde{\chi}^0_1}\lsim 800$ GeV). 
This reduces considerably the relic abundance and
place it inside the bounds. Point {\bf A} in the figure
corresponds to this situation.
Nevertheless, the 
neutralino annihilation in the galactic center is dominated by the $A-$Higgs exchange,
but being far from the $A-$pole region ($ 2 m_{\tilde \chi^0_1} \sim m_A$) the
gamma-ray flux is not very large compared with the ones produced in
the other regions to be discussed 
below (See Fig.~\ref{fig:GLAST}a below, where this narrow corridor can be
clearly distinguished).

Now,
for a given value of $m$, and from large to small $M$,
$m_A$ decreases. When $m_A$ becomes close to $2m_{\tilde{\chi}^0_1}$
the annihilation cross section increases and therefore the
relic density decreases entering in an allowed corridor (see e.g. point {\bf B}) with
$0.094<\Omega_{\tilde{\chi}^0_1}h^2<0.3$.
Large gamma-ray fluxes dominated by $b \overline{b}$ final
states through $A-$exchange are produced. 
For slightly smaller values of $M$ one finally arrives to the $A-$pole region where
$m_A\sim 2m_{\tilde{\chi}^0_1}$ and
the relic density is too small.
Smaller values of $m_A$ close this region, producing a decrease in the
annihilation cross section, which allows entry into  
the second corridor with the relic density inside the
observational bounds (see e.g. point {\bf C}). For $m_A$ too small this second corridor is closed,
and the relic density is too large,
$\Omega_{\tilde{\chi}^0_1}h^2>0.3$.

Finally, there is the narrow Higgsino (focus-point) corridor
(see e.g. point {\bf D})
close to the no electroweak symmetry breaking (nEWSB) region
where  $\mu^2$ becomes negative.
In particular, the value of the relic density 
%This affects the relic density
$\Omega_{\tilde{\chi}^0_1}$ is affected due to the increase of the
Higgsino
components of ${\tilde{\chi}^0_1}$ with respect to the dominant
bino component of most of the parameter space. 
Thus the relic abundance is placed
inside the astrophysical bounds.

The neutralino being mainly Higgsino couples strongly to the $Z$ boson 
(see Fig.~\ref{fig:feynmandetail}) giving rise to a large gamma-ray flux.
 For these points the neutralino is lighter than in the previous three cases.
 A full analysis of neutralino annihilation in the universal case parameter space can be found in \cite{Bertin:2002ky}.

Let us finally remark that qualitatively similar allowed corridors 
are found for the non-universal cases of (\ref{3cases}).
A detailed discussion can be found in Ref.~\cite{Mambrini:2004ke}.

Concerning the experimental constraints,
the lower
bounds on the masses of SUSY particles and on the
lightest Higgs have been implemented, as well as the experimental
bounds on the branching ratio of the \bsg\ process and on 
$\asusy$. We also 
illustrate these constraints in Fig.~\ref{fig:mSUGRAtb50scan}.
Note that we are using $\mu > 0$.
We will not consider in the calculation the opposite sign of $\mu$
because this would produce a negative contribution for the $g_{\mu}-2$,
and, as will be discussed below, we are mainly interested in positive
contributions. Recall that the sign of the contribution is basically
given by $\mu M_2$, and that $M$, and therefore $M_2$, can always
be made positive after performing an $U(1)_R$ rotation.

%============ FIGURE 2 : Univ.  tb5     =====================================

\begin{figure}[!]
    \begin{center}
%\vskip -3.5truecm
       \epsfig{file=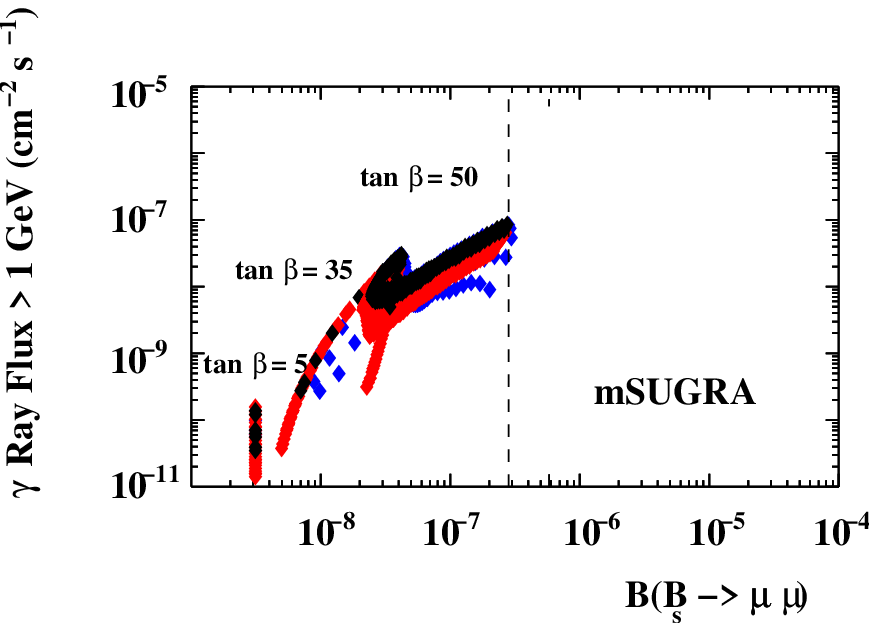,width=0.4\textwidth}\hskip
       1cm

       \epsfig{file=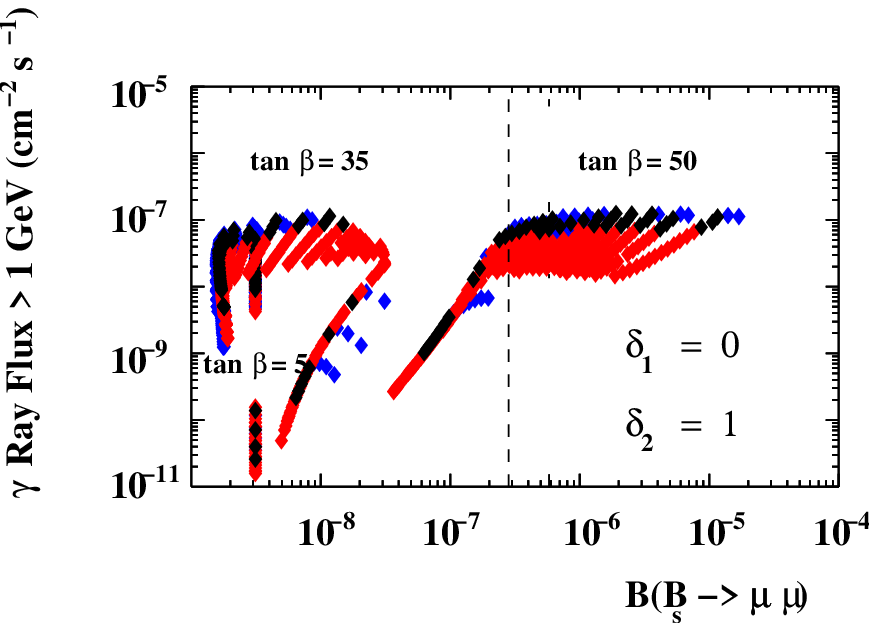,width=0.4\textwidth}\hskip 1cm
       \epsfig{file=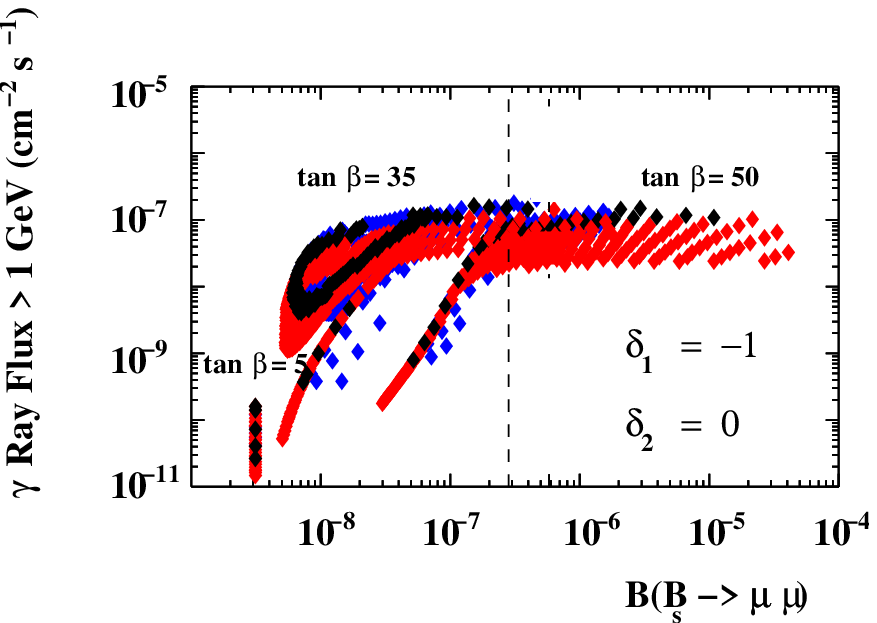,width=0.4\textwidth}%\hskip 1cm
          \caption{{\footnotesize
Gamma-ray flux for a threshold of 1 GeV  versus $B(B_s \to \mu^+ \mu^-)$ in
case  {\it a)} mSUGRA of Eq.~(\ref{3cases}) in the text,
and non universal Higgs cases 
{\it b)} $\delta_{1}=0, \delta_{2}=1$, and 
{\it c)} $\delta_{1}=-1, \delta_{2}=0 $.
Points depicted with light grey (red) points have 
$0.129<\Omega_{\tilde{\chi}^0_1}h^2<0.3$, 
with black points have
$0.094<\Omega_{\tilde{\chi}^0_1} h^2<0.129$,
and finally with dark grey (blue) points have
$0.03<\Omega_{\tilde{\chi}^0_1} h^2<0.094$ 
with the appropriate rescaling of the density of neutralinos
in the galaxy as discussed in the text. 
}}
        \label{fig:bsmumu}
    \end{center}
\end{figure}

For $\asusy$, we have taken into account the
recent experimental result for the muon
anomalous magnetic moment \cite{g-2}, as well as the most recent
theoretical evaluations of the Standard Model contributions
\cite{newg2}. It is found that when $e^+e^-$ data
are used the experimental excess in $(g_\mu-2)$ would constrain a
possible SUSY contribution to be
$\asusy=(27.1 \pm 10)\times 10^{-10}$. 
At $2\sigma$ level this implies 
$7.1\times 10^{-10}\lsim\asusy\lsim 47.1\times 10^{-10}$.
It is worth noticing here that when tau data are used a smaller
discrepancy with the experimental measurement is found.
In order not to exclude the latter possibility we will discuss the 
relevant value
$\asusy= 7.1\times 10^{-10}$ 
%in the figures 
throughout the paper. For example, in Fig.~\ref{fig:mSUGRAtb50scan}
the region of the parameter space to the right of this (black dashed) line
will be forbidden (allowed) if one consider electron (tau) data.

On the other hand,
the measurements of $B\to X_s\gamma$ decays 
at 
CLEO \cite{cleo} 
and BELLE \cite{belle},
lead to bounds on the branching ratio 
$b\to s\gamma$. In particular we impose on our computation
$2.33\times 10^{-4}\leq BR(b\to s\gamma)\leq 4.15\times 10^{-4}$, where the evaluation 
is carried out using the routine provided by
the program micrOMEGAs \cite{micromegas}.
This program is also used for our evaluation of $\asusy$ and 
relic neutralino density.

Finally, we have considered in the computation the experimental limit on the
$B_s \to \mu^+ \mu^-$
branching ratio \cite{bmumuexp}.
Although the upper bound on this process 
B($B_s \to \mu^+ \mu^-$) $< 2.9\times 10^{-7}$
does not constrain the parameter space of mSUGRA,
it has been stressed recently that in the non-universal cases
giving rise to large direct detection cross sections
the constraints can be very important \cite{ko,ko2}.
The reason being that there is a strong correlation between both
observables.
The branching ratio as well as the cross section increase for
low values of the Higgs masses (and large $\tan\beta$).

However, in the case of indirect detection of neutralinos through
gamma-ray fluxes the correlation is more diluted.
In particular, when the neutralino has an important
Higgsino component,
the $Z-$exchange channel dominates and as a consequence 
there is no a direct link between the value of the
gamma-ray flux and B($B_s \to \mu^+ \mu^-$).
We can see this in 
Fig.~\ref{fig:bsmumu},
where the gamma-ray flux versus the branching ratio is shown for 
cases {\it a), b), c)} of (\ref{3cases}).
As mentioned above, for mSUGRA the parameter space is not constrained.
Although this is not the case for the non-universal examples,   
note that still large values of the gamma-ray fluxes can be obtained
fulfilling the upper bound for B($B_s \to \mu^+ \mu^-$).

\section{Comparison with the experiments}
\label{comparison}

In this section we will compare the theoretical predictions for the
indirect detection of neutralinos through gamma-ray fluxes, with
the sensitivity of present and future experiments.
As discussed in the Introduction, several experiments have already
data that apparently cannot be explained with the usual gamma-ray background.
Given this situation the comparison between the theoretical
computation
%of gamma rays produced by neutralino annihilation,
and the experimental data is crucial.
To carry this out we will apply the formulas of Sect.~\ref{gammaray}
to the halo models with baryonic compression discussed in Sect.~\ref{crucial},
in the context of a general SUGRA theory where non-universal soft
terms can be present.
We will discuss first the case of space-based detectors such as 
EGRET and GLAST, and finally
atmospheric telescopes such as CANGAROO and HESS.

\subsection{EGRET}

The EGRET telescope 
on board of the Compton Gamma-Ray Observatory 
has carried out the first all-sky survey in high-energy
gamma-rays ($\approx$ 30 MeV -- 30 GeV) 
%up to an energy of about 20 GeV 
over a period of 5 years,
from April 1991 
until September 1996.
As a result of this survey, it has detected
a signal \cite{EGRET} 
above about 1 GeV,
with a value for the flux of about
$10^{-8}$ cm$^{-2}$\ s$^{-1}$,
that apparently cannot be explained with the 
usual gamma-ray 
background\footnote{Although alternative explanations
for this result have been proposed 
modifying conveniently  
the standard theory of galactic gamma ray \cite{modi}, these
turn out to be also controversial \cite{modi2}.}. 
The source, 
possibly diffuse rather than pointlike, is located within the $1.5^o$
($\Delta \Omega \sim 10^{-3}$ sr) of the galactic center.
Due to the lack
of precision data in the high energy bins, it seems impossible however to distinguish
any annihilation channel leading to this photon excess.

In recent papers  \cite{UllioEGRET,Bottino}
studies of the perspective of interpreting the spectral features of
EGRET data as produced by neutralino annihilation in the galactic 
center\footnote{see also \cite{Wboer} for a general analysis.}, 
were carried out.
In particular, in Ref.~\cite{UllioEGRET} an interesting analysis was carried
out in the context of a toy dark matter model assuming an annihilation
cross section $\sigma v \sim 3\times 10^{-26} \mathrm{cm}^3 \mathrm{s}^{-1}$
scaling with the inverse of the relic abundance. 
Also neutralino dark matter in the mSUGRA context was considered, 
including experimental
constraints coming from accelerator physics. 
In  Ref.~\cite{Bottino} the analysis was carried out in the framework
of an effective MSSM with non-universal gaugino masses,
obtaining that the EGRET excess cannot be reproduced with a  NFW profile.

We will show in this section that it is possible to
reproduce the spectral features of EGRET data
in the framework of general SUGRA, 
when a NFW profile including adiabatic compression, NFW$_c$, is considered.

\begin{figure}[!]
    \begin{center}
%\centerline{
\vskip -4.truecm
       \epsfig{file=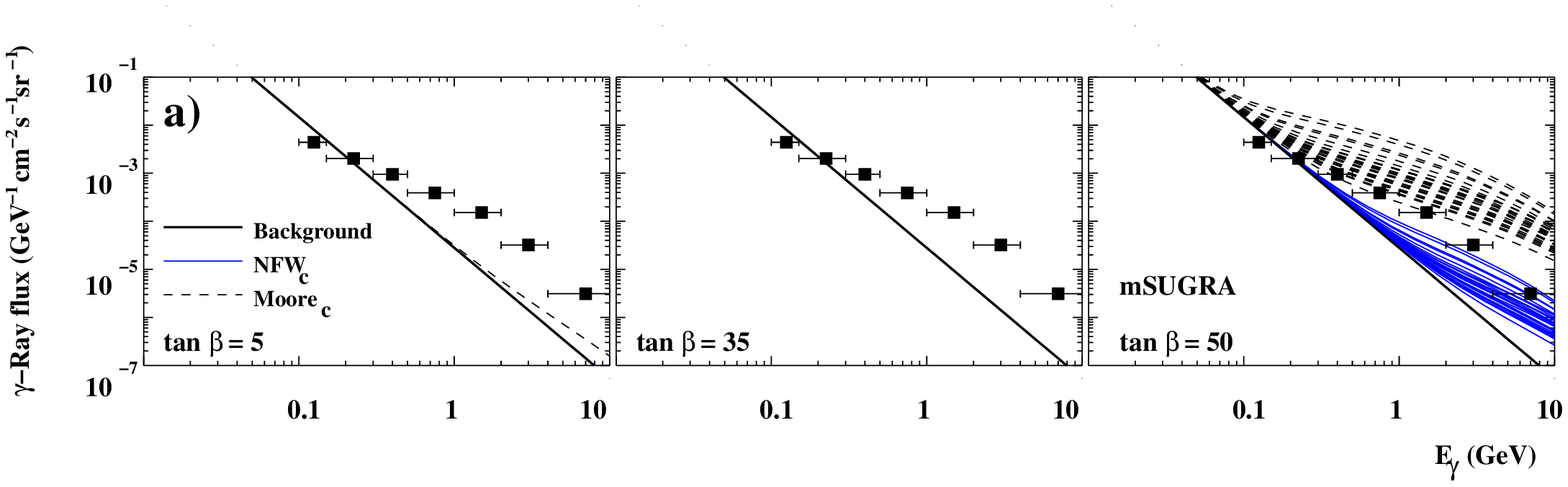,width=1.\textwidth}%\hskip 1cm
	\vskip -0.5cm
       \epsfig{file=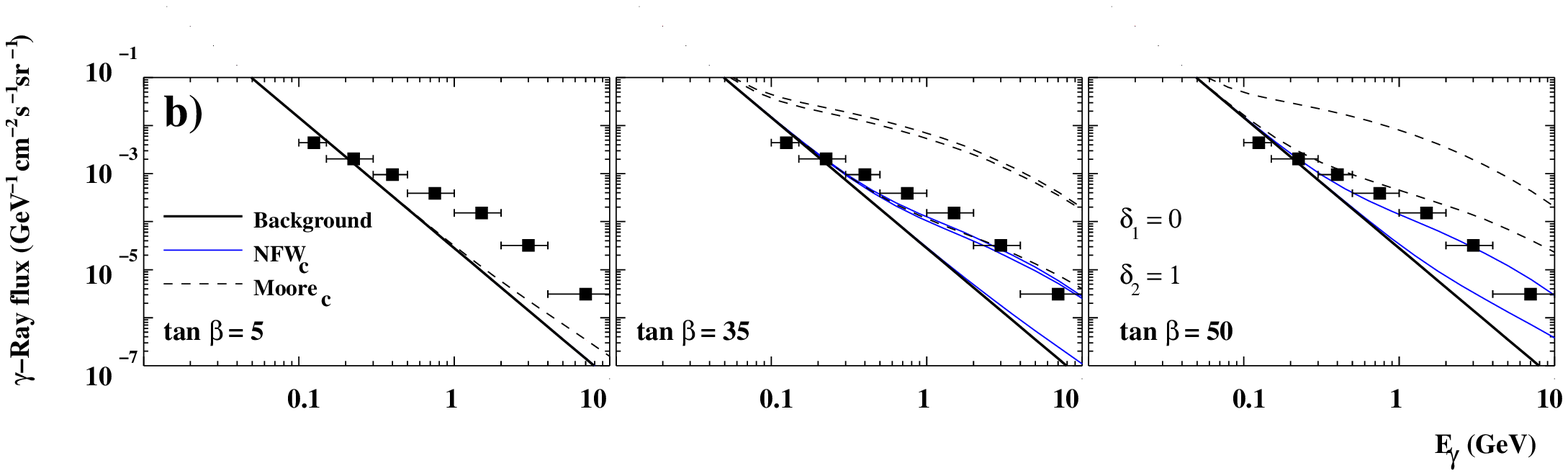,width=1.\textwidth}%\hskip 1cm
	\vskip -0.5 cm
       \epsfig{file=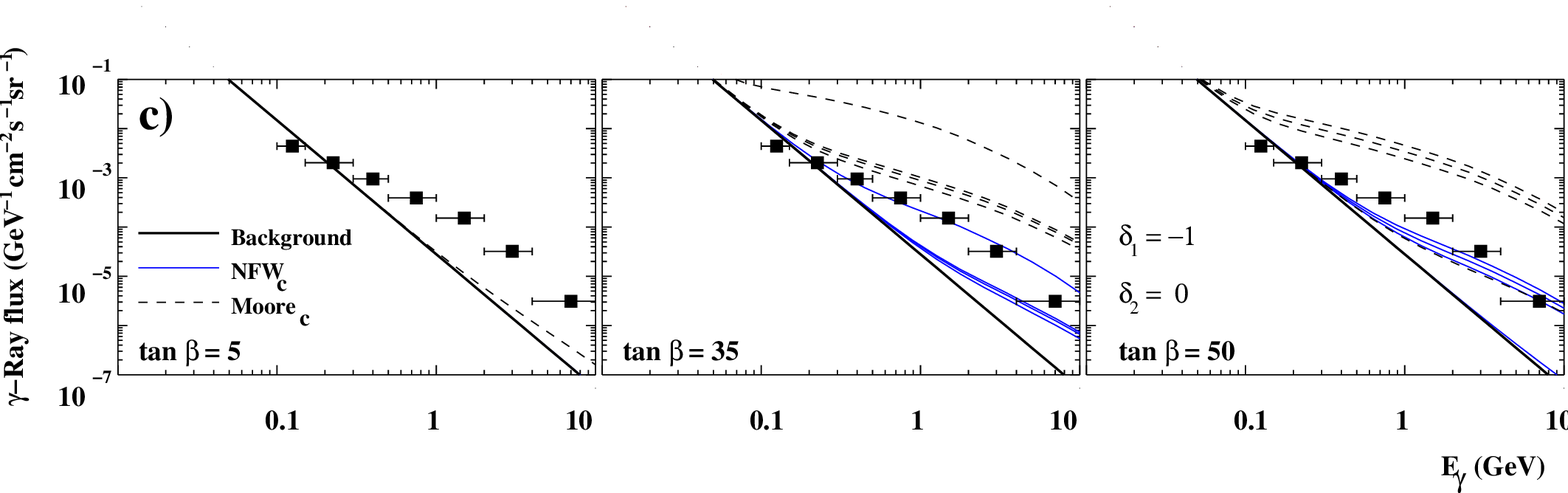,width=1.\textwidth}%\hskip 1cm
	\vskip -0.5 cm
       \epsfig{file=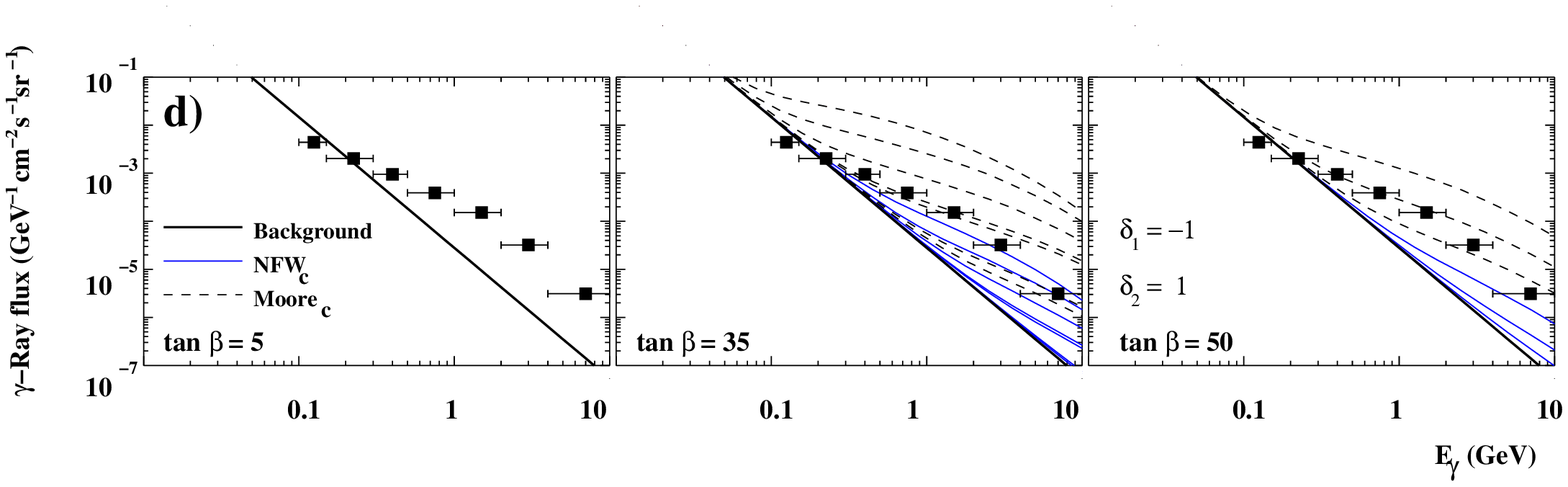,width=1.\textwidth}%\hskip 1cm
	\vskip -0.5cm
       \epsfig{file=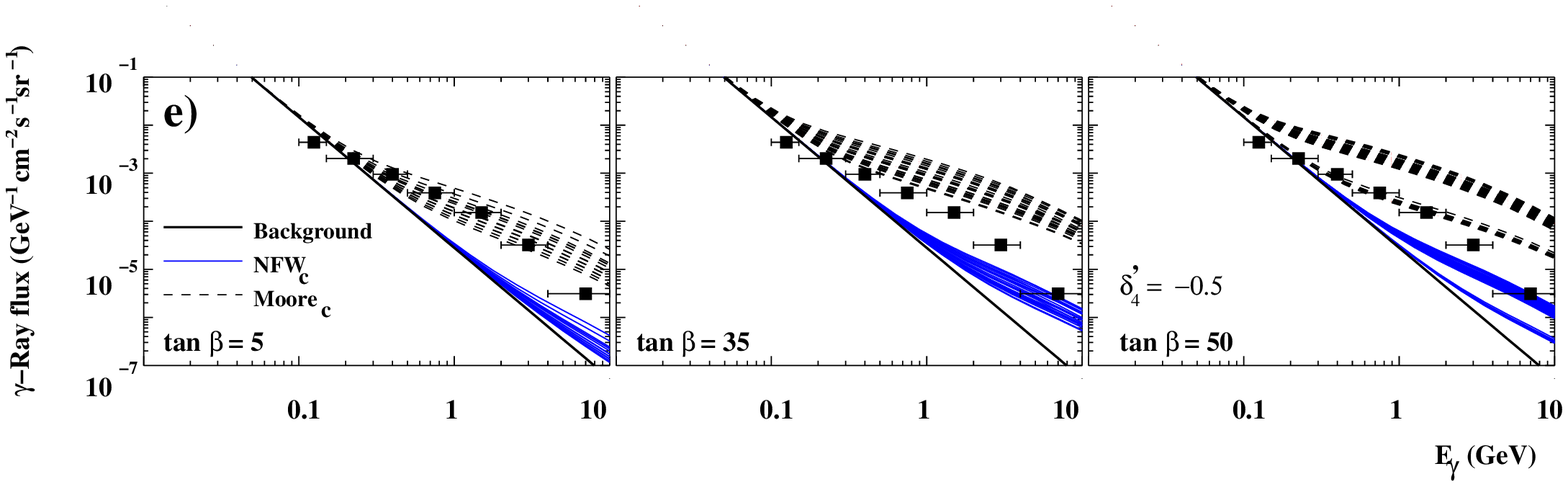,width=1.\textwidth}%\hskip 1cm

%       }
          \caption{{\footnotesize
Gamma--ray spectra $\Phi_{\gamma}(E_{\gamma})$
from the galactic center as functions of the photon energy
for the 
SUGRA 
cases discussed in Eq.~(\ref{3cases})
for  $\tan \beta=5, 35, 50$, $A=0$
and $\mu > 0$, compared with data from the EGRET experiment. 
NFW and Moore et al. profiles with
adiabatic compression are used with $\Delta \Omega \sim 10^{-3}$ sr.
All points shown after a scan on $m$ and $M$ (0 -- 2000 GeV) 
fulfil 
%all 
the accelerator constraints discussed in the text, 
and WMAP bounds.
%(BUT g-2 IS NOT INCLUDED ?? {\bf [Yann : No Carlos, g-2 not included.
%I am not really partisan to exclude the point which doesn't respect
%the 2 sigma deviation to g-2. I am ok to plot this limit, like
%that, readers chose to take into account or not this constraint (is 2 sigma
%an excess?) but not to exclude directly the points from an analysis.
%I think it is a little bit too strong.]}). 
}}
        \label{fig:EGRET}
    \end{center}

\end{figure}

The results can be seen in Fig.~\ref{fig:EGRET},
where several cases of Eq.~(\ref{3cases}) have been studied
for a scan on $m$ and $M$ from 0 to 2 TeV and three values of 
tan$\beta$ (5, 35 and 50). Only points of the parameter space
producing interesting gamma-ray fluxes
concerning the EGRET experimental data are shown. 
Concerning accelerator constraints, these points fulfil 
the lower bounds on the masses
of SUSY particles and Higgs boson, as well as the experimental
bounds on the branching ratios of the \bsg\ 
and $B_s \to \mu^+ \mu^-$ processes.
They also fulfil WMAP bounds discussed in the previous section.
In particular, the thin solid (blue) line corresponds to a NFW$_c$ density
profile. 
Concerning the background of the diffuse gamma-ray flux of
the inner galaxy at the energy range of interest for our analysis 
(from 100 MeV to 10 GeV), the main production of gamma-rays is
the interaction of cosmic rays (mainly protons and helium nuclei) with the interstellar
medium (atomic and molecular hydrogen, and helium). Neutral pions $\pi^0$ 
produced in this process radiate gamma-rays with a spectrum peaked at $\sim 70$ MeV
($m_{\pi}/2$), dropping at high energies with an energy power law
which follows the initial cosmic ray spectrum of the form $E^{-\alpha}$ with $\alpha\sim 2.7$
\cite{UllioEGRET, modi2}. The normalization factor 
has been fixed to have the best fit
in agreement to the standard scenario, 
$\phi_{b}=2\times 10^{-6} \mathrm{cm}^{-2} \mathrm{s}^{-1} 
\mathrm{GeV}^{-1} \mathrm{sr}^{-1}$ 
for 2 GeV $<E_{bin}<$ 4 GeV \cite{UllioEGRET}.

As one can see, with small values of tan$\beta$ we are not able
to find points reproducing EGRET data. However, interesting points can be found
for larger values of tan$\beta$. In particular, we show in the figure
the cases tan$\beta$=35,50 and one can see that with non-universal
cases EGRET data can be reproduced.
This is obtained for neutralino masses between 
150 and 600 GeV.

Let us remark that it is possible to differentiate each point of the parameter space ($m$, $M$) by
its gamma-ray spectrum. Indeed, just a look at Fig. \ref{fig:EGRET}b for tan$\beta = 35$
shows two kind of spectra. The two highest lines correspond
to points with ($m$, $M$) = (533 GeV, 400 GeV) and (733 GeV, 533 GeV),  
and they are located inside the Higgsino 
region dominated by $f \bar{f}$ final states through
$Z-$exchange (see e.g. point {\bf D} of Fig. \ref{fig:mSUGRAtb50scan}). On the other hand, 
the lower line corresponds to ($m$, $M$) = (266 GeV, 666 GeV),
which lies inside the stau coannihilation region (see e.g. point {\bf A} of
Fig. \ref{fig:mSUGRAtb50scan}), 
where the annihilation is weak but 
dominated by $A$-exchange, and the spectrum is thus a $b \overline{b}$ one.
In the case of Fig. \ref{fig:EGRET}c with tan$\beta = 35$, the higher fluxes
corresponds to the closing of the $A-$pole (see e.g. point {\bf C} of Fig. \ref{fig:mSUGRAtb50scan}) 
whereas the lower flux spectrum is obtained through the opening of
the $A-$pole (see e.g. point {\bf B} of Fig. \ref{fig:mSUGRAtb50scan}). 
The same remarks can be made for the other cases illustrated in Fig. \ref{fig:EGRET}.

We have also carried out the same 
analysis for a Moore et al. profile including adiabatic compression,
Moore$_c$,
as shown in the figure with a dashed line for the different points of
the parameter space. Clearly,
due to the very singular behaviour of this profile
many points are constrained by EGRET data.

\subsection{GLAST}

\begin{figure}[!]
    \begin{center}
\vskip -4truecm
        \epsfig{file=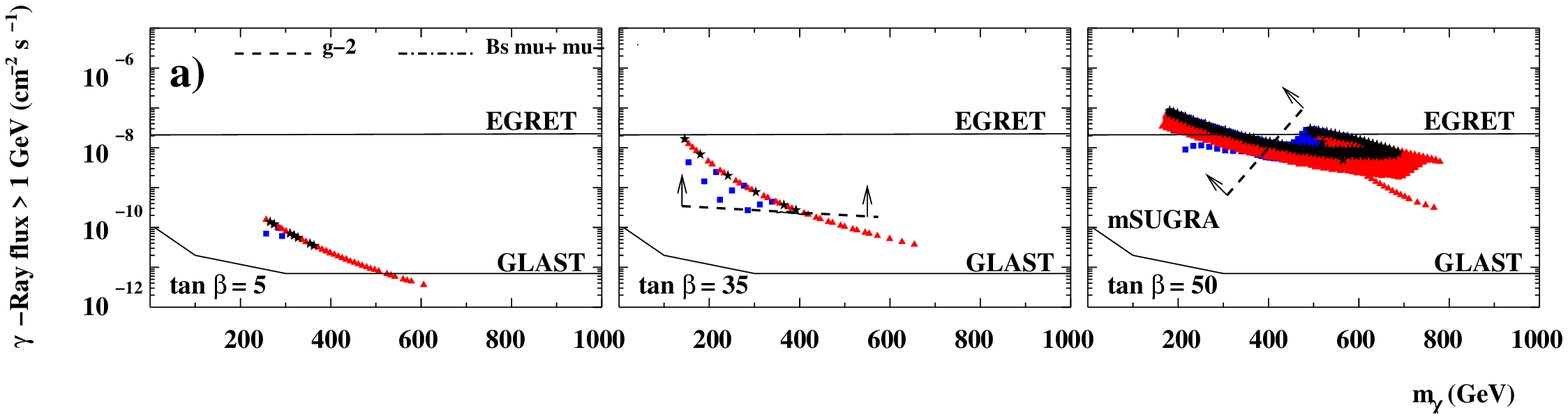,width=1.\textwidth}%\hskip 1cm
	\vskip -0.3cm       
	\epsfig{file=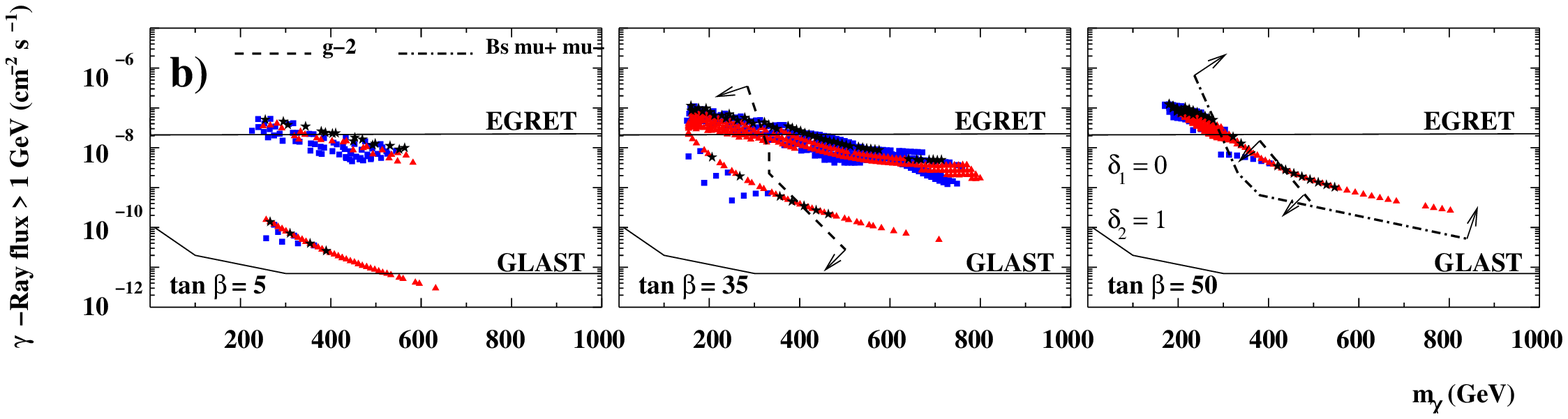,width=1.\textwidth}%\hskip 1cm
       	\vskip -0.3cm
	\epsfig{file=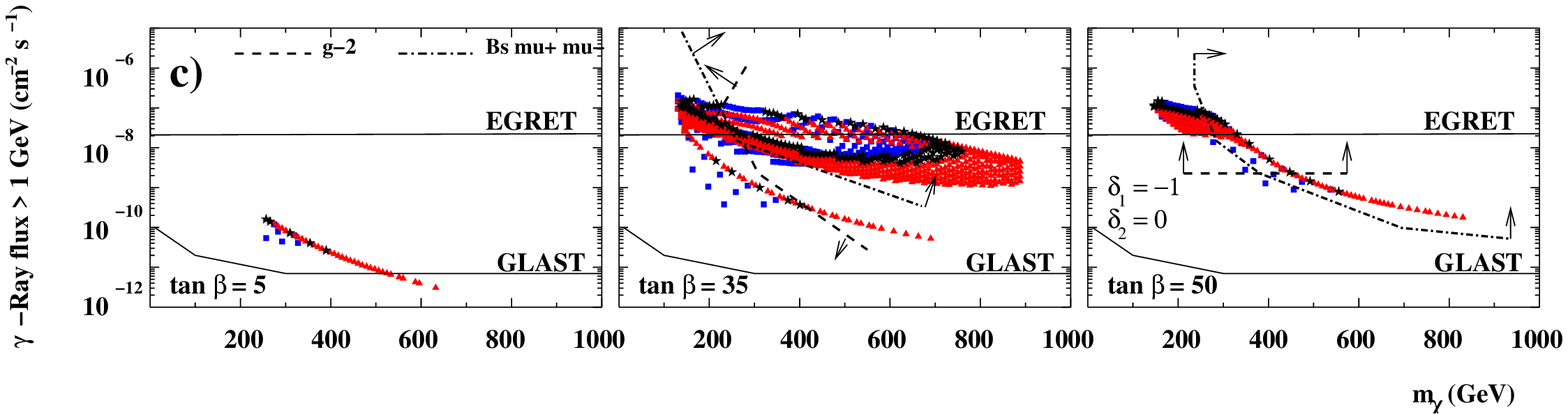,width=1.\textwidth}%\hskip 1cm
	\vskip -0.3cm
       \epsfig{file=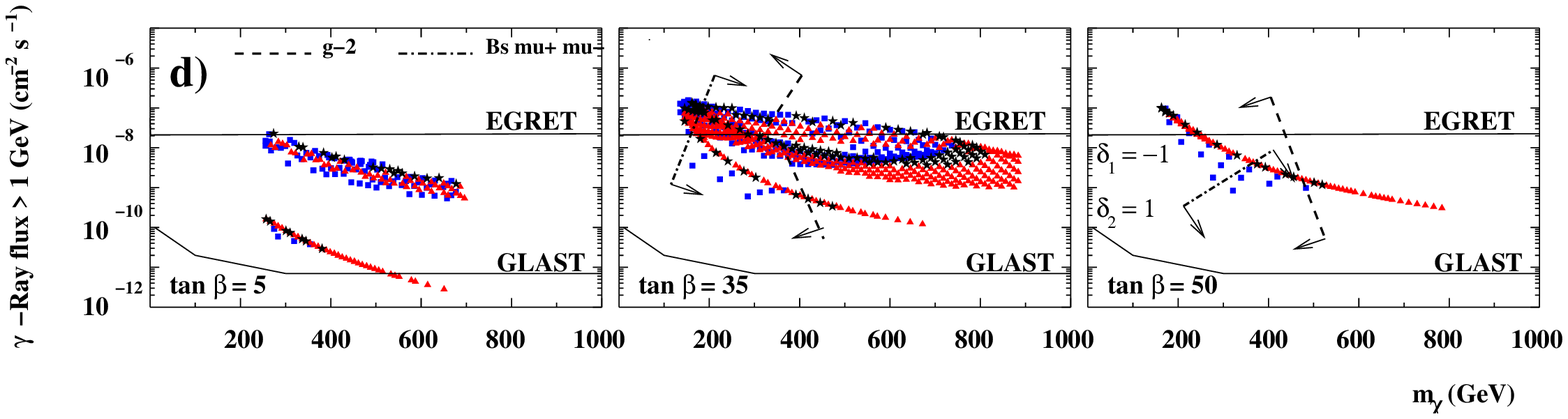,width=1.\textwidth}%\hskip 1cm
	\vskip -0.3cm
       \epsfig{file=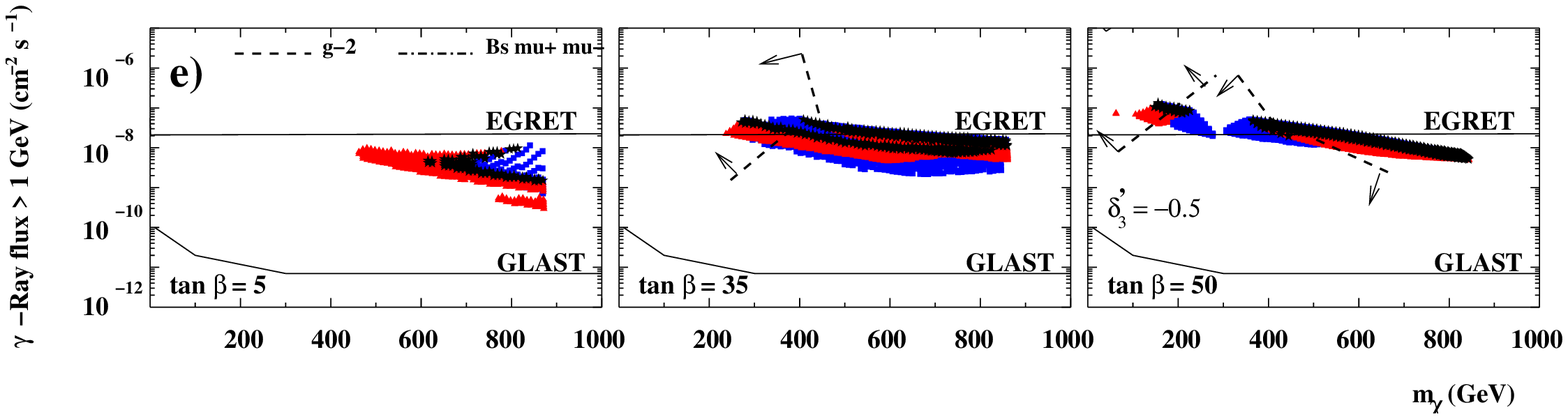,width=1.\textwidth}%\hskip 1cm
          
\caption{{
\footnotesize
Scatter plot of the gamma-ray flux $\Phi_{\gamma}$ 
for a threshold of 1 GeV
as a function of the neutralino mass $m_{\chi}$ 
for the SUGRA 
cases discussed in Eq.~(\ref{3cases}), and for the same
parameter space used in Fig.~\ref{fig:EGRET}.
A NFW profile with
adiabatic compression is used with $\Delta \Omega \sim 10^{-5}$ sr.
All points in the figure
fulfil the \bsg\ bounds. 
For  $\tan\beta$=5 all points 
have $\asusy < 7.1 \times 10^{-10}$. For the other values of  $\tan\beta$,
$\asusy > 7.1 \times 10^{-10}$ is fulfilled for the points located on
the arrow side of the dashed lines, and 
B($B_s \to \mu^+ \mu^-$) $< 2.9 \times 10^{-10}$ for those on 
the arrow side of the dot-dashed lines.
Points 
depicted with light grey (magenta) triangles have 
$0.129<\Omega_{\tilde{\chi}^0_1}h^2<0.3$, 
with black stars have
$0.094<\Omega_{\tilde{\chi}^0_1} h^2<0.129$,
and finally with dark grey (blue) boxes have
$0.03<\Omega_{\tilde{\chi}^0_1} h^2<0.094$ 
with the appropriate rescaling of the density of neutralinos
in the galaxy as discussed in the text. 
Solid lines denote the $5\sigma$ sensitivity curves for satellites.
From top to bottom, 
the first solid line
corresponds to the signal reported by EGRET, and
the upper area bounded by the second solid line will be analyzed by 
GLAST experiment.
}}
        \label{fig:GLAST}
    \end{center}

\end{figure}

Crucial information on the origin and nature of the EGRET excess at the galactic center
will be provided by another space-based detector, GLAST. 
This is scheduled for launch
in 2007, and 
will perform an all-sky survey covering a larger energy range 
($\approx$ 10 -- 300 GeV) than EGRET,
with a wider effective area and a better resolution in energy. 
The experiment will also allow to search for an angular distribution of
the dark matter halo. Indeed, with an angular resolution
of $0.1^o$ ($\Delta \Omega \sim 10^{-5}$ sr),
GLAST will be able to point and analyse the inner center of the Milky
Way ($\sim 7$ pc). 
Even if alternative explanations 
solve the data problem pointed out by EGRET, 
GLAST will be able to detect a much smaller flux of gamma rays from
dark matter $\sim 10^{-11}$ cm$^{-2}$\ s$^{-1}$.

We summarize the results of our analysis in 
Fig.~\ref{fig:GLAST}, using
the same parameter space as in Fig.~\ref{fig:EGRET}, and a NFW$_c$ profile. 
There, the values of the total gamma-ray flux 
$\Phi_{\gamma}$ for a threshold of 1 GeV 
allowed by all experimental constraints
as a function of the neutralino mass 
$m_{\tilde{\chi}^0_1}$ are depicted.
Concerning the predicted sensitivity for the 
experiment, we follow Ref.~\cite{Fornengo}
and show on the plot the $5 \sigma$ 
sensitivity curves assuming an angular resolution
of $0.1^o$ ($\Delta \Omega = 10^{-5}$), and a total effective pointing time
of 30 days for the satellite. 
We observe that, basically, the whole parameter space of general SUGRA
will be tested by GLAST. Note that even 
for the mSUGRA case points corresponding
to a value of tan$\beta$ as small as 5
will be reached  by GLAST.
This is a remarkable result if we realise that interesting direct
detection experiments, such as e.g. EDELWEIS-II or CDMS-II, will
only be able to cover small regions of the parameter space.
More poweful detectors such as GENIUS will cover larger regions,
but still not too large compared with the parameter space of general
SUGRA. In contrast, we have observed that indirect detection
experiments
will be able to test the parameter space of 
SUGRA\footnote{See also Ref.~\cite{Mambrini:2004kv} 
for a comparison between direct
and
indirect dark matter search without using compressed halo models.}.

It is also worth noticing
that many points are already constrained by EGRET data.
See e.g. case {\it c)} with $\tan\beta=35$.
Let us remark
that in Fig.~\ref{fig:GLAST} EGRET line corresponds to
$\Delta \Omega \sim 10^{-3}$ sr, and therefore for its study 
all points in the figure
should be rescaled by a factor 
$\bar{J}(10^{-3}  {\rm sr}) 10^{-3}/ \bar{J}(10^{-5}  {\rm sr}) 10^{-5}  \simeq 1.5.$
Of course, a Moore$_c$ profile will give rise to larger fluxes.

\subsection{CANGAROO}

Recently,  the CANGAROO--II atmospheric Cherenkov telescope has made a
significant detection of gamma rays from the galactic center 
region \cite{Cangaroo1}.
In particular, 
the collaboration has published the spectrum obtained in six energy bins, from
200 GeV to 3 TeV.
Observations taken during 2001 and 2002 have detected a statistically
significant excess at energies greater than 
$\sim 250$ GeV, with an integrated flux of
$\sim 2 \times 10^{-10} 
~ \mathrm{photons}~ \mathrm{cm}^{-2}~ \mathrm{s}^{-1}$.
These measurements indicate a very soft spectrum $\propto E^{-4.6 \pm 0.5}$.
As discussed in Ref.~\cite{HooperSilk}, this signal could be fitted with a spectrum
of a TeV-dark matter candidate. It is interesting to see whether it is possible
to obtain such a candidate in SUGRA scenarios imposing the accelerator
and WMAP constraints, and withing the framework of adiabatically
compressed halos.

As we can see in Fig.~\ref{fig:mSUGRAtb50scan}, most of the regions of
the parameter space 
of mSUGRA which satisfy WMAP constraints give mainly a sub-TeV spectrum in scalar and
gaugino sectors. 
It is thus very difficult to find a dark matter candidate fitting the observations made by 
the CANGAROO-II experiment.
Nevertheless, it is possible to extend the allowed region if we release the gaugino
universality. One way to achieve this is to decrease $M_2$ while
keeping $M_1 = M_3$ 
at $M_{GUT}$. 
This allows to reduce significantly the relic abundance of the wino--like neutralino,
through its coannihilation with the lightest 
chargino $\chi^+_1$. Such an scenario,
well motivated e.g. in anomaly mediation models, predicts higher values for
the gaugino masses fulfilling WMAP constraints than those
of universal cases, 
typically around the TeV scale, which is precisely 
the region favoured by CANGAROO-II results.

\begin{figure}[!]
    \begin{center}
\centerline{
       \epsfig{file=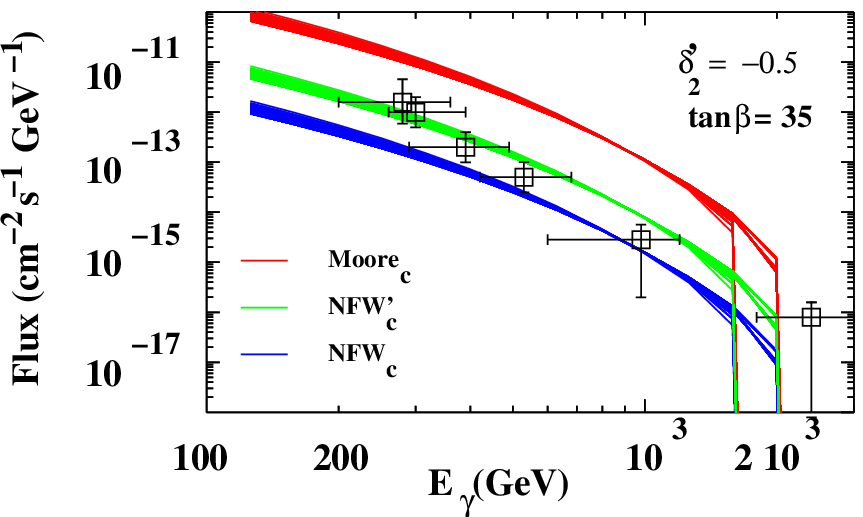,width=0.5\textwidth}
       \epsfig{file=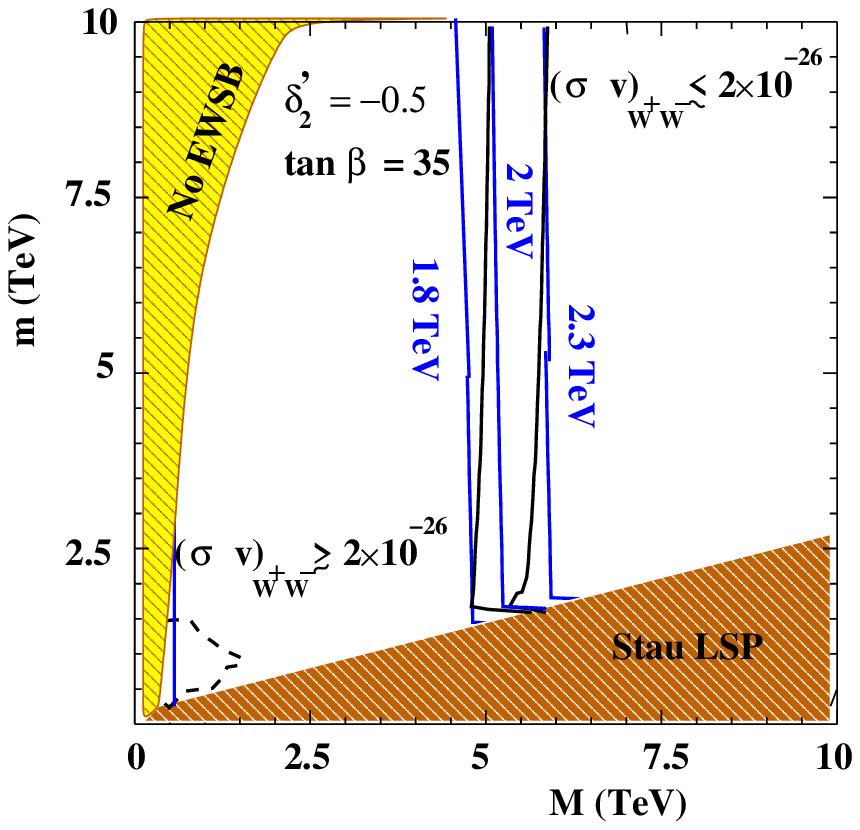,width=0.4\textwidth}
       }
          \caption{{
\footnotesize 
Left frame:
Gamma-ray spectra
from the galactic center as functions of the photon energy
for the case {\it f)} discussed in Eq.~(\ref{3cases})
with non-universal gaugino masses $M_2=0.5 M$, for  $\tan \beta=35$, $A=0$
and $\mu > 0$, compared with data from the CANGAROO--II experiment. 
NFW and Moore et al. profiles, as well as the average profile defined
in Sect.~\ref{crucial} with
adiabatic compression are used with $\Delta \Omega \sim 10^{-5}$ sr.
All points shown after a scan on $m$ and $M$ (0--10 TeV) 
fulfil the accelerator constraints discussed in the text,
and WMAP bounds.
% (BUT g-2 IS NOT INCLUDED).
Right frame:
Experimental and astrophysical bounds in the parameter space ($m$,
$M$).
The conventions and constraints are the same as in 
Fig. \ref{fig:mSUGRAtb50scan}. Blue lines are contours of neutralino masses.
}}
        \label{fig:CANGAROOcase5}
    \end{center}
\end{figure}
\begin{figure}[!]
    \begin{center}
\centerline{
       \epsfig{file=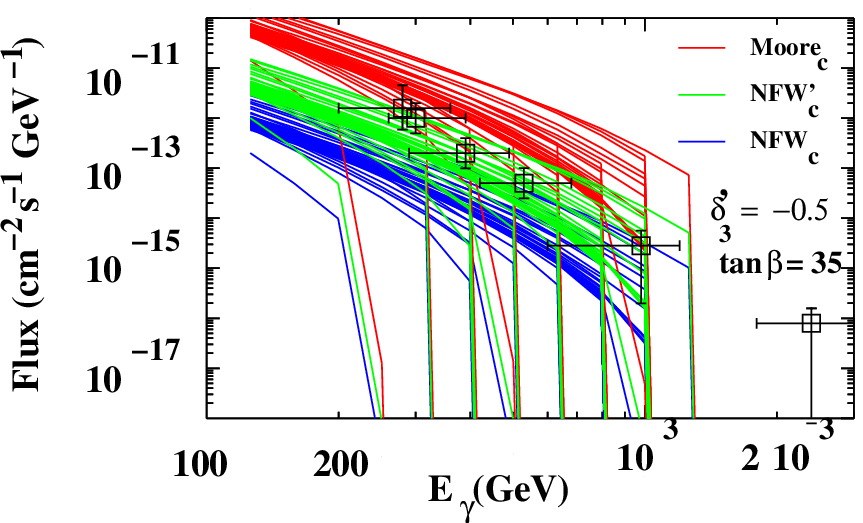,width=0.5\textwidth}
       \epsfig{file=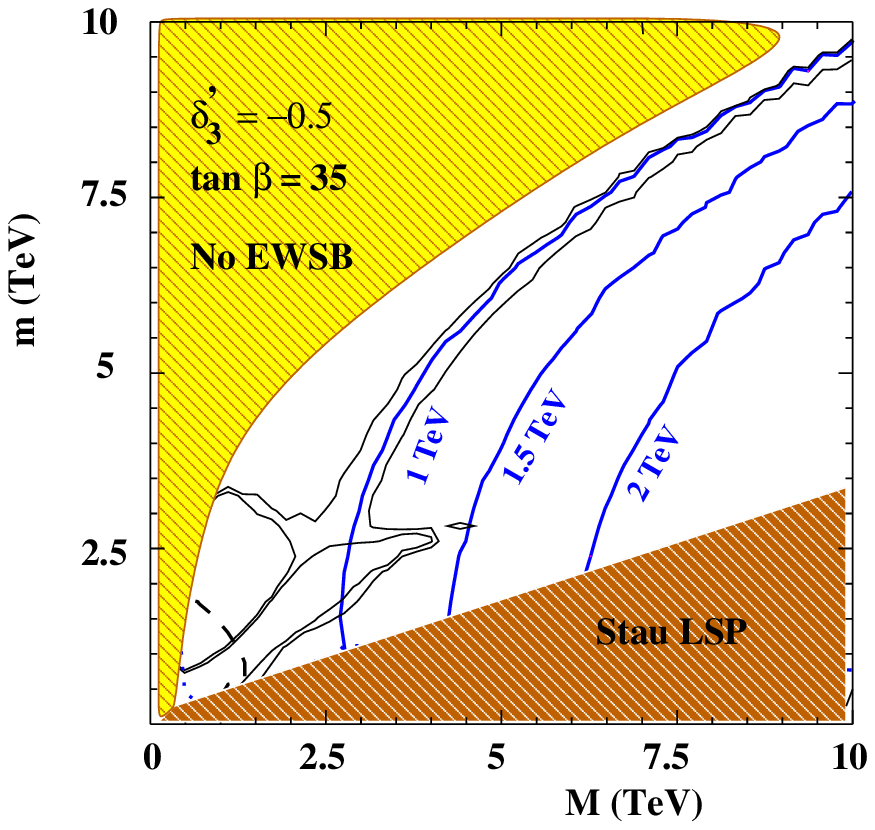,width=0.4\textwidth}
       }
          \caption{{
\footnotesize 
The same as in Fig. \ref{fig:CANGAROOcase5} but for the non-universal
case {\it e)} with $M_3=0.5 M$. 
}}
        \label{fig:CANGAROOcase4}
    \end{center}
\end{figure}

As an illustration, let us consider the case {\it f)} in
Eq.~(\ref{3cases}) giving rise to $M_2=0.5M$.
We show this case in Fig. \ref{fig:CANGAROOcase5}.
There all points fulfil 
the lower bounds on the masses
of SUSY particles and Higgs boson, as well as the experimental
bounds on the branching ratios of the \bsg\ and $B_s \to \mu^+ \mu^-$
processes.
They also fulfil the WMAP bounds.
As we can see in the left and right frames of the figure, 
%Fig. \ref{fig:CANGAROOcase5}, 
there is an interesting
region of the parameter space fulfilling the WMAP constraints and
giving rise to
a sufficiently large flux to fit the data from CANGAROO-II. 
Let us remark that the highest 
energy bin shown (of about 2.5 TeV) is less than 1 $\sigma$ in excess of a null 
result, so it should be considered only as an upper limit.  Excluding this
point from the analysis, we see that this non-universal
model can fit CANGAROO data and WMAP bounds in halos models with profiles
between NFW$_c$ and Moore$_c$.
In particular, this is the case
of the scenario
discussed in Sect.~\ref{crucial},
$\rm{NFW'_c}$, 
%NFW'$_c$,
with a value of $\bar J$ given by 5 $\times \bar J_{\rm{NFW_c}}$,
as can be seen in the left frame of the figure. 

It is worth noticing that 
the parameter space where 
WMAP constraints are fulfilled corresponds to a narrow range 
of the neutralino mass ($\sim 2$ TeV), as can be seen
in the right frame of Fig. \ref{fig:CANGAROOcase5}. 
This region is mainly independent of tan$\beta$
and the value of $\delta'_2$ as far as the neutralino is wino like 
($\delta'_2 < -0.45$). In this region of the parameter space,
the relic density calculation is governed by the $\chi_1^0-\chi_1^+$ coannihilation
into $W^+ W^-$ ($m_{\chi^0_1} \sim m_{\chi^+_1} \sim M_2$), whereas the flux is largely
 dominated by the annihilation process $\chi^0_1 ~ \chi^0_1 \to W^+ ~ W^-$ 
through $t-$chanel $\chi^+_1$ exchange 
($\sigma v |_{WW} \sim 10^{-26}\mathrm{cm}^{2}\mathrm{s}^{-1}$).

Another possibility is to produce a neutralino-LSP with a higher
higgsino component. 
Indeed, as discussed in Ref.~\cite{Mambrini:2004ke}, 
decreasing the value of $M_3$ while
keeping $M_2 = M_1$ at $M_{GUT}$ increases the Higgsino fraction of 
the neutralino
through the renormalization group equations (RGEs),
and, as a consequence, its coupling to the $Z$ boson. In this case, 
the relic density constraint can be satisfied more easily because the
contribution from the annihilation into gauge bosons channel is more important
and the focus point region is much wider. Another consequence of the decreasing
of $M_3$ at $M_{GUT}$ is the decreasing of the pseudoscalar mass $m_A$ through
$m_{H_2}$. 
In the left frame of Fig. \ref{fig:CANGAROOcase4} we show
the case {\it e)} of 
Eq.~(\ref{3cases}) with $M_3=0.5M$.
For tan$\beta$ larger than about 10, the neutralino annihilation into 
fermions (mainly $b \overline{b}$) starts 
to dominate. Eventually, the focus point
and Higgs annihilation regions merge, for example for tan$\beta=$ 35 and 
$M \approx 2$ TeV, as was already pointed out by the authors of 
Ref.~\cite{Belangernonouniv}.
This region is clearly visible in 
the right frame of Fig. \ref{fig:CANGAROOcase4}.  
%{\bf 
%In fact, we can calculate an approximation for the mass of a wino--like neutralino 
%respecting WMAP constraint. Indeed, in the simplest situation, where the annihilation
%cross--section is independent of velocity, the relic abundance is approximately given
%by
%
%\begin{equation}
%\Omega_{\chi^0_1} h^2 \sim 
%\left(
%\frac{<\sigma v>}{3 \times 10^{-27} \mathrm{cm^3 s^{-1}}}
%\right)^{-1}
%\end{equation}
%
%\noindent
Whereas in the previous analised case {\it f)}
the WMAP constraint was fulfilled through one main chanel
($\chi^0_1 ~ \chi^+_1$ coannihilation), the non--universality of $M_3$ offers more
possibilities through the exchange of Higgses. 
That explains why the region allowed by
WMAP constraint is broader in the right frame of
Fig. \ref{fig:CANGAROOcase4} than in
Fig. \ref{fig:CANGAROOcase5}.
%}

Of course in these models the SUSY spectrum is very heavy.
For example, 
for a point with $m = 5 $ TeV, $M = 5$ TeV reproducing
CANGAROO-II data, the spectrum in case {\it e)} ({\it f)} consists of
heavy squarks, $m_{\tilde q} \sim 8$ TeV (10 TeV), neutralinos and charginos
$m_{\chi^0_1} \sim m_{\chi^+_1} \sim 1$ TeV (2 TeV), and also heavy higgses 
$m_A \sim 5$ TeV (6 TeV). 
If we do not take into account the potential fine-tuning problem
associated with such models, and consider them reliable,
this would be a difficult situation for any hope
of discovering SUSY 
at colliders, since the only detectable particle would be the lightest Higgs. 
We have plotted for information  the 2$\sigma$ standard deviation 
$a_{\mu}^{\mathrm{SUSY}} = 7.1 \times 10^{-10}$ with dashed line in
the right frames of 
Figs. \ref{fig:CANGAROOcase5} and \ref{fig:CANGAROOcase4}. Only an
unambiguous evidence for a non-zero contribution to $\delta a_{\mu}$ would restrict
this class of models with heavy sparticles. 

Let us finally mention
that we have also try to reproduce the flux measured by CANGAROO-II
using cases {\it b), c)} and {\it d)} with non universal scalars.
They give rise to a neutralino too light to be able to reproduce those data.

%{\bf
%*) Concerning the Higgsino part : M3 = 0.6 M is working but M3 = 0.5 M no, try
%to find an explanation to that. The fact is that The relic density is always
%too much when we try to have a 2 TeV neutralino. Lower relic density require:
%Increasing M0, but we decrease mneut. Or lowering Mhalf. But we lower mneut
%too... For instance, m0 = 4100 GeV, mhalf = 6550 GeV gives Omega = 0.4, Mneut =1934
%(see the sheet of paper with Manu remark on heavy higgsino neutralino)
%
%*) If we take -0.4 for deltaprime3, we have (it seems with the point I checked)
%that we have the same maximum mass limit. This will be great, and a kind
%of "model independent" limit.  
%}

\subsection{Combining EGRET and CANGAROO}

\begin{figure}
    \begin{center}
\centerline{
       \epsfig{file=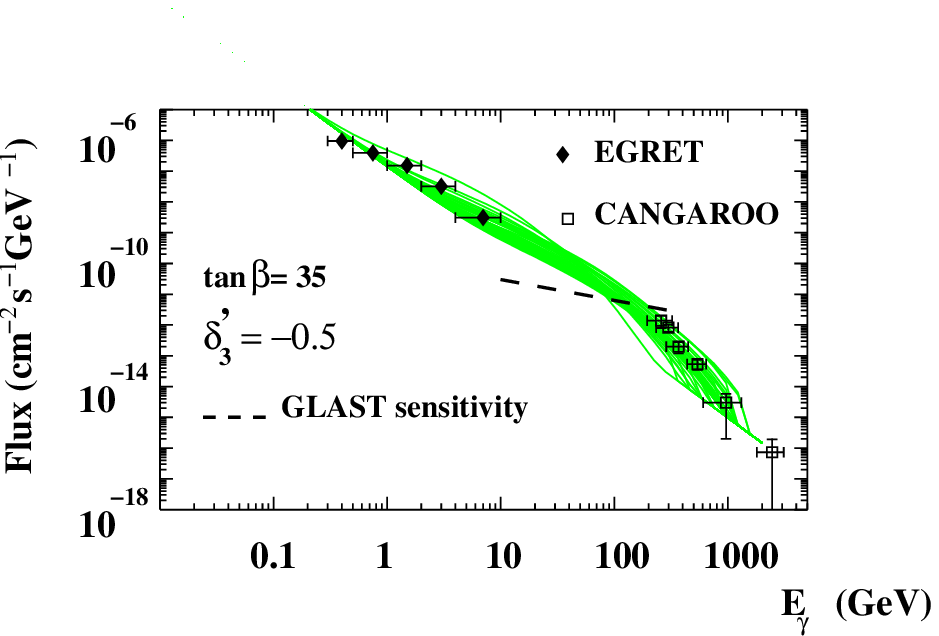,width=0.5\textwidth}
       }
          \caption{{
\footnotesize 
The same as in the left frame of Fig. \ref{fig:CANGAROOcase5} 
but for the non--universal case {\it e)} with $M_3=0.5 M$,
compared with data for EGRET and CANGAROO--II experiments and
the expected GLAST sensitivity. Here only the average profile
defined in Sect.~\ref{crucial} with adiabatic compression, 
$\rm{NFW'_c}$,
%NFW'$_c$, 
is used.  
}}
        \label{fig:EGRETCANGAROO}
    \end{center}
\end{figure}

After analysing EGRET and CANGAROO-II data, it seems natural to
try to fit both experiments with only one non--universal scenario. 
We notice here that a similar exercise has been carried out by the authors of
Ref.~\cite{Fornengo}.
The baryonic cooling effect on the fluxes gives us the order of
magnitude needed 
to fit with both data with a 1 TeV neutralino in the 
non--universal case {\it e)} with $M_3=0.5 M$.
It is worth noticing that the CANGAROO-II collaboration in \cite{Cangaroo1}
pointed out already that the EGRET and 
CANGAROO-II data can be relatively smoothly 
connected with a cutoff energy of 1--3 TeV.   
In our case, for that we need to use 
the NFW'$_c$ scenario discussed in Sect.~\ref{crucial}. Typical points
of the parameter space fullfilling all experimental constraints and fitting
both set of data lie between ($m=800$ GeV, $M= 800$ GeV) and
($m=3$ TeV, $M=3$ TeV).

It is also interesting to see the complementarity
of GLAST with EGRET 
and CANGAROO. GLAST will perform an all-sky survey detection of fluxes 
with energy from 1 GeV to 300 GeV, 
exactly filling the actual lack of experimental data 
in this energy range (see Fig. \ref{fig:EGRETCANGAROO}),
and checking the CANGAROO results. Indeed, we have calculated
that the integrated gamma ray flux for such a signal will be around 
$5 \times 10^{-11} ~ 
%\mathrm{ph} ~ 
\mathrm{cm^{-2}} ~ \mathrm{s^{-1}}$. 
We have shown this sensitivity curve in Fig. \ref{fig:EGRETCANGAROO} 
for  $\Delta \Omega = 10^{-5}$, which is the typical detector acceptance, following
the prescriptions given in \cite{Fornengo}. We clearly see that GLAST 
will help to cover
the entire spectrum. 

We have also tried to reproduce the observed data using a wino-like
neutralino
arising from the non-universal case {\it f)} with $M_2 = 0.5 M$.
But any model of this type fulfilling WMAP bounds gives rise to a  too heavy neutralino-LSP
($\sim 2$ TeV) whose spectral features cannot explain at the same time both the CANGAROO-II
and EGRET excess. 

%{\bf manu : it seems to me that with M2=0.55*M1 instead of
%  0.5 we can have a 1TeV WMAP bino and slightly wino neutralino (3000 2000 0 35 1)
%  annihilating into ttbar (but the relic is driven by coannihilation), the
%  reslting fluxes are too low ... maybe need to check a bit more}

\subsection{HESS}

The HESS Cherenkov telescope experiment has recently published new
data on gamma rays, detecting a signal from the galactic center
\cite{HESS}. The measured flux and spectrum differ substantially from
previous results, in particular those reported by the CANGAROO
collaboration, exhibiting a much harder power--law energy spectrum with 
spectral index of about $-2.2$ and extended up to 9 TeV.
The authors of \cite{HESS} already pointed out that if we assume that
the observed gamma rays represent a continuum annihilation spectrum, 
we expect $m_{\chi} \gsim 12$ TeV. Actually such a heavy neutralino-LSP is 
not natural in the framework of a consistent supergravity model when we 
impose the renormalisation group equations and radiative electroweak symmetry
breaking.

Although we performed some scans in all non universality directions using 
numerical 
dichotomy methods, no point in the parameter space in any non-universal case 
studied was able to give a several 10 TeV 
neutralino satisfying 
WMAP constraint but this can be sensitive to the RGE and relic density
calculation codes. 

% {\bf manu : do you prefer this ? : 
%Using dedicated algorythms and non universality directions, it might be
%possible to find a several 10 TeV neutralino satisfying 
%WMAP constraint but this can be sensitive to the RGE and relic density
%calculation codes for the extrem required parameter values. 

%I think I prefer ... or we can remove this paragraph ? ...

Nevertheless, without RGE and taking soft parameters at the electroweak scale,
the constraints are easier to evade. On top of that, in a very effective  
approach using  completely free parameters and couplings in cross sections, 
neutralinos with 
$m_{\chi}\gtrsim 10$ TeV and $\Omega_{\chi} h^2 \sim WMAP$ may certainly be 
fine tuned. We did not adopt such approaches since the MSSM is motivated by 
high energy and theoretical considerations.
In this section, we analyze  in a quite model-independant way the
  conditions 
required on the particle
physics field to fit with 
the HESS data thanks to dark matter annihilation.

\begin{figure}
\begin{center}
\begin{tabular}{cccc}
a)&
\psfrag{dNdx}[rb][ct]{\tiny $dN/d(E_{\gamma}/m_{\chi})$}
\psfrag{Egammamx}[c][c]{\tiny $E_{\gamma}/m_{\chi}$}
\psfrag{mass}[l][l]{\tiny $m_{\chi}=1$ TeV }
\psfrag{process}[l][l]{\tiny $\chi \chi \to b \bar{b}$ }
\psfrag{salati}[l][l]{\tiny {\green Bengtsson et al}}
\psfrag{scopel}[l][l]{\tiny {\darkred Fornengo et al}}
\psfrag{manu}[l][l]{\tiny {\darkblue $ 1.2 e^{-10x}/x^{1.5} $  }}

 \includegraphics[width=0.3\textwidth]{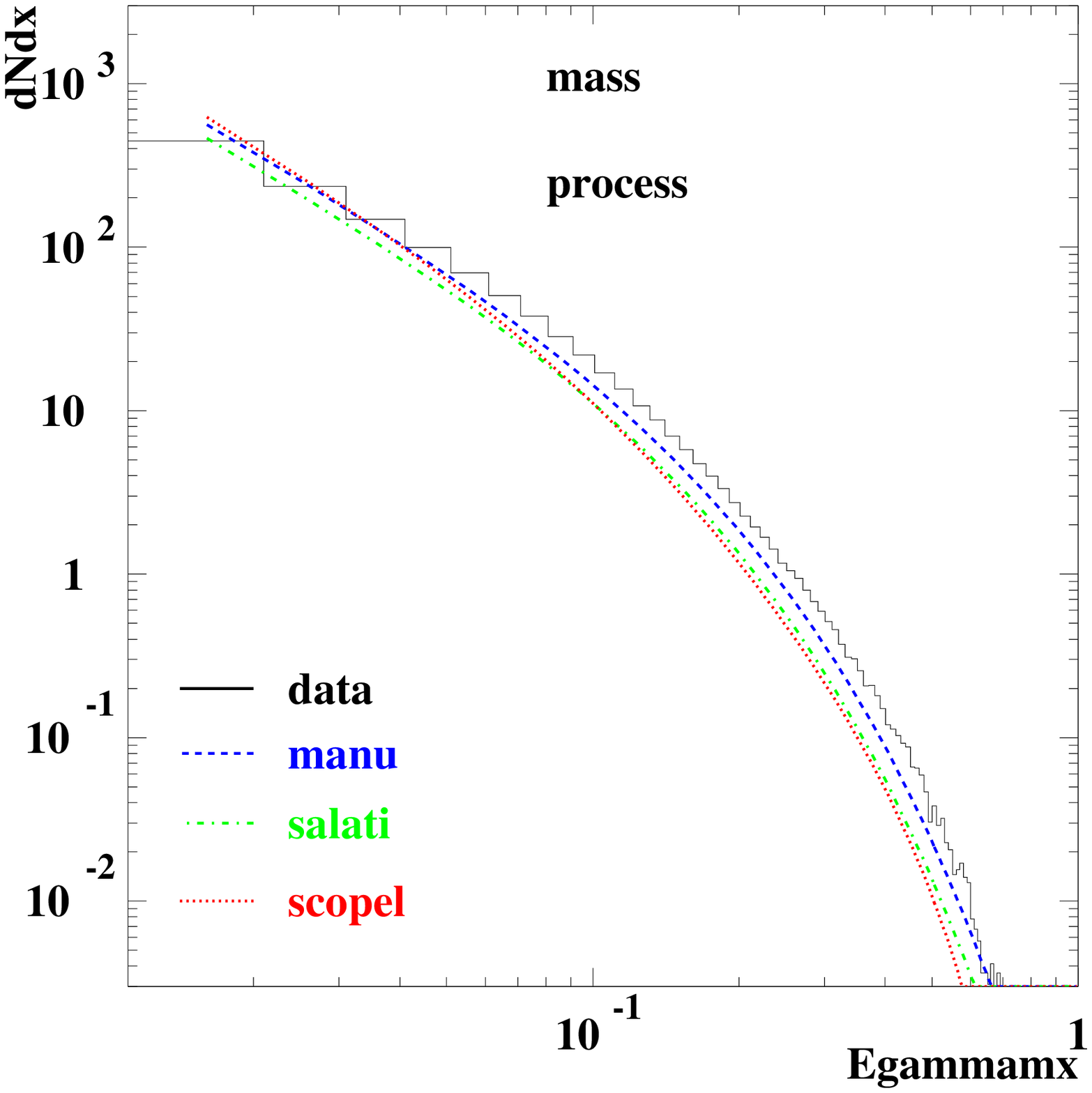}&

\psfrag{dNdx}[rb][ct]{\tiny $dN/d(E_{\gamma}/m_{\chi})$}
\psfrag{Egammamx}[c][c]{\tiny $E_{\gamma}/m_{\chi}$}
\psfrag{mass}[l][l]{\tiny $m_{\chi}=15$ TeV }
\psfrag{process}[l][l]{\tiny $\chi \chi \to b \bar{b}$ }
\psfrag{salati}[l][l]{\tiny {\green Bengtsson et al}}
\psfrag{scopel}[l][l]{\tiny {\darkred Fornengo et al}}
\psfrag{manu}[l][l]{\tiny {\darkblue $ 1.2 e^{-10x}/x^{1.5} $  }}

 \includegraphics[width=0.3\textwidth]{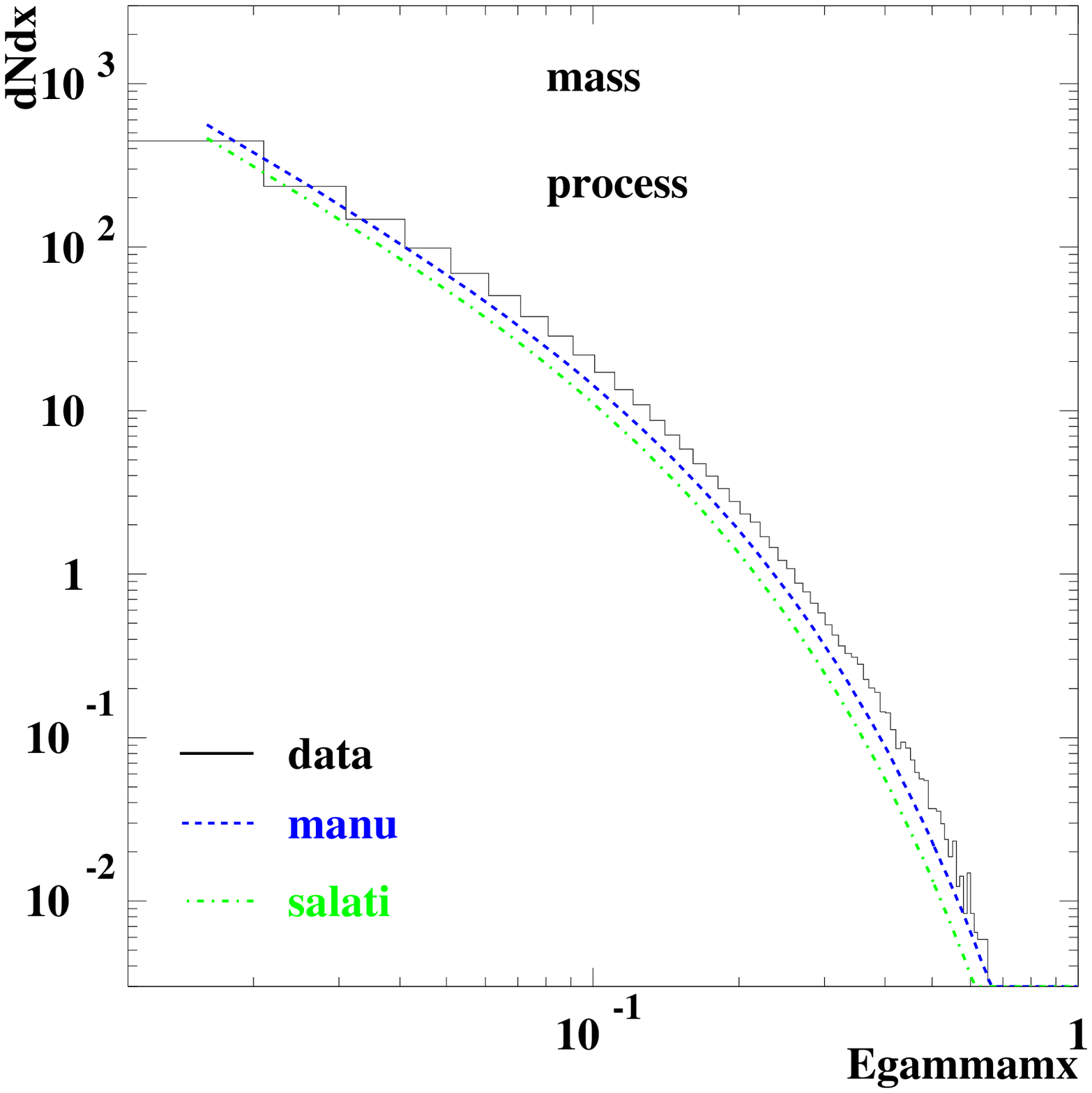}&

\psfrag{dNdx}[rb][ct]{\tiny $dN/d(E_{\gamma}/m_{\chi})$}
\psfrag{Egammamx}[c][c]{\tiny $E_{\gamma}/m_{\chi}$}
\psfrag{mass}[l][l]{\tiny $m_{\chi}=30$ TeV }
\psfrag{process}[l][l]{\tiny $\chi \chi \to b \bar{b}$ }
\psfrag{salati}[l][l]{\tiny {\green Bengtsson et al}}
\psfrag{scopel}[l][l]{\tiny {\darkred Fornengo et al}}
\psfrag{manu}[l][l]{\tiny {\darkblue $ 1.2 e^{-10x}/x^{1.5} $  }}

 \includegraphics[width=0.3\textwidth]{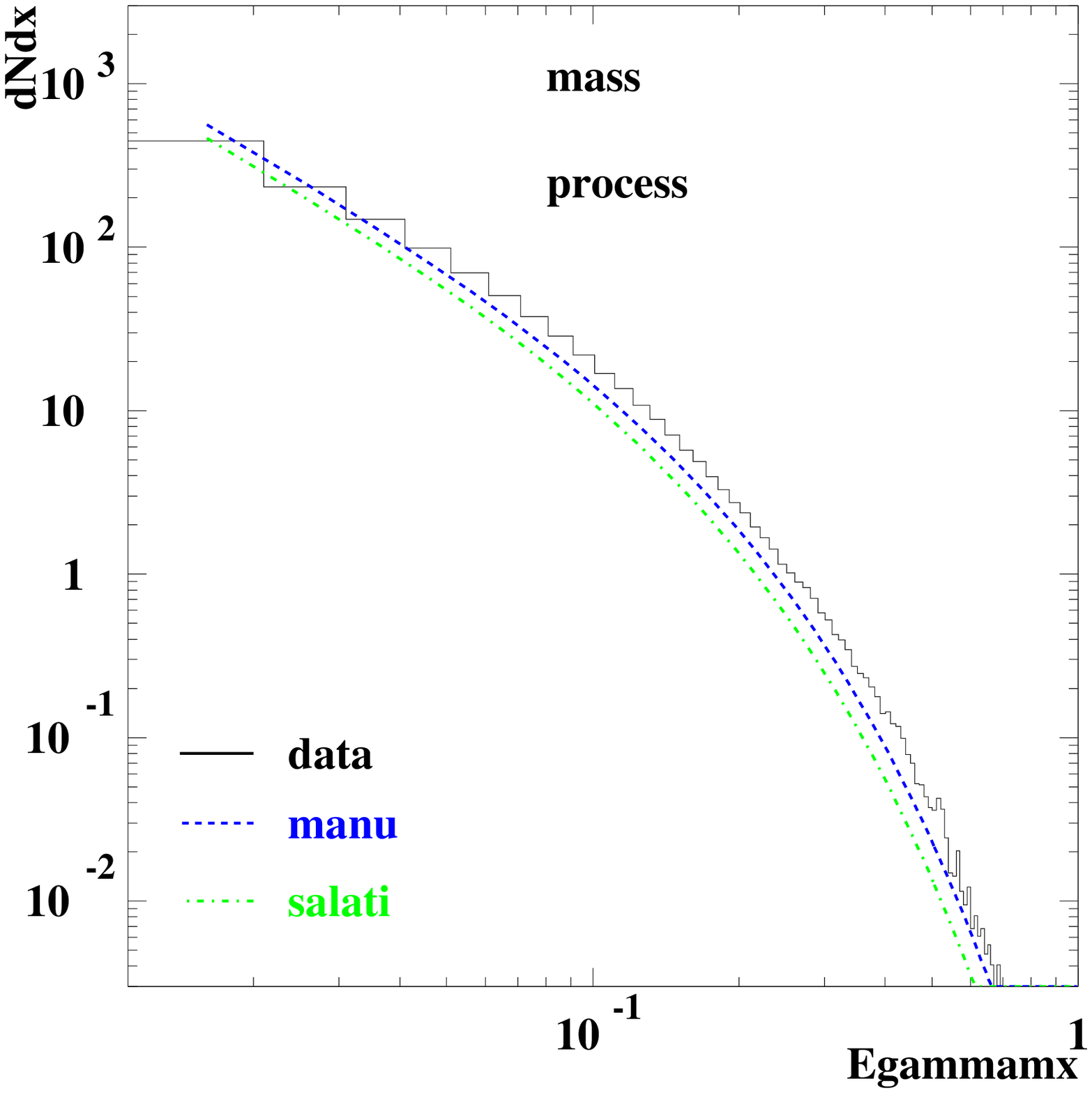}\\
b)&
\psfrag{dNdx}[rb][ct]{\tiny $dN/d(E_{\gamma}/m_{\chi})$}
\psfrag{Egammamx}[c][c]{\tiny $E_{\gamma}/m_{\chi}$}
\psfrag{mass}[l][l]{\tiny $m_{\chi}=1$ TeV }
\psfrag{process}[l][l]{\tiny $\chi \chi \to t \bar{t}$ }
\psfrag{salati}[l][l]{\tiny {\green Bengtsson et al}}
\psfrag{scopel}[l][l]{\tiny {\darkred Fornengo et al}}
\psfrag{manu}[l][l]{\tiny {\darkblue $ 1.27 e^{-14x}/x^{1.5} $  }}

 \includegraphics[width=0.3\textwidth]{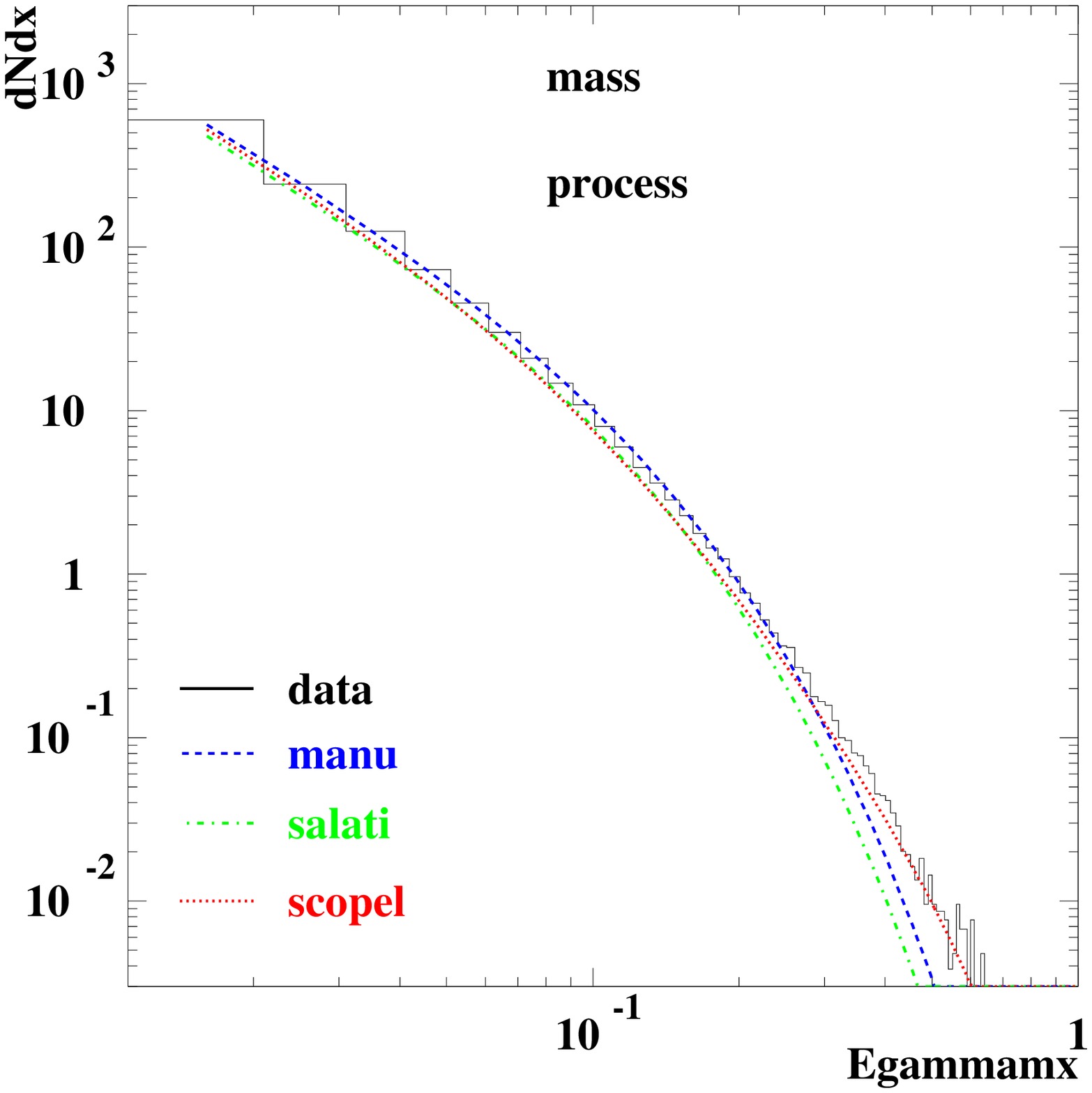}&

\psfrag{dNdx}[rb][ct]{\tiny $dN/d(E_{\gamma}/m_{\chi})$}
\psfrag{Egammamx}[c][c]{\tiny $E_{\gamma}/m_{\chi}$}
\psfrag{mass}[l][l]{\tiny $m_{\chi}=15$ TeV }
\psfrag{process}[l][l]{\tiny $\chi \chi \to t \bar{t}$ }
\psfrag{salati}[l][l]{\tiny {\green Bengtsson et al}}
\psfrag{scopel}[l][l]{\tiny {\darkred Fornengo et al}}
\psfrag{manu}[l][l]{\tiny {\darkblue $ 1.27 e^{-14x}/x^{1.5} $  }}

 \includegraphics[width=0.3\textwidth]{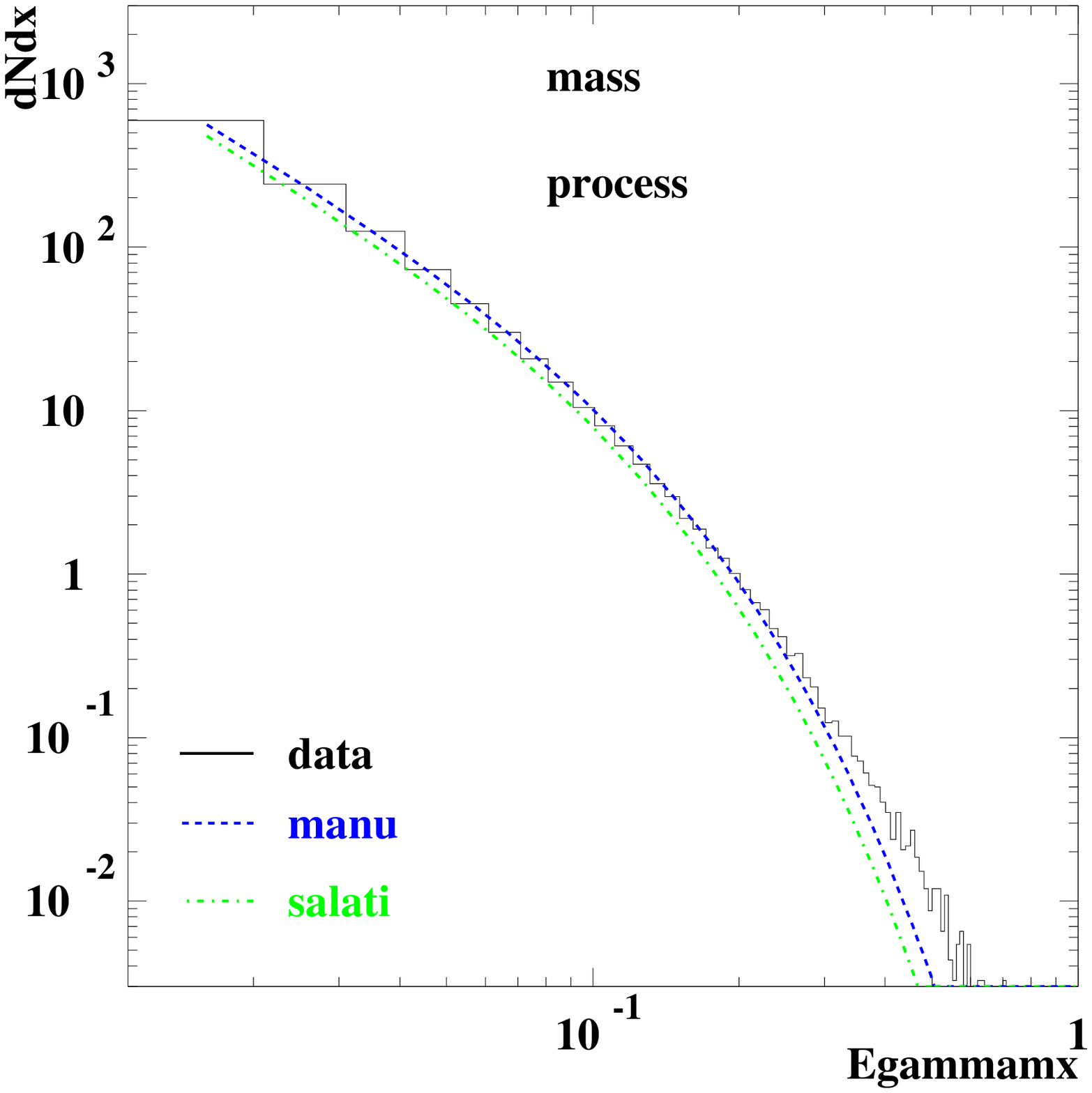}&

\psfrag{dNdx}[rb][ct]{\tiny $dN/d(E_{\gamma}/m_{\chi})$}
\psfrag{Egammamx}[c][c]{\tiny $E_{\gamma}/m_{\chi}$}
\psfrag{mass}[l][l]{\tiny $m_{\chi}=30$ TeV }
\psfrag{process}[l][l]{\tiny $\chi \chi \to t \bar{t}$ }
\psfrag{salati}[l][l]{\tiny {\green Bengtsson et al}}
\psfrag{scopel}[l][l]{\tiny {\darkred Fornengo et al}}
\psfrag{manu}[l][l]{\tiny {\darkblue $ 1.27 e^{-14x}/x^{1.5} $  }}

 \includegraphics[width=0.3\textwidth]{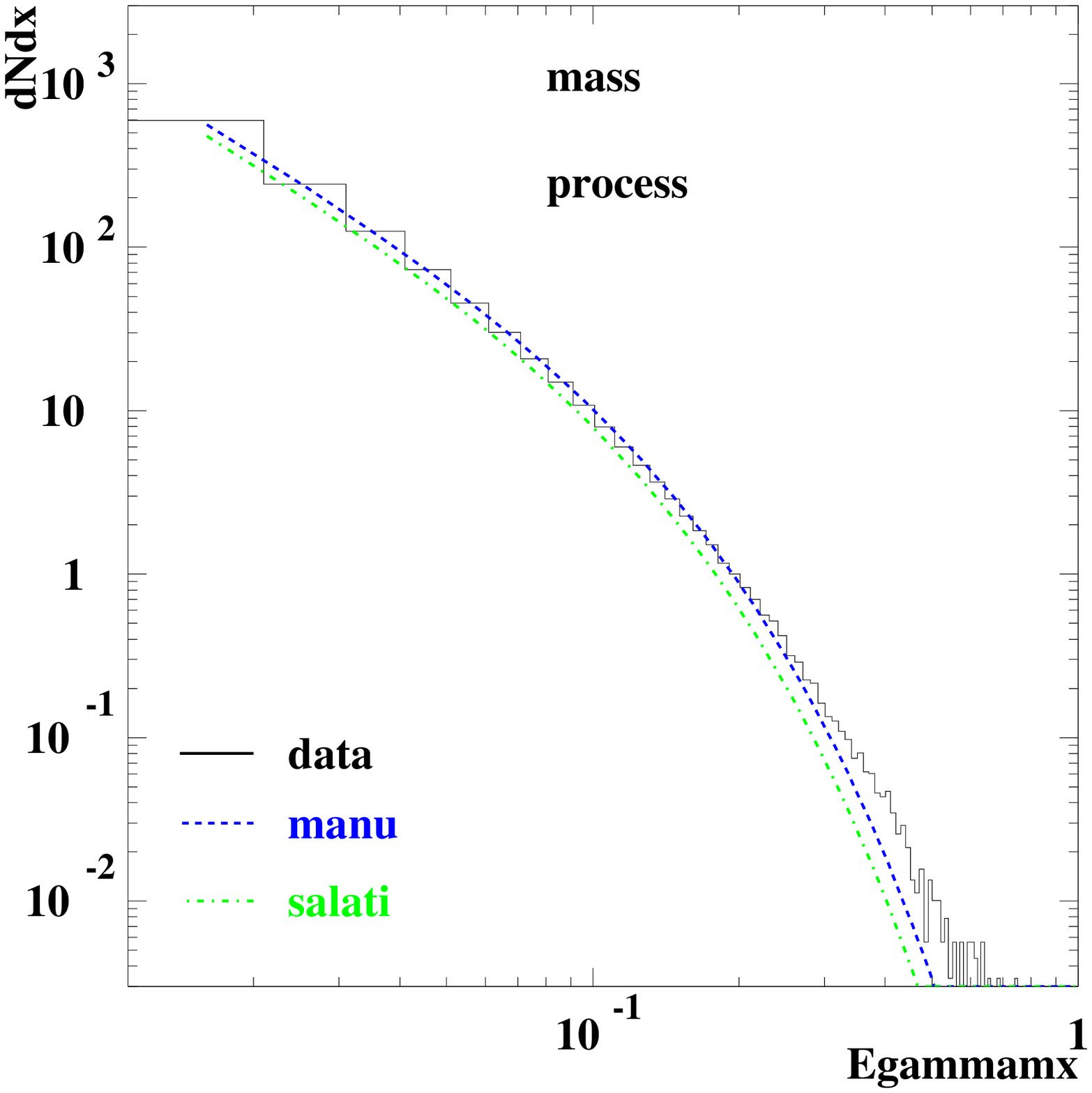}\\

c)&
\psfrag{dNdx}[rb][ct]{\tiny $dN/d(E_{\gamma}/m_{\chi})$}
\psfrag{Egammamx}[c][c]{\tiny $E_{\gamma}/m_{\chi}$}
\psfrag{mass}[l][l]{\tiny $m_{\chi}=1$ TeV }
\psfrag{process}[l][l]{\tiny $\chi \chi \to W^+ W^-$ }
\psfrag{salati}[l][l]{\tiny {\green Bengtsson et al}}
\psfrag{scopel}[l][l]{\tiny {\darkred Fornengo et al}}
\psfrag{manu}[l][l]{\tiny {\darkblue $ 0.82 e^{-8.25x}/x^{1.5} $  }}

 \includegraphics[width=0.3\textwidth]{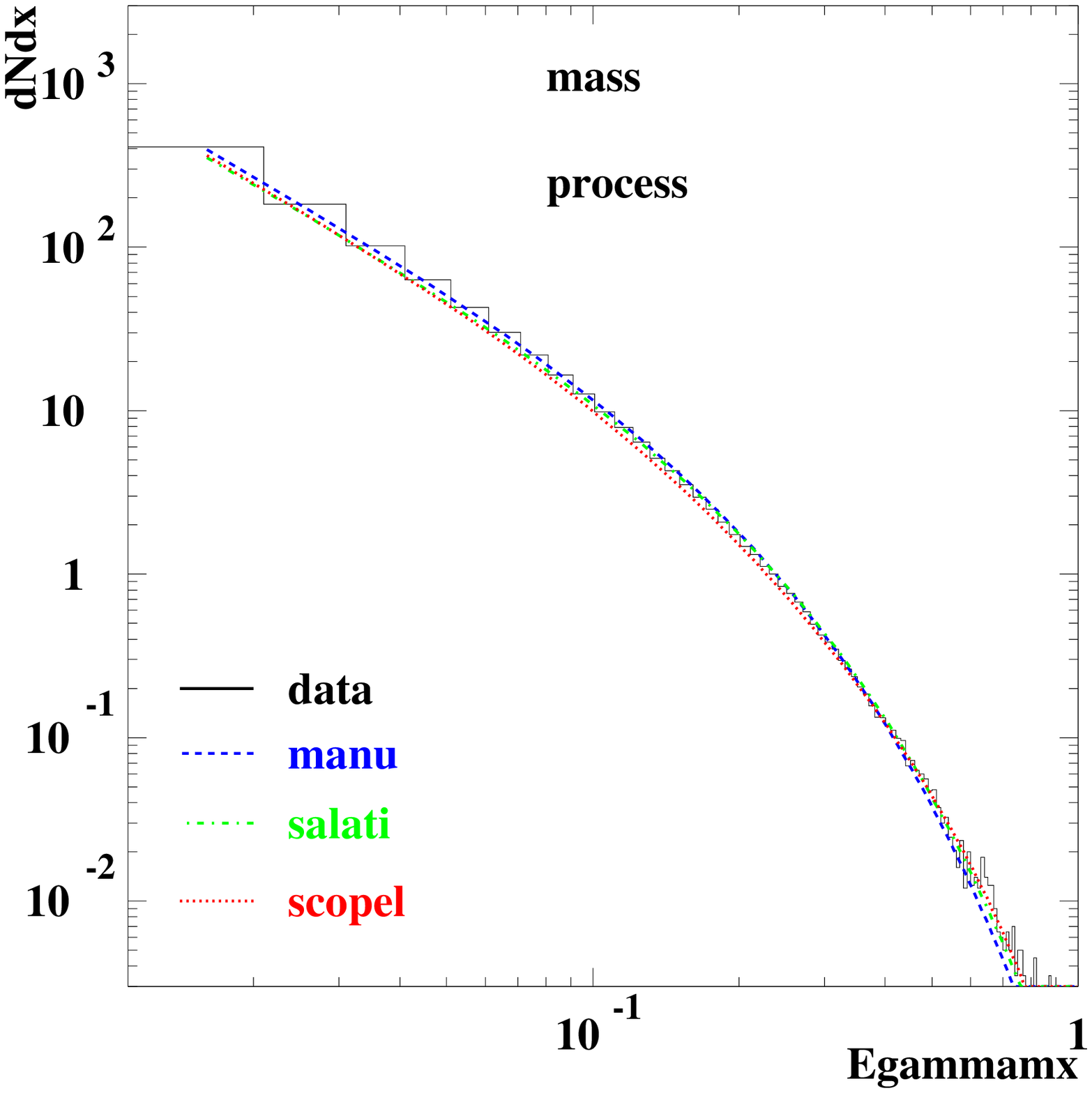}&

\psfrag{dNdx}[rb][ct]{\tiny $dN/d(E_{\gamma}/m_{\chi})$}
\psfrag{Egammamx}[c][c]{\tiny $E_{\gamma}/m_{\chi}$}
\psfrag{mass}[l][l]{\tiny $m_{\chi}=15$ TeV }
\psfrag{process}[l][l]{\tiny $\chi \chi \to W^+ W^-$ }
\psfrag{salati}[l][l]{\tiny {\green Bengtsson et al}}
\psfrag{scopel}[l][l]{\tiny {\darkred Fornengo et al}}
\psfrag{manu}[l][l]{\tiny {\darkblue $ 0.82 e^{-8.25x}/x^{1.5} $  }}

 \includegraphics[width=0.3\textwidth]{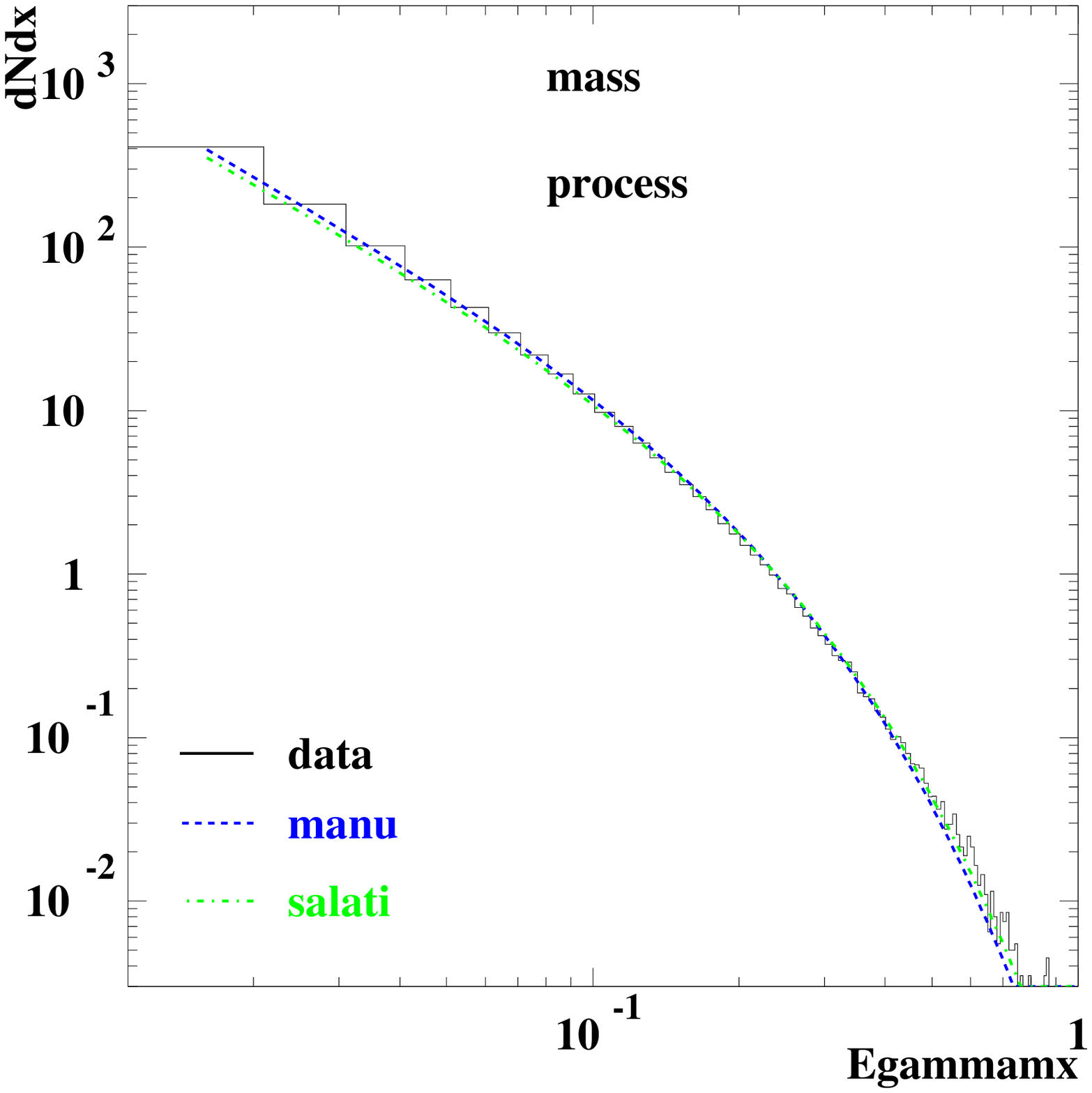}&

\psfrag{dNdx}[rb][ct]{\tiny $dN/d(E_{\gamma}/m_{\chi})$}
\psfrag{Egammamx}[c][c]{\tiny $E_{\gamma}/m_{\chi}$}
\psfrag{mass}[l][l]{\tiny $m_{\chi}=30$ TeV }
\psfrag{process}[l][l]{\tiny $\chi \chi \to W^+ W^-$ }
\psfrag{salati}[l][l]{\tiny {\green Bengtsson et al}}
\psfrag{scopel}[l][l]{\tiny {\darkred Fornengo et al}}
\psfrag{manu}[l][l]{\tiny {\darkblue $ 0.82 e^{-8.25x}/x^{1.5} $  }}

 \includegraphics[width=0.3\textwidth]{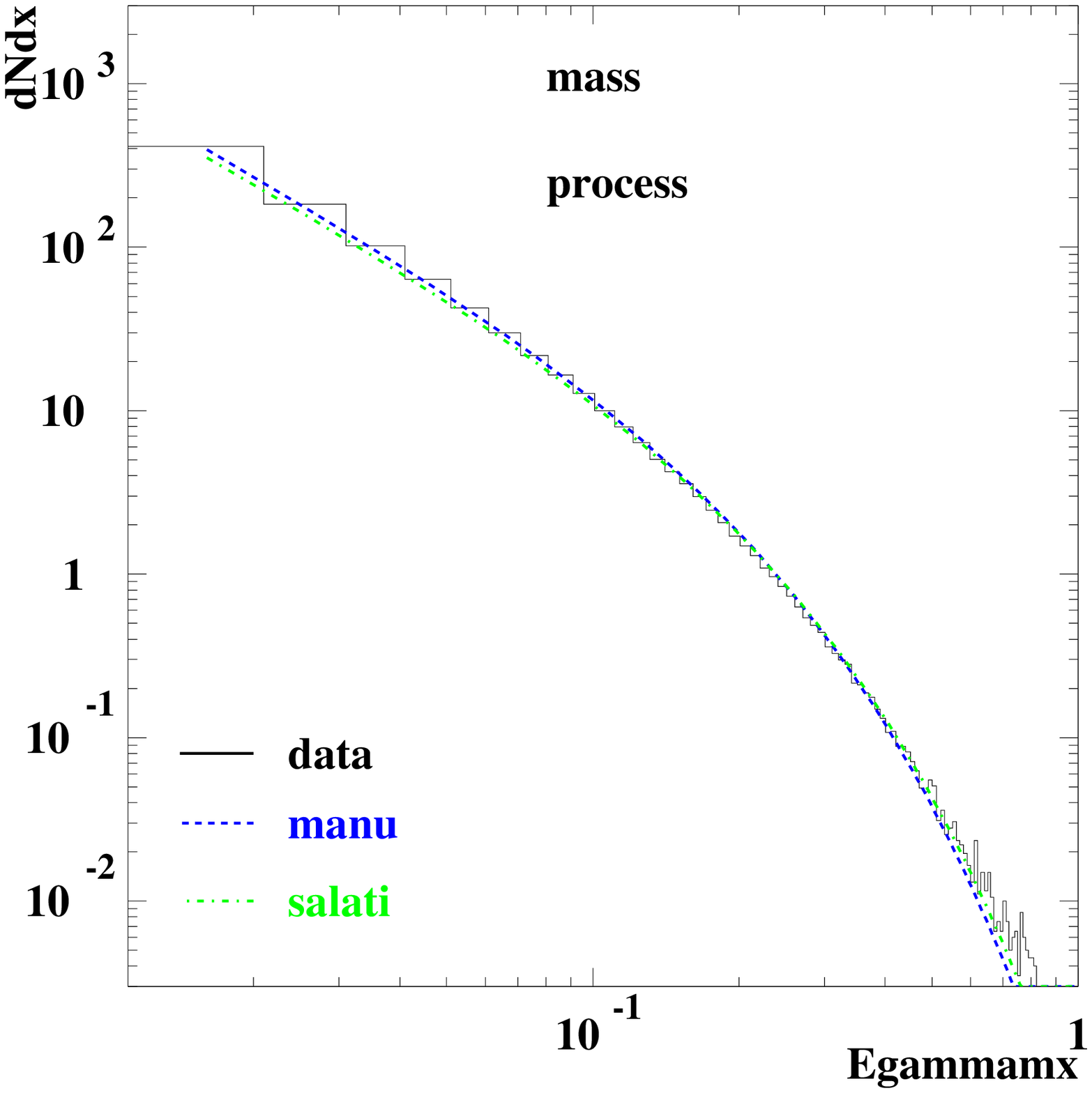}\\
d)&
\psfrag{dNdx}[rb][ct]{\tiny $dN/d(E_{\gamma}/m_{\chi})$}
\psfrag{Egammamx}[c][c]{\tiny $E_{\gamma}/m_{\chi}$}
\psfrag{mass}[l][l]{\tiny $m_{\chi}=1$ TeV }
\psfrag{process}[l][l]{\tiny $\chi \chi \to \tau^+ \tau^-$ }
\psfrag{salati}[l][l]{\tiny {\green Bengtsson et al}}
\psfrag{scopel}[l][l]{\tiny {\darkred Fornengo et al}}
\psfrag{manu}[l][l]{\tiny {\darkblue $ \frac{ e^{-5x}}{ x^{{\strut 1.15}}} (10x-9x^2-2x^3)$  }}

 \includegraphics[width=0.3\textwidth]{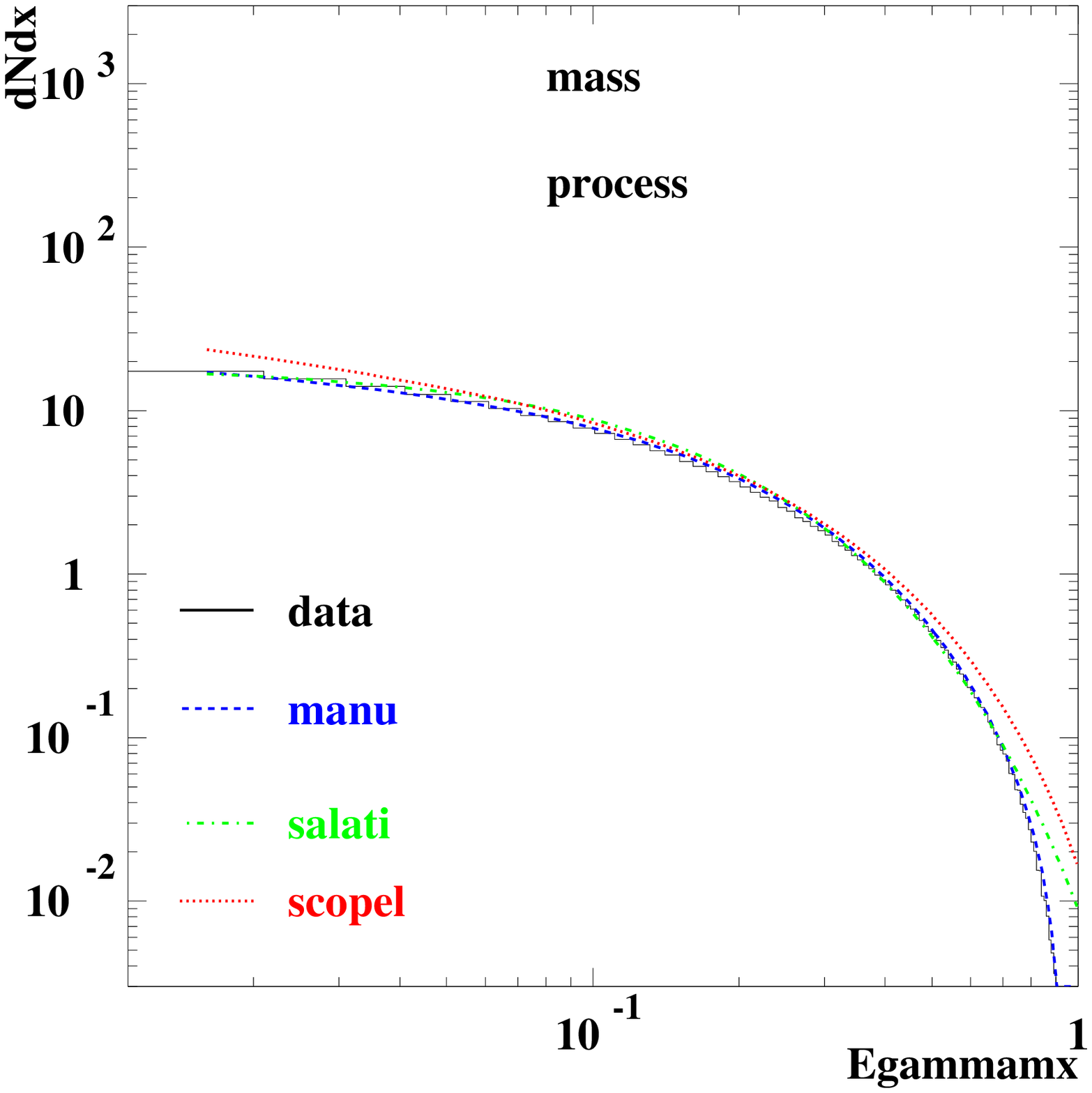}&

\psfrag{dNdx}[rb][ct]{\tiny $dN/d(E_{\gamma}/m_{\chi})$}
\psfrag{Egammamx}[c][c]{\tiny $E_{\gamma}/m_{\chi}$}
\psfrag{mass}[l][l]{\tiny $m_{\chi}=15$ TeV }
\psfrag{process}[l][l]{\tiny $\chi \chi \to \tau^+ \tau^-$ }
\psfrag{salati}[l][l]{\tiny {\green Bengtsson et al}}
\psfrag{scopel}[l][l]{\tiny {\darkred Fornengo et al}}
\psfrag{manu}[l][l]{\tiny {\darkblue $\frac{e^{-5x}}{x^{{\strut 1.15}}} (10x-9x^2-2x^3)$  }}

 \includegraphics[width=0.3\textwidth]{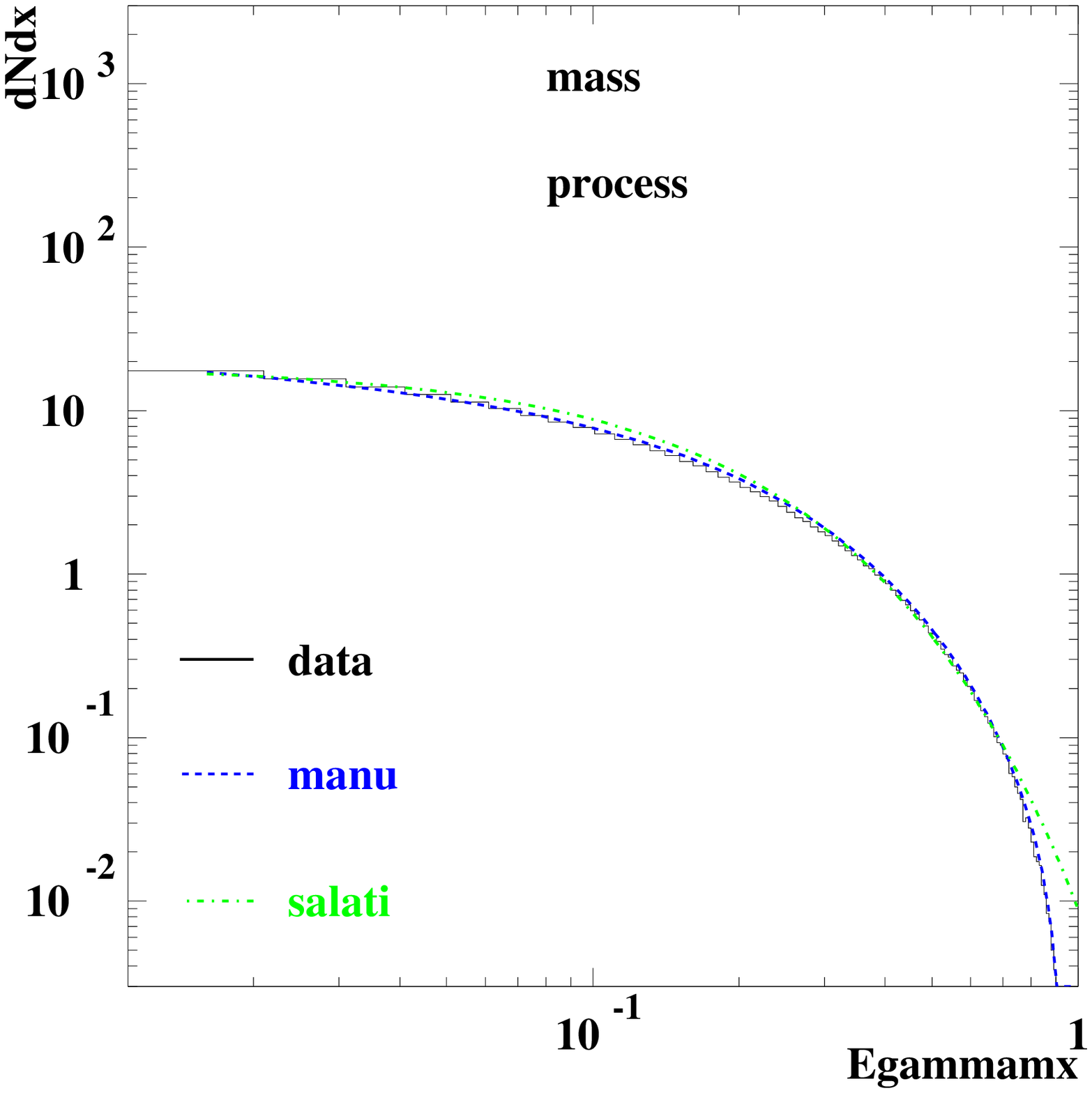}&

\psfrag{dNdx}[rb][ct]{\tiny $dN/d(E_{\gamma}/m_{\chi})$}
\psfrag{Egammamx}[c][c]{\tiny $E_{\gamma}/m_{\chi}$}
\psfrag{mass}[l][l]{\tiny $m_{\chi}=30$ TeV }
\psfrag{process}[l][l]{\tiny $\chi \chi \to \tau^+ \tau^-$ }
\psfrag{salati}[l][l]{\tiny {\green Bengtsson et al}}
\psfrag{scopel}[l][l]{\tiny {\darkred Fornengo et al}}
\psfrag{manu}[l][l]{\tiny {\darkblue $\frac{e^{-5x}}{x^{{\strut 1.15}}} (10x-9x^2-2x^3)$  }}

 \includegraphics[width=0.3\textwidth]{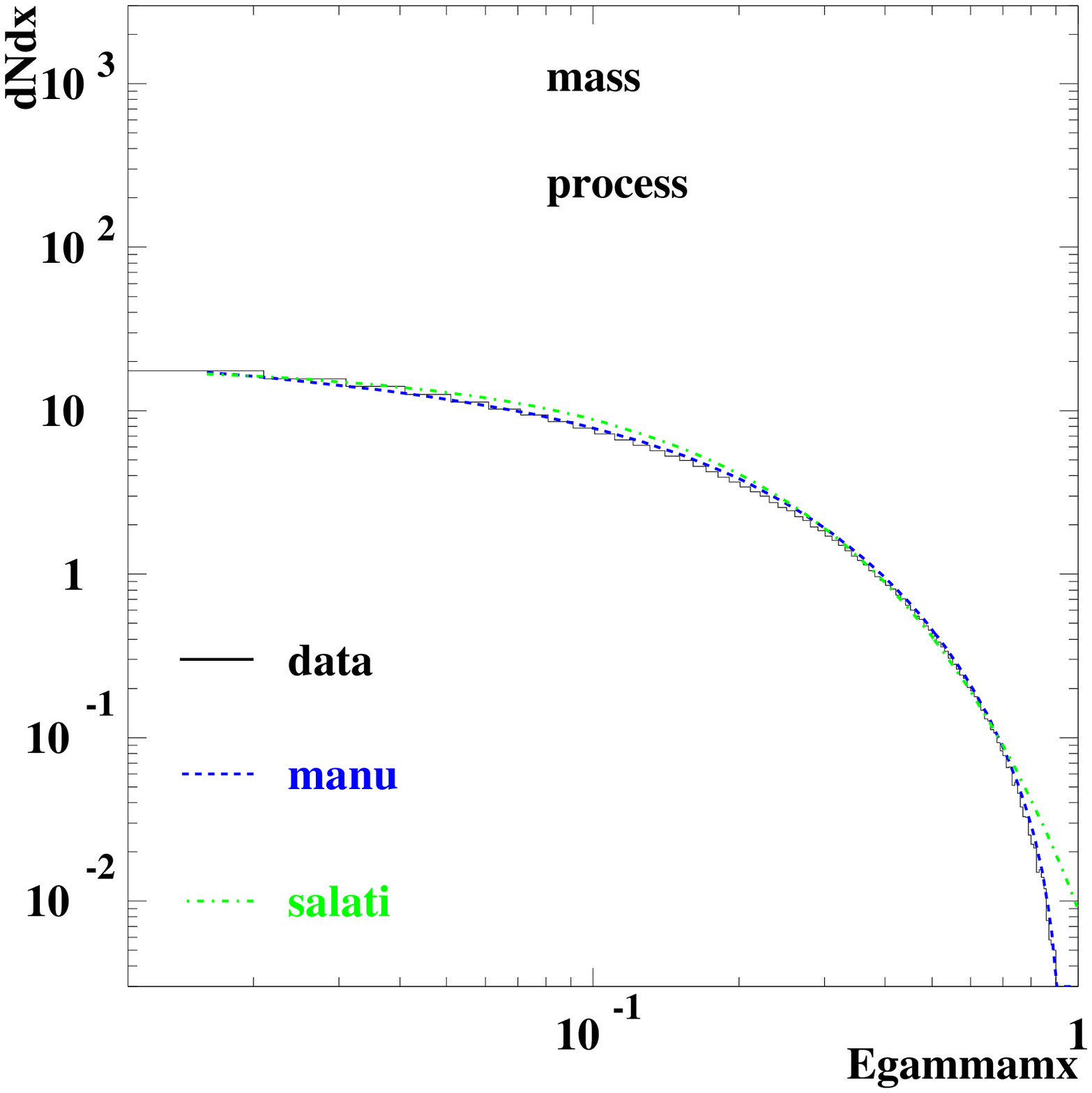}\\

\end{tabular}
\caption{\small PYTHIA simulations for gamma spectrum coming from the decays
  of dark matter particle annihilation products ( a) $b\bar{b}$, b) $t\bar{t}$,
  c) $W^+W^-$, d) $\tau^+\tau^-$ ) for three dark matter particle
  masses: 1,15 and 30 TeV. We show the data, our fit results and also compare with previous studies \cite{Salati,Fornengo}.}
\label{fig:simupythia}
\end{center}
\end{figure}

We performed a simulation with the PYTHIA \footnote{We used january 2005
  release of PYTHIA (version 6.317).} package \cite{pythia} to evaluate the
  gamma-ray fluxes from hadronisation and radiative processes. We consider a
 dark matter particle annihilating at rest in  the $b\bar{b}$, $t\bar{t}$,
  $W^+W^-$ and $\tau^+\tau^-$ final states. We first require a 1 TeV mass
  ({\it i.e.} $E_{cm}$=2 TeV) to check our results
  with previous studies \cite{Salati,Fornengo} and then generate two relevant
  masses for HESS data: 15 TeV and 30 TeV ({\it i.e.} respectively $E_{cm}=$30
  and 60 TeV). The results are shown on figure
  \ref{fig:simupythia}.  Even if more precise fits can be obtained with more
  parameters, the annihilation spectra can
be quite simply approximated by :

\begin{equation}
 \frac{d N^{i}_{\gamma}}{d
  x}=\frac{a_i e^{-b_ix}}{x^{1.5}}
\label{eq:spectrumbtW}  
\end{equation}
where $x=E_{\gamma}/m_{\chi}$ and we obtained $(a,b)=(1.2,10.)$ for $b\bar{b}$
 \footnote{Despite the Hill function, the $b\bar{b}$ case can suggest to relaxe the -1.5 exponent to get better fit.},
 $(a,b)=(1.27,14)$ for $t\bar{t}$ and $(a,b)=(0.82,8.25)$ for $W^+W^-$. For
 $\tau^+\tau^-$ we used the function proposed in \cite{Fornengo} 
\begin{equation}
 \frac{d N^{i}_{\gamma}}{d
  x}= x^{a_{\tau}} (b_{\tau} x + c_{\tau} x^2 + d_{\tau} x^3) e^{-e_{\tau}x}
\label{eq:spectrumtau}  
\end{equation}
to well reproduce the harder $\tau^+\tau^-$ spectrum and found $(a_{\tau},b_{\tau},c_{\tau},d_{\tau},e_{\tau})=(-1.15,10,-9.-2,5)$.

The gamma ray flux was expressed in Eq. (\ref{Eq:totflux}),
with $\langle \sigma_i v \rangle= BR_i \langle \sigma v \rangle$, $BR_i$ being
the branching ratio of annihilation channel $i$.

\begin{figure}
\begin{center}
\begin{tabular}{cc}
\psfrag{NFWcomp}[l][l]{\scriptsize ${\rm NFW_c}$}
\psfrag{LE2dNdE}[c][c][1][90]{\scriptsize $\log_{10}{[E^2 dN/dE \ ({\rm
      m^{-2}s^{-1} TeV})]}$}
%\psfrag{LogETeV}[c][b][1][1]{\scriptsize $\log_{10}{[E({\rm TeV})]}$}
\psfrag{LogETeV}[c][b][1][1]{\scriptsize $ E\ ({\rm TeV})$}
\psfrag{sigmav255}[c][l]{\tiny $\langle\sigma v \rangle=3\times 10^{-26}\ {\rm cm^3.s^{-1}}$}
\psfrag{15}[c][c]{\tiny $m_{\chi}=15\ {\rm TeV}$}
\psfrag{TeV}{}
\psfrag{tautau}[c][c]{\tiny {\purple $\tau^+ \tau^-$}}
\psfrag{WW}[c][r]{\tiny {\green $W^+ W^-$}}
\psfrag{bbbar}[c][c]{\tiny {\darkblue $b \bar{b}$}}
\psfrag{ttbar}[c][c]{\tiny {\darkred $t \bar{t}$}}
 \includegraphics[width=0.5\textwidth]{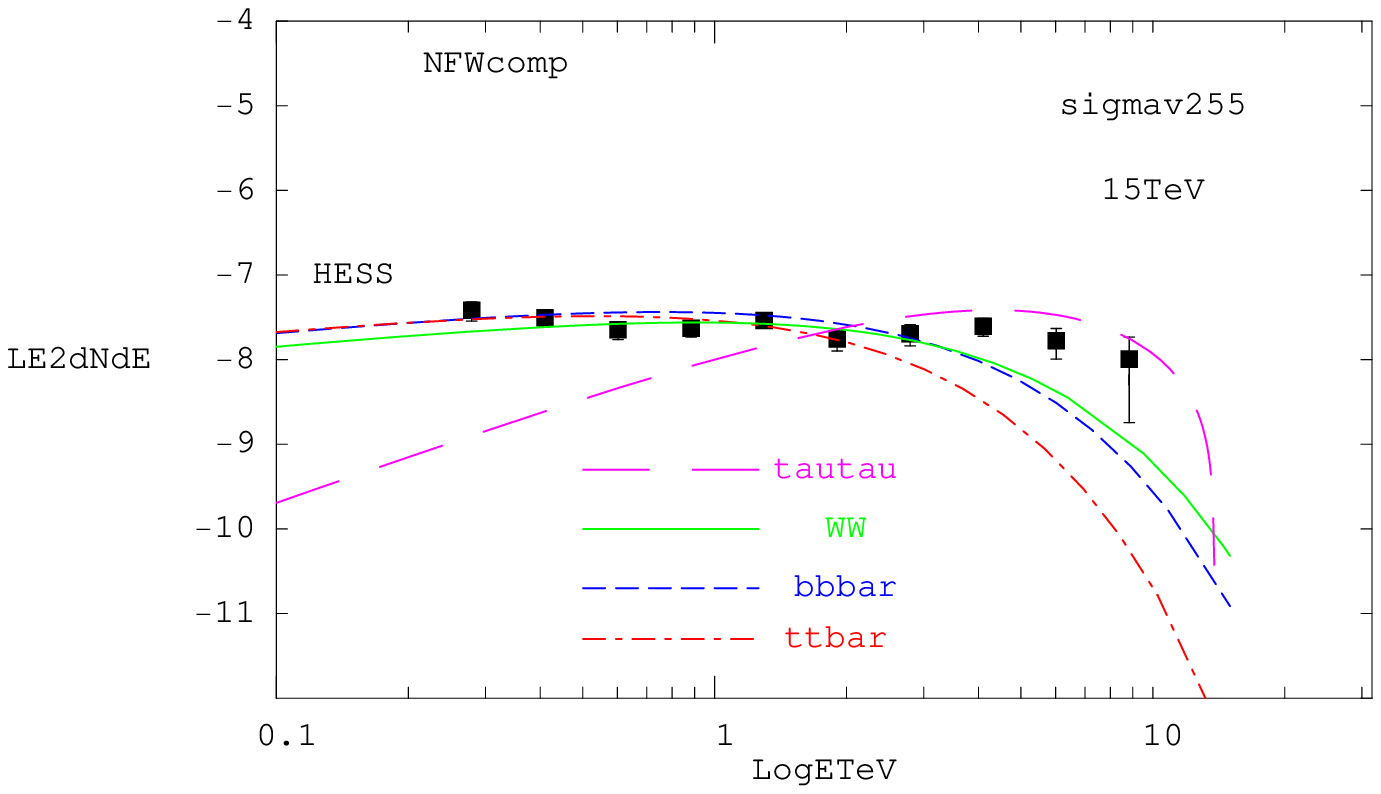}&
\psfrag{NFWcomp}[l][l]{\scriptsize ${\rm NFW_c}$}
\psfrag{LE2dNdE}[c][c][1][90]{\scriptsize $\log_{10}{[E^2 dN/dE \ ({\rm m^{-2}s^{-1} TeV})]}$}
%\psfrag{LogETeV}[c][b][1][1]{\scriptsize $\log_{10}{[E({\rm TeV})]}$}
\psfrag{LogETeV}[c][b][1][1]{\scriptsize $ E\ ({\rm TeV})$}
\psfrag{sigmav25}[c][l]{\tiny $\langle\sigma v \rangle=10^{-25}\ {\rm cm^3.s^{-1}}$}
\psfrag{30}[c][c]{\tiny $m_{\chi}=30\ {\rm TeV}$}
\psfrag{TeV}{}
\psfrag{tautau}[c][c]{\tiny {\purple $\tau^+ \tau^-$}}
\psfrag{WW}[c][r]{\tiny {\green $W^+ W^-$}}
\psfrag{bbbar}[c][c]{\tiny {\darkblue $b \bar{b}$}}
\psfrag{ttbar}[c][c]{\tiny {\darkred $t \bar{t}$}}
\includegraphics[width=0.5\textwidth]{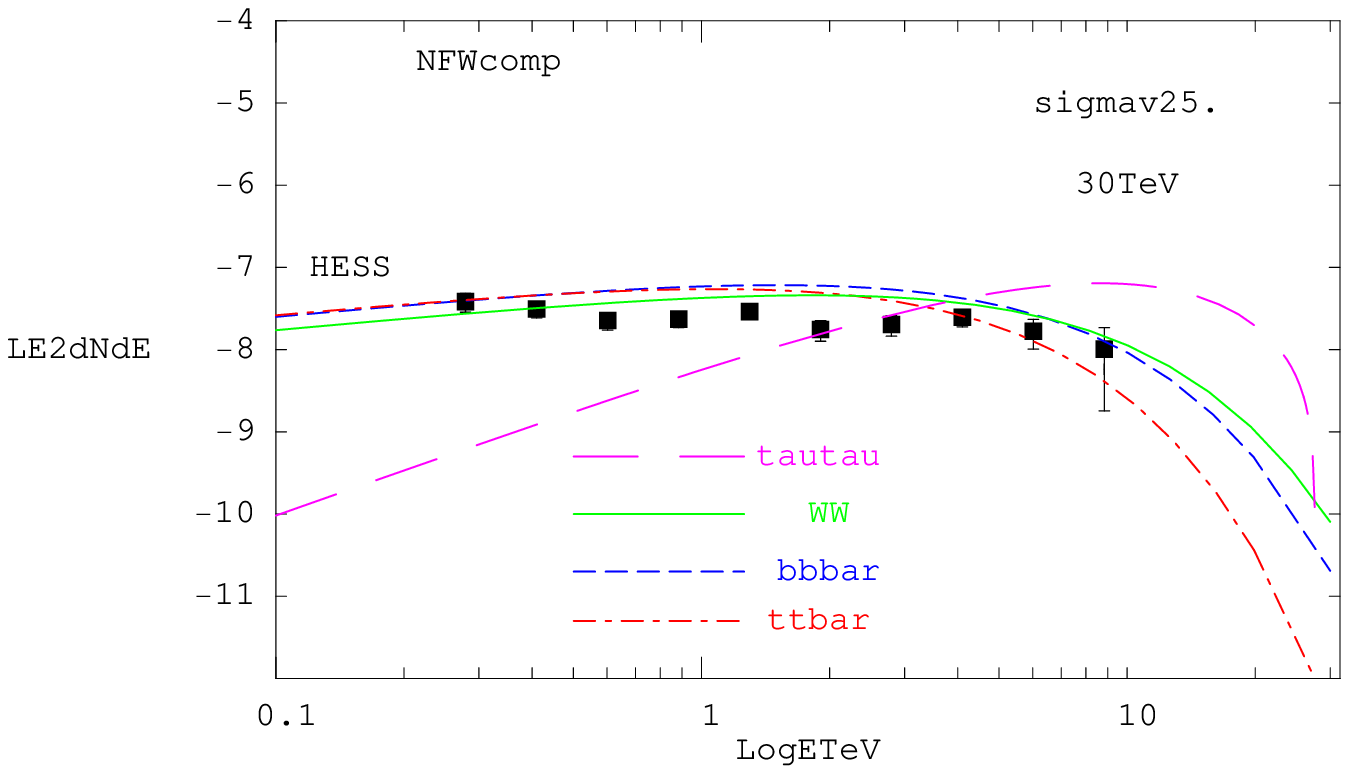}\\
%a) $\langle\sigma v
%  \rangle=10^{-26}\ {\rm cm^3.s^{-1}}$ & b) $\langle\sigma v
%  \rangle=10^{-25}\ {\rm cm^3.s^{-1}}$\\
a)&b)\\
&\\
\psfrag{Moorecomp}[l][l]{\scriptsize ${\rm Moore_c}$}
\psfrag{LE2dNdE}[c][c][1][90]{\scriptsize $\log_{10}{[E^2 dN/dE \ ({\rm
      m^{-2}s^{-1} TeV})]}$}
%\psfrag{LogETeV}[c][b][1][1]{\scriptsize $\log_{10}{[E({\rm TeV})]}$}
\psfrag{LogETeV}[c][b][1][1]{\scriptsize $ E\ ({\rm TeV})$}
\psfrag{sigmav27}[c][l]{\tiny $\langle\sigma v \rangle=10^{-27}\ {\rm cm^3.s^{-1}}$}
\psfrag{15}[c][c]{\tiny $m_{\chi}=15\ {\rm TeV}$}
\psfrag{TeV}{}
\psfrag{tautau}[c][c]{\tiny {\purple $\tau^+ \tau^-$}}
\psfrag{WW}[c][r]{\tiny {\green $W^+ W^-$}}
\psfrag{bbbar}[c][c]{\tiny {\darkblue $b \bar{b}$}}
\psfrag{ttbar}[c][c]{\tiny {\darkred $t \bar{t}$}}
 \includegraphics[width=0.5\textwidth]{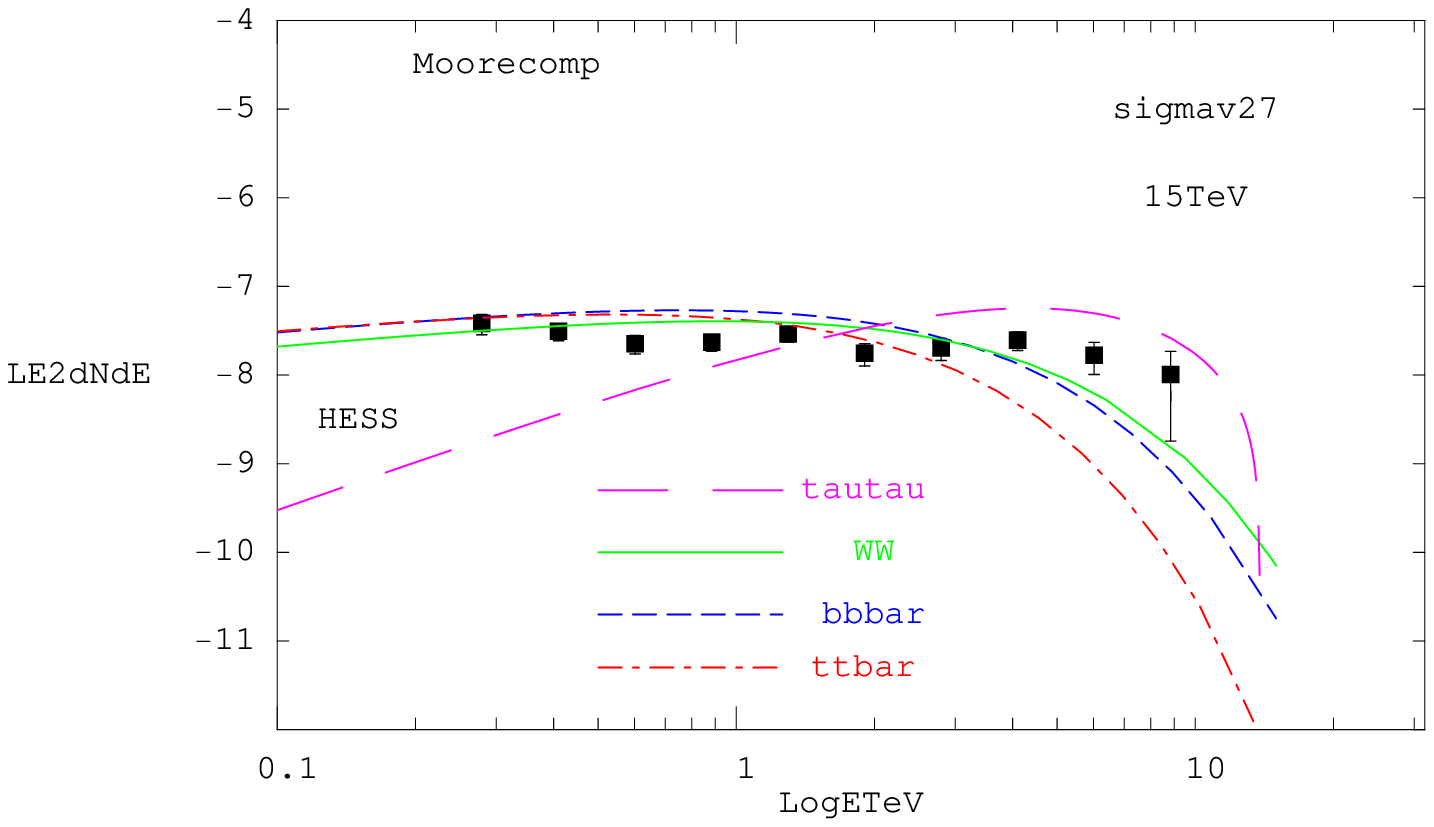}&
\psfrag{Moorecomp}[l][l]{\scriptsize ${\rm Moore_c}$}
\psfrag{LE2dNdE}[c][c][1][90]{\scriptsize $\log_{10}{[E^2 dN/dE \ ({\rm m^{-2}s^{-1} TeV})]}$}
%\psfrag{LogETeV}[c][b][1][1]{\scriptsize $\log_{10}{[E({\rm TeV})]}$}
\psfrag{LogETeV}[c][b][1][1]{\scriptsize $ E\ ({\rm TeV})$}
\psfrag{sigmav27}[c][l]{\tiny $\langle\sigma v \rangle=10^{-27}\ {\rm cm^3.s^{-1}}$}
\psfrag{30}[c][c]{\tiny $m_{\chi}=30\ {\rm TeV}$}
\psfrag{TeV}{}
\psfrag{tautau}[c][c]{\tiny {\purple $\tau^+ \tau^-$}}
\psfrag{WW}[c][r]{\tiny {\green $W^+ W^-$}}
\psfrag{bbbar}[c][c]{\tiny {\darkblue $b \bar{b}$}}
\psfrag{ttbar}[c][c]{\tiny {\darkred $t \bar{t}$}}
\includegraphics[width=0.5\textwidth]{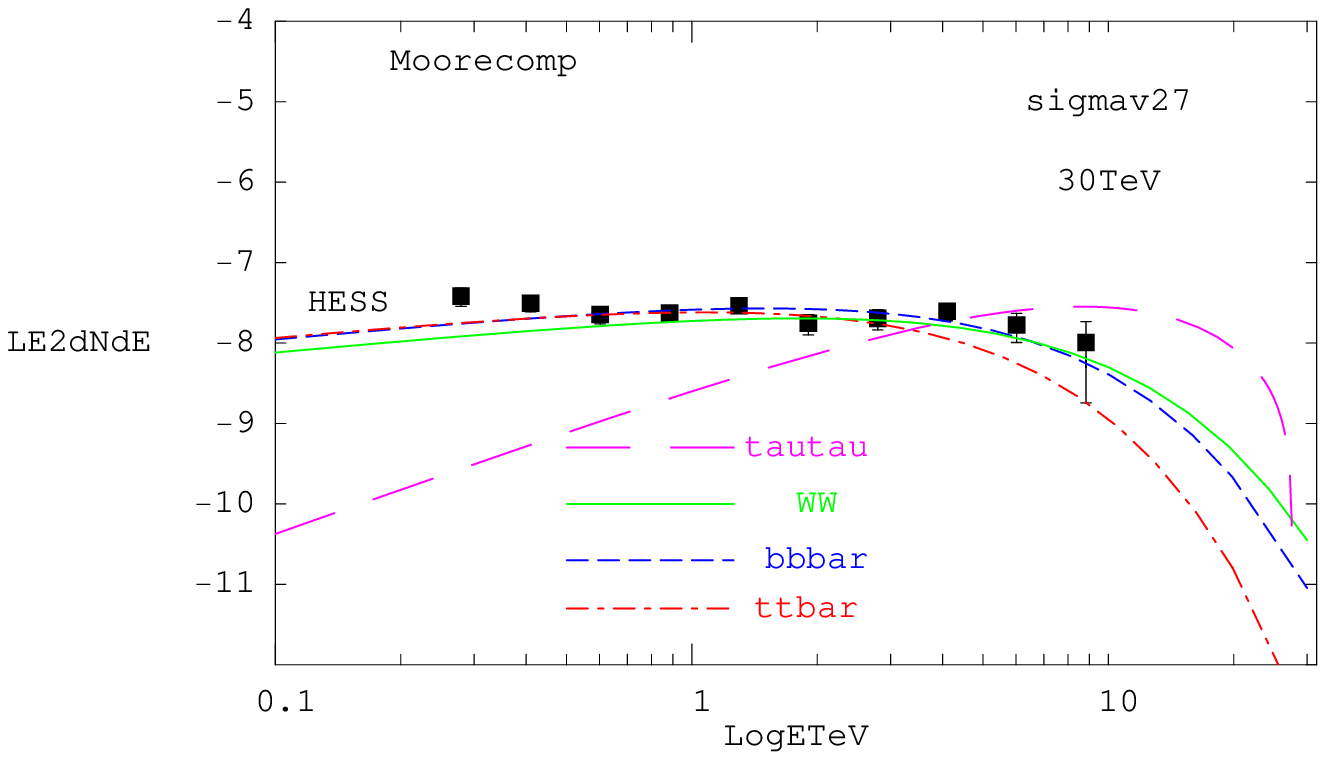}\\
%c) $\langle\sigma v
%  \rangle=10^{-26}\ {\rm cm^3.s^{-1}}$ & d) $\langle\sigma v
%  \rangle=10^{-26.7}\ {\rm cm^3.s^{-1}}$ \\
c)&d)\\
\end{tabular}
\caption{\small Dark matter annihilation versus HESS data for :  NFW compressed in a) and b),
  Moore et al. compressed  in c) and d).}
\label{fig:HESSmodelsanddatas}
\end{center}
\end{figure}

\noindent
With this parameterization, the gamma flux can be rewritten as

\begin{equation}
\frac{d\Phi_{\gamma}}{dE_{\gamma}} ({\rm cm^{-2} s^{-1} GeV^{-1}})= \frac{0.94 \times 10^{-13}}{ m_{\chi}({\rm GeV})}\sum_i  \frac{d N^{i}_{\gamma}}{d
  x} BR_i \left( \frac{\langle \sigma v \rangle}{10^{-29}{\rm cm^3
      s^{-1}}} \right) \left(\frac{100 {\rm GeV}}{m_{\chi}} \right)^2
\bar{J}(\Delta \Omega) \Delta \Omega , 
\end{equation}

\noindent
where $m_{\chi}$, $\langle \sigma v \rangle$, $BR_i$ 
are now 
considered as free parameters 
and the astrophysics contribution $\bar{J}$ 
is fixed by our halo density profiles.
The results are shown in Fig. \ref{fig:HESSmodelsanddatas}. 

We point the
fact that we do not only
 compare the spectrum shape of the signal with possible dark matter 
annihilation explanation. Indeed, it should be noticed that our compressed
halo profiles give rise to absolute gamma fluxes within the HESS data order of
magnitude with $\langle \sigma v \rangle$ values in agreement with the WMAP
requirement.

According to our analysis, interpreting the HESS data in terms of dark
matter particle annihilation is possible for a mass in the range
10 TeV $\lesssim m_{\chi} \lesssim$ 30 TeV and annihilation should be
dominated by $b\bar{b}$, $t\bar{t}$ or $W^+W^-$ final states with a possible
$\tau^+\tau^-$ contribution 
\footnote{In the MSSM,
  $\frac{BR_{\tau^+\tau^-}}{BR_{b\bar{b}}} \sim
  \frac{1}{3}\frac{m^2_{\tau}}{m^2_b}\sim 0.15$ for pseudo scalar exchange.}. 
The values for $\langle \sigma v \rangle$ required for NFW compressed and Moore
compressed are respectively around $10^{-25}\ {\rm cm^3
      s^{-1}}$ and $10^{-27}\ {\rm cm^3
      s^{-1}}$.

We also tried to see if it was possible to explain at the same time both
HESS and EGRET excess by invoking a very heavy neutralino. We can easily
predict the differential flux around 1 GeV for a 15 TeV neutralino.
Using Eq. (\ref{eq:spectrumbtW}) for a $t \bar{t}$ distribution (the one
producing the less energetic photons) normalised with the HESS data
 in the 2.6 TeV energy bin 
($2. \times 10^{-16} \rm{cm^{-2} ~ s^{-1} GeV^{-1}}$) 
we obtain :
$dN_{\gamma}^t / dE_{\gamma} (1 \mathrm{GeV}) \sim 7.1 \times 10^{-12} 
\rm{cm^{-2} ~ s^{-1} GeV^{-1}}$, which is far
below the EGRET excess data in the 1 GeV energy bin 
($1.52 \times 10^{-7}\rm{cm^{-2} ~ s^{-1} GeV^{-1}}$). 
At the same time, we have numerically checked that such a heavy neutralino cannot be invoked
to explain the EGRET excess in any of its energy bin. 
During this work, we saw that in a recent note added in \cite{Fornengo},
the authors reach the same conclusion.

\section{Conclusions}

We have analysed the effect of the compression of the dark matter
due to the infall of baryons to the galactic center on the
gamma-ray flux. In addition, we have also consider the effect of non-universal 
soft terms.
This analysis shows that neutralino dark matter 
annihilation can give rise to signals largely reachable by
future experiments like GLAST. 
In particular, 
we observe that, basically, the whole parameter space of general SUGRA
will be tested by GLAST. 
%Note that 
Even 
for the mSUGRA case, points corresponding
to a value of tan$\beta$ as small as 5
will be tested. 
%reached  by GLAST.
This is a remarkable result if we realise that 
%interesting 
direct
detection experiments
%, such as e.g. EDELWEIS-II or CDMS-II, 
will
only be able to cover a small region of the parameter space.
%More poweful detectors such as GENIUS will cover larger regions,
%but still not too large compared with the parameter space of general
%SUGRA. 
In contrast, we have observed that indirect detection
experiments
will be able to test the parameter space of 
SUGRA.
%\footnote{See also Ref.~\cite{Mambrini:2004kv} 
%for a comparison between direct
%and
%indirect dark matter search without using compressed halo models.}.
Let us point out that in this analysis we have
taken into account all the recent 
experimental constraints, such as the lower bound on the Higgs mass,
the \bsg\ branching ratio, the muon $g-2$
and the recently improved upper bound on $B(B_s \to \mu^+ \mu^-)$,
as well as the astrophysical (WMAP) bounds on the dark matter density.

Actually, in this SUGRA framework we have also been able to  
fit present excess from EGRET and CANGAROO using different
non-universal scenarios,
and even fit the data from both experiments with only one scenario.
We have also studied the recent HESS data implying a neutralino
heavier than 12 TeV. Because of such a heavy neutralino,
it is not natural to find solutions in the SUGRA framework.
Nevertheless we have carried out a quite model-independent
analysis, and found the conditions required on the particle physics
side to fit the HESS data thanks to dark matter annihilation.

In any case, we must keep in mind that the current data obtained by the 
different gamma-rays observations from the galactic
center region by these three experiments
%EGRET, CANGAROO or HESS 
do not allow us to conclude about a dark
matter annihilation origin rather than other, less exotic astrophysics sources. 
Fortunately, this situation may change with 
the improvement of angular 
resolution and energy sensitivity 
of future detectors like GLAST, which will be a complementary
source of gamma-ray data.
% after EGRET and HESS, 
%as every signal seems to agree reasonably with a particle physics
%spectrum. 

%{\bf manu : 
%
%\noindent
%1) do we have to say here that data are in strong conflict with cusp ?
%
%\noindent
%2) HESS next futur data ...  unpublished not ... official : 
%
%\noindent
%- points at higher energy with same spectral index ... not too
%  bad for dark matter ... 
%
%\noindent
%- points at lower energy 100-200 GeV with same spectral index ... more (too ?) 
%difficult for dark matter
%
%
%\noindent
%-extension of the source in the galactic plane ... very bad for/the end of(?)
%dark matter signal interpretation  (still preliminary)
%
%
%}

\vspace*{1cm}
\noindent{\bf Acknowledgements}

Y. Mambrini thanks specially J. Paul
for useful discussions.
E. Nezri warmly thanks C. Woodswolf-Castanier for his essential 
help and  F.-S. Ling for usefull discussions.
The work of Y. Mambrini was financially supported 
by the DESY Theory Group. E. Nezri work was supported by the I.I.S.N. and the 
Belgian Federal Science Policy (return grant and IAP 5/27).
The work of C. Mu\~noz was supported 
in part by the Spanish DGI of the
MEC under Proyectos Nacionales BFM2003-01266 and FPA2003-04597;
also by the European Union under the RTN program  
MRTN-CT-2004-503369, and under the ENTApP Network of the ILIAS
project
RII3-CT-2004-506222; and also by KAIST under the Visiting Professor Program.

%\newpage

\nocite{}
\bibliography{bmn}

\begin{thebibliography}{99}



\bibitem{Persic} For reviews, see e.g.
  M. Persic, P. Salucci and F. Stel, 
  `The universal rotation curve of spiral galaxies: 
  I. The dark matter connection',
  {\it Mon. Not. Roy. Astron. Soc.} {\bf 281} (1996) 27
  [arXiv:astro-ph/9506004];
  %%CITATION = ASTRO-PH 9506004;%%

  \noindent 
  Y. Sofue and V. Rubin,
  `Rotation curves of spiral galaxies',
  [arXiv:astro-ph/0010594];
  %%CITATION = ASTRO-PH 0010594;%%
  
  \noindent
  M. Roncadelli,
  `Searching for dark matter',
  [arXiv:astro-ph/0307115];
  %%CITATION = ASTRO-PH 0307115;%%
  
  \noindent
  P. Salucci and M. Persic, 
  `Dark matter halos around galaxies',
  [arXiv:astro-ph/9703027];
  %%CITATION = ASTRO-PH 9703027;%%
  
  \noindent
  E. Battaner and E. Florido, 
  `The rotation curve of spiral galaxies and its cosmological
  implications', {\it Fund. Cosmic Phys.} {\bf 21} (2000) 1
  [arXiv:astro-ph/0010475].
  %%CITATION = ASTRO-PH 0010475;%%
   
  
\bibitem{Freedman} For a review, see  
  W.~L. Freedman,  
  `Determination of cosmological parameters',
  {\it Phys. Scripta} {\bf T85} (2000) 37
  [arXiv:astro-ph/9905222].
  %%CITATION = ASTRO-PH 9905222;%%
   
    
\bibitem{wmap03-1}
  WMAP Collaboration,
  C.~L. Bennett et al., `First year wilkinson microwave anisotropy probe
  (WMAP) observations: preliminary maps and basic results',
  {\it Astrophys. J. Suppl.} {\bf 148} (2003) 1 [arXiv:astro-ph/0302207];
  D.~N. Spergel et al., `First year wilkinson microwave anisotropy 
  probe (WMAP) observations: determination of cosmological parameters',
  {\it Astrophys. J. Suppl.} {\bf 148} (2003) 175 [arXiv:astro-ph/0302209];
  %%CITATION = ASTRO-PH 0302209;%%
  L. Verde et al., `First year Wilkinson microwave anisotropy
  probe (WMAP) observations: parameter estimation methodology',
  {\it Astrophys. J. Suppl.} 148 (2003) 195 [arXiv:astro-ph/0302218].
  %%CITATION = ASTRO-PH 0302218;%%
 
%\bibitem{mireview} For a recent review, see
%  C. Mu-b-b\~A\^A\pm oz,-A-A
%  `Dark matter detection in the light of recent experimental
%  results', 
%to appear in {\it Int. J. Mod. Phys.} {\bf A}, [arXiv:hep-ph/0309346].
%  %%CITATION = HEP-PH 0309346;%%


\bibitem{nuevas}
J.~E. Gunn, B.~W. Lee, I. Lerche, D.~N. Schramm and G. Steigman,
`Some astrophysical consequences of the existence of a heavy stable
neutral lepton', 
{\it Astrophys. J.} {\bf 223} (1978) 1015;

\noindent 
F.~W. Stecker,
`The cosmic gamma-ray background from the annihilation of primordial
stable neutral heavy leptons', 
{\it Astrophys. J.} {\bf 223} (1978) 1032; 

\noindent
Y. Zeldovich, A. Klypin, M. Khlopov and V. Chechetkin
`Astrophysical bounds on the mass of heavy stable neutral leptons', 
{\it Yadernaya Fizika} (1980) {\bf 31}, 1286. [English translation: 
{\it Sov. J. Nucl. Phys.}
(1980) {\bf 31}, 664];

\noindent 
M. Srednicki, S. Theisen and J. Silk,
`Cosmic quarkonium: A probe of dark matter'
{\it Phys. Rev. Lett.} {\bf 56} (1986) 263, 
Erratum-ibid. {\bf 56} (1986) 1883;

\noindent
S. Rudaz,
`Cosmic production of quarkonium?',
{\it Phys. Rev. Lett.} {\bf 56} (1986) 2128; 

\noindent
M.~S. Turner,
`Probing the structure of the galactic halo with gamma rays produced
by annihilations of weakly interacting massive particle',
{\it Phys. Rev.} {\bf D34} (1986) 34;

\noindent
L. Bergstrom and H. Snellman,
`Observable monochromatic photons from cosmic photino annihilation',
{\it Phys. Rev.} {\bf D37} (1988) 3737; 




\noindent 
F.~W. Stecker, 
{\it Phys. Lett.} {\bf B201} (1988) 529;

\noindent F.~W. Stecker and A.~J. Tylka,
`Spectra, fluxes and observability of gamma-rays from dark matter
annihilation in the galaxy',
{\it Astrophys. J.} {\bf 343} (1989) 169;

\noindent 
A. Bouquet, P. Salati and J. Silk,
`Gamma-ray lines as a probe for a cold dark matter halo', 
{\it Phys. Rev.} {\bf D40} (1989) 3168; 


\noindent S. Rudaz and F.~W. Stecker,
`On the observability of the gamma-ray line flux from dark matter 
annihilation'
{\it Astrophys. J.} {\bf 368} (1991) 406.

\bibitem{center}
M. Urban, A. Bouquet, B. Degrange, P. Fleury, J. Kaplan, A.~L. Melchior
and E. Pare, 
`Searching for TeV dark matter by atmospheric Cherenkov techniques',
{\it Phys. Lett.} {\bf B293} (1992) 136 [arXiv:hep-ph/9208255];

V.~S. Berezinsky, A.~V. Gurevich and K.~P. Zybin,
`Distribution of dark matter in the galaxy and the lower limits
for the masses of supersymmetric particles',
{\it Phys. Lett.} {\bf B294} (1992) 221;

\noindent V. Berezinsky, A. Bottino and G. Mignola, 
`High-energy gamma radiation from the galactic center due
to neutralino annihilation', 
{\it Phys. Lett.} {\bf B325} (1994) 136 [arXiv:hep-ph/9402215].



\bibitem{halo2}
L. Bergstrom, J. Edsjo and  P. Ullio,
`Possible indications of a clumpy dark matter halo',
{\it Phys. Rev.} {\bf D58} (1998) 083507 [arXiv:astro-ph/9804050];

L. Bergstrom, J. Edsjo, P. Gondolo and  P. Ullio,
`Clumpy neutralino dark matter',
{\it Phys. Rev.} {\bf D59} (1999) 043506 [arXiv:astro-ph/9806072].

% \bibitem{nearby}
% E.~Baltz, C. Briot, P. Salati, R. Taillet and J. Silk,
% `Detection of neutralino annihilation photons from external galaxies',
% {\it Phys. Rev.} {\bf D61} (2000) 023514 [arXiv:astro-ph/9909112];
   

% \noindent 
% A. Falvard et al.,
% %, E. Giraud, A. Jacholkowska, J. Lavalle, E. Nuss,
% %F. Piron, M. Sapinski, P. Salati, R. Taillet , 
% %K. Jedamzik, G. Moultaka (LPMT-Montpellier)
% `Supersymmetric dark matter in M31: can one see 
% neutralino annihilation with CELESTE?',
% {\it Astropart. Phys.} {\bf 20} (2004) 467;
 

% \noindent 
% L. Pieri and E. Branchini,
% `On dark matter annihilation in the local group',
% {\it Phys. Rev.} {\bf D69} (2004) 043512 [arXiv:astro-ph/0307209];


% \noindent 
% S. Peirani, R. Mohayaee and J.~A. de Freitas Pacheco,> 

% `Indirect search for dark matter: prospects for GLAST',
% arXiv:astro-ph/0401378.



\bibitem{Feng} 
%For a review, see e.g.
J.~L. Feng, K.~T. Matchev, F. Wilczek,
`Prospects for indirect detection of neutralino dark matter',
  {\it Phys. Rev.} {\bf D63} (2001) 045024 [arXiv:astro-ph/0008115].



\bibitem{nearby2}
P. Ullio, L. Bergstrom, J. Edsjo and C. Lacey,
`Cosmological dark matter annihilations into gamma-rays - a closer
look',
{\it Phys. Rev.} {\bf D66} (2002) 123502 [astro-ph/0207125].

  


\bibitem{Ullio3} L. Bergstrom, P. Ullio and J.~H. Buckley, 
`Observability of gamma rays from dark matter neutralino annihilations in
the Milky Way halo', 
{\it Astropart. Phys.} {\bf 9} (1998) 137 [arXiv:astro-ph/9712318].


\bibitem{Ullio1} L. Bergstrom and P. Ullio, 
`Full one-loop calculation of neutralino annihilation into two photons',   
{\it Nucl. Phys.} {\bf B504} (1997) 27 [arXiv:hep-ph/9706232];

\noindent 
%\bibitem{Gondolo1} 
Z. Bern P. Gondolo and M. Perelstein, 
`Neutralino annihilation into two photons',
{\it Phys. Lett.} {\bf B411} (1997) 86 [arXiv:hep-ph/9706538].

\bibitem{Ullio2} 
P. Ullio and L. Bergstrom, 
`Neutralino annihilation into a photon and a $Z$ boson', 
{\it Phys. Rev.} {\bf D57} (1998) 1962 [arXiv:hep-ph/9707333].

\bibitem{Berezinsky} 
V.~ S. Berezinsky, A. Bottino and V. de Alfaro, 
`Is it possible to detect the gamma ray line from neutralino-neutralino
 annihilation?', 
{\it Phys. Lett.} {\bf B274} (1992) 122. 


\bibitem{Salati} 
H.~U.~ Bengtsson, P.~ Salati and J.~ Silk,
`Quark flavours and the $\gamma$-ray spectrum from halo dark matter annihilations', 
{\it Nucl. Phys.}, {\bf B346} (1990) 129.


\bibitem{Bertone} G. Bertone, G. Servant and G. Sigl,
`Indirect detection of Kaluza-Klein dark matter',
{\it Phys. Rev.} {\bf D68} (2003)  044008 [arXiv:hep-ph/0211342].
 

\bibitem{UllioEGRET}
A. Cesarini, F. Fucito, A. Lionetto, A. Morselli and P. Ullio,
`The galactic center as a dark matter gamma-ray source',
{\it Astropart. Phys.} {\bf 21} (2004) 267 [arXiv:astro-ph/0305075].

\bibitem{Boer}
W. de Boer, M. Herold, C. Sander and V. Zhukov,
`Indirect evidence for the supersymmetric nature of dark matter from 
the combined data on galactic positrons, antiprotons and gamma rays',
arXiv:hep-ph/0309029.

\bibitem{Wang}
D. Hooper and L.-T. Wang,
`Direct and indirect detection of neutralino dark matter in 
selected supersymmetry breaking scenarios',
{\it Phys. Rev. } {\bf D69} (2004) 035001 [arXiv:hep-ph/0309036].

\bibitem{farrill}
H. Baer, A. Belyaev, T.Krupovnickas and J.~O'Farrill,
`Indirect, direct and collider detection of neutralino
dark matter',  {\it JCAP} {\bf 08} (2004) 005
[arXiv:hep-ph/0405210].



% \cite{Edsjo:2004pf}
 \bibitem{Edsjo:2004pf}
J.~Edsjo, M.~Schelke and P.~Ullio,
`Direct versus indirect detection in mSUGRA with self-consistent halo
models',
arXiv:astro-ph/0405414.
% %CITATION = ASTRO-PH 0405414;%%





\bibitem{hooper} D. Hooper and B. Dingus,
`Improving the angular resolution of EGRET and new limits on 
supersymmetric dark matter near the galactic center',
arXiv:astro-ph/0212509.

\bibitem{Bottino} 
A. Bottino, F. Donato, N. Fornengo and S. Scopel, 
`Indirect signals from light neutralinos in supersymmetric models
without gaugino mass unification', 
{\it Phys. Rev. } {\bf D70} (2004) 015005
[arXiv:hep-ph/0401186]. 
 

\bibitem{Mannheim} D. Elsasser and K. Mannheim,
`Supersymmetric dark matter and the extragalactic gamma ray
background', 
{\it Phys. Rev. Lett.} {\bf 94} (2005) 171302
[arXiv:astro-ph/0405325]. 

%\cite{Bertone:2004ag}
\bibitem{Bertone:2004ag}
  G.~Bertone, E.~Nezri, J.~Orloff and J.~Silk,
  %``Neutrinos from dark matter annihilations at the galactic centre,''
  {\it Phys.\ Rev.\ } {\bf D70}, 063503 (2004)
  [arXiv:astro-ph/0403322].
  %%CITATION = ASTRO-PH 0403322;%%


\bibitem{Mambrini} 
G. Bertone, P. Binetruy, Y. Mambrini and E. Nezri,
`Annihilation radiation of dark matter in heterotic orbifold models', 
arXiv:hep-ph/0406083. 
 

\bibitem{Mambrini2}
P. Binetruy, A. Birkedal-Hansen, Y. Mambrini and B.D. Nelson,
`Phenomenological aspects of heterotic orbifold models at one loop',
arXiv:hep-ph/0308047.

%\cite{Mambrini:2004ke}
\bibitem{Mambrini:2004ke}
Y.~Mambrini and C.~Mu\~noz,
`Gamma-ray detection from neutralino annihilation in non-universal SUGRA
scenarios', arXiv:hep-ph/0407158.
%%CITATION = HEP-PH 0407158;%%

\bibitem{BaerNUMH}
H. Baer, A. Mustafayev, S. Profumo, A. Belyaev, X. Tata,
`Direct, Indirect and Collider Detection of Neutralino Dark Matter In Susy 
Models with Non--universal Higgs Masses', arXiv:hep-ph/0504001.


%\cite{Mambrini:2004kv}
\bibitem{Mambrini:2004kv}
Y.~Mambrini and C.~Mu\~noz,
`A comparison between direct and indirect dark matter search',
{\it JCAP} {\bf 10} (2004) 003
[arXiv:hep-ph/0407352].
%%CITATION = HEP-PH 0407352;%%

\bibitem{Wboer}
W. de Boer, M. Herold, C. Sander, V. Zhukov, A.V. Gladyshev and
D.I. Kazakov,
`Excess of EGRET galactic gamma ray data interpreted as dark matter
annihilation',
arXiv:astro-ph/0408272;

\noindent
W. de Boer,
`Evidence for dark matter annihilation from galactic gamma rays?',
arXiv:hep-ph/0408166.

\bibitem{Khlopov}
K. Belotsky, D. Fargion, M. Khlopov and R. Konoplich,
`May heavy neutrinos solve underground and cosmic ray puzzles?',
arXiv:hep-ph/0411093.

%\bibitem{dama}
%  DAMA Collaboration, R. Bernabei et al., 
%  `Search for WIMP annual modulation signature: results from
%  DAMA/NaI-3 and DAMA/NaI-4 and the global combined analysis', 
%  {\it Phys. Lett. } {\bf B480} (2000) 23;
%  `Dark matter search', 
%  {\it Riv. Nuovo Cim.} {\bf 26} (2003) 1 [arXiv:astro-ph/0307403].
%  %%CITATION = ASTRO-PH 0307403;%%


%\bibitem{halo} P. Belli, R. Cerulli, N. Fornengo and S. Scopel,
%  `Effect of the galactic halo modeling on the DAMA/NaI annual modulation 
%  result: an extended analysis of the data for WIMPs 
%  with a purely spin-independent coupling',
%  {\it Phys. Rev. } {\bf D66} (2002) 043503 [arXiv:hep-ph/0203242]. 
%  %%CITATION = HEP-PH 0203242;%%


%\bibitem{experimento2} CDMS Collaboration,
%  R. Abusaidi et al., `Exclusion limits on the 
%  WIMP nucleon cross-section from the cryogenic dark matter search', 
%  {\it Phys. Rev. Lett.} {\bf 84} (2000) 5699 [arXiv:astro-ph/0002471];
%  %%CITATION = ASTRO-PH 0002471;%%
%  D. Abrams et al.,
%  `Exclusion limits on the 
%  WIMP nucleon cross-section from the cryogenic dark matter search',
%  {\it Phys. Rev. } {\bf D66} (2002) 122003 [arXiv:astro-ph/0203500];
%  %%CITATION = ASTRO-PH 0203500;%%
%  D.~S. Akerib et al., 
%  `New results from the cryogenic dark matter search experiment',
%{\it Phys. Rev.} {\bf D68} (2003) 082002 [arXiv:hep-ex/0306001].
%  %%CITATION = HEP-EX 0306001;%%
 

%\bibitem{edelweiss} EDELWEISS Collaboration,  
%  A. Benoit {\em et al.}, 
%  `First results of the EDELWEISS WIMP search using a 320-g
%  heat-and-ionization Ge detector',
%  {\it Phys. Lett. } {\bf B513} (2001) 15 [arXiv:astro-ph/0106094];
%  %%CITATION = ASTRO-PH 0106094;%%
%  `Improved exclusion limits from the EDELWEISS WIMP search', 
%  {\it Phys. Lett. } {\bf B545} (2002) 43 [arXiv:astro-ph/0206271].
%  %%CITATION = ASTRO-PH 0206271;%%
  

%\bibitem{IGEX3}
%  A. Morales,
%  `Searching for WIMP dark matter: the case for Germanium ionization
%  detectors', arXiv:hep-ex/0111089.
%  %%CITATION = HEP-EX 0111089;%%

  
%\bibitem{HDMS2} GENIUS Collaboration,
%  H.~V. Klapdor-Kleingrothaus et al., 
%  `GENIUS - a Supersensitive Germanium Detector System for Rare Events',
%  arXiv:hep-ph/9910205.
%  %%CITATION = HEP-PH 9910205;%%

\bibitem{EGRET} EGRET Collaboration, 
S.~D. Hunger et al., `EGRET observations of the
diffuse gamma-ray emission from the galactic plane',
{\it Astrophys. J.} {\bf 481} (1997) 205; 
H.~A. Mayer-Hasselwander et al., 
`High-Energy Gamma-Ray Emission from the Galactic Center'
{\it Astron. \& Astrophys.} {\bf 335} (1998) 161.

\bibitem{Cangaroo1} CANGAROO-II Collaboration,
K.~Tsuchiya et al.,
`Detection of sub-TeV gamma-rays from the galactic center direction by
CANGAROO-II',
{\it Astrophys.\ J.\ } {\bf 606} (2004) L115
[arXiv:astro-ph/0403592].



\bibitem{HESS} HESS Collaboration,
F. Aharonian et al., `Very high energy gamma rays from
the direction of Sagittarius $A^*$', 
{\it Astron. \& Astrophys.} {\bf L13} (2004) 425
[arXiv:astro-ph/0408145].





\bibitem{GLAST} N. Gehrels, P. Michelson,
`GLAST: the next generation high-energy gamma-ray astronomy mission',
{\it Astropart. Phys.} {\bf 11} (1999) 277;

\noindent 
See also the web page http://www-glast.stanford.edu


\bibitem{mireview} For a recent review, see C. Mu\~noz, 
`Dark matter detection in the light of recent experimental results', 
{\it Int. J. Mod. Phys.} {\bf A19} (2004) 3093 [arXiv:hep-ph/0309346].


%\cite{Prada:2004pi}
\bibitem{Prada:2004pi}
F.~Prada, A.~Klypin, J.~Flix, M.~Martinez and E.~Simonneau,
`Astrophysical inputs on the SUSY dark matter annihilation
detectability', 
arXiv:astro-ph/0401512.
%%CITATION = ASTRO-PH 0401512;%%



%\bibitem{HESS} HESS Collaboration,
%J.~A. Hinton et al.,
%`The status of the HESS project',
%{\it New Astron. Rev.} {\bf 48} (2004) 331 [arXiv:astro-ph/0403052].



\bibitem{Navarro:1996he} J.~F. Navarro, C.~S. Frenk and S.~D.~M. White, 
`The structure of cold dark matter halos', 
{\it Astrophys. J.} {\bf 462} (1996) 563 [arXiv:astro-ph/9508025].

%\cite{Moore:1999gc}
\bibitem{Moore:1999gc}
B.~Moore, T.~Quinn, F.~Governato, J.~Stadel and G.~Lake,
`Cold collapse and the core catastrophe',
{\it Mon. Not. Roy. Astron. Soc.} {\bf 310} (1999) 1147
[arXiv:astro-ph/9903164].
%%CITATION = ASTRO-PH 9903164;%%

\bibitem{ko}
S. Baek, Y.G. Kim and P. Ko,
`Neutralino dark matter scattering and $B_s \to \mu^+ \mu^-$ in SUSY
models', 
{\it J. High Energy Phys.} {\bf 02} (2005) 067
[arXiv:hep-ph/0406033].

\bibitem{ko2}
S. Baek, D.G. Cerde\~no, Y.G. Kim, P. Ko and C. Mu\~noz,
`Direct detection of neutralino dark matter in supergravity',
{\it J. High Energy Phys.} {\bf 06} (2005) 017 [arXiv:hep-ph/0505019].
%`Neutralino dark matter scattering and $B_s \to \mu^+ \mu^-$ in SUSY
%models', 
% {\it J. High Energy Phys.} {\bf 02} (2005) 067
% [arXiv: hep-ph/0406033].



\bibitem{model} L. Hernquist,
`An analitical model for spherical galaxies
and bulges', {\it Astrophys. J.} {\bf 356} (1990) 359.

\bibitem{Hansen} S.H. Hansen, 
`Dark matter density profiles from the Jeans equation',
{\it Mon. Not. Roy. Astron. Soc.} {\bf 352} (2004) 
L41 [arXiv:astro-ph/0405371].



%\cite{Blumenthal:1985qy}
\bibitem{Blumenthal:1985qy}
G.~R.~Blumenthal, S.~M.~Faber, R.~Flores and J.~R.~Primack,
``Contraction of dark matter galactic halos due to baryonic infall,''
{\it Astrophys.\ J.}  {\bf 301} (1986) 27.
%%CITATION = ASJOA,301,27;%%

\bibitem{Jesseit}
R.~Jesseit, T.~Naab and A.~Burkert,
``The validity of the adiabatic contraction approximation for dark
matter halos''
{\it Astrophys.\ J.\ Lett.}  {\bf 571} (2002) L89 [arXiv:astro-ph/0204164].


\bibitem{Gnedin}
O.~Y.~Gnedin, A.~V.~Kravtsov, A.~A.~Klypin and D.~Nagai,
``Response of dark matter halos to condensation of baryons: cosmological
simulations and improved adiabatic contraction model,''
{\it Astrophys.\ J.\ }  {\bf 616} (2004) 16
[arXiv:astro-ph/0406247].



%\cite{Diemand:2004wh}
\bibitem{Diemand:2004wh}
J.~Diemand, B.~Moore and J.~Stadel,
`Convergence and scatter of cluster density profiles',
{\it Mon.\ Not.\ Roy.\ Astron.\ Soc.\ } {\bf 353} (2004) 624
[arXiv:astro-ph/0402267].
%%CITATION = ASTRO-PH 0402267;%%


\bibitem{merrit04}
D.~Merrit, `Evolution of the dark matter distribution at the galactic center,'
{\it Phys.\ Rev.\ Lett.\ } {\bf 92} (2004) 201304. 


\bibitem{gfandmerrit}
G.~Bertone and D. Merritt, 
`Time-dependent models for dark matter at the galactic center',
arXiv:astro-ph/0501555.



%\cite{Athanassoula:2005dw}
\bibitem{Athanassoula:2005dw}
E.~Athanassoula, F.~S.~Ling and E.~Nezri,
`Halo geometry and dark matter annihilation signal',
%``Halo geometry and dark matter annihilation signal,''
arXiv:astro-ph/0504631.
  %%CITATION = ASTRO-PH 0504631;%%


%\cite{manunonU}
 \bibitem{manunonU}
V.~Bertin, E.~Nezri and J.~Orloff,
 `Neutralino dark matter beyond CMSSM universality,'
 {\it J. High Energy Phys.} {\bf 02} (2003) 046
 [arXiv:hep-ph/0210034].


 \bibitem{Birkedal-Hansen:2002am}
 A.~Birkedal-Hansen and B.~D.~Nelson,
 `Relic neutralino densities and detection rates with nonuniversal 
 gaugino masses,'
 {\it Phys. Rev.} {\bf D67} (2003) 095006
 [arXiv:hep-ph/0211071].


\bibitem{darksusynew}
P. Gondolo, J. Edsjo, P. Ullio, L. Bergstrom, M. Schelke and E.~A. Baltz,
`DarkSUSY: Computing supersymmetric
dark matter properties numerically',
arXiv:astro-ph/0406204;

\noindent
See also the web page
http://www.physto.se/\char126edsjo/darksusy


\bibitem{Suspect}
A. Djouadi, J.~L. Kneur and G. Moultaka,
`SuSpect: a Fortran code for the supersymmetric and Higgs particle
spectrum in the MSSM', arXiv:hep-ph/0211331;

\noindent See also the web page
 http://www.lpm.univ-montp2.fr:6714/\char126kneur/suspect.html

%\cite{Bertin:2002ky}
\bibitem{Bertin:2002ky}
  V.~Bertin, E.~Nezri and J.~Orloff,
  %``Neutrino indirect detection of neutralino dark matter in the CMSSM,''
  Eur.\ Phys.\ J.\ C {\bf 26}, 111 (2002)
  [arXiv:hep-ph/0204135].
  %%CITATION = HEP-PH 0204135;%%


\bibitem{g-2} Muon g-2 Collaboration, G.~W.~Bennett et al.,
`Measurement of the negative muon anomalous magnetic moment to
0.7-ppm', {\it Phys. Rev. Lett.} {\bf 92} (2004) 161802
[arXiv:hep-ex/0401008].
  %%CITATION = HEP-EX 0401008;%%
  

\bibitem{newg2}
  M.~Davier, S.~Eidelman, A.~Hocker and Z.~Zhang,
  `Updated estimate of the muon magnetic moment using revised results
  from $e^+ e^-$ annihilation',
  {\it Eur. Phys. J. C} {\bf 31} (2003) 503
  [arXiv:hep-ph/0308213];
  %%CITATION = HEP-PH 0308213;%%
  
  \noindent
  K.~Hagiwara, A.~D.~Martin, D.~Nomura and T.~Teubner,
  `Predictions for $g-2$ of the muon and $\alpha_{QED}(M_Z^2)$',
{\it Phys. Rev.} {\bf D69} (2004) 093003
[arXiv:hep-ph/0312250];
  %%CITATION = HEP-PH 0312250;%%
  
  \noindent
  J.~F.~de Troconiz and F.~J.~Yndurain,
  `The hadronic contributions to the anomalous magnetic moment of the
  muon', 
  [arXiv:hep-ph/0402285].
  %%CITATION = HEP-PH 0402285;%%




\bibitem{cleo} 
CLEO Collaboration, S. Chen et al., 
`Branching fraction and photon energy spectrum for
$b\to s\gamma$',
{\it Phys. Rev. Lett.} {\bf 87} (2001) 251807
%Phys. Rev. Lett. 74 (1995) 2885.
%S. Chen et al., CLEO CONF 01-16.
[arXiv:hep-ex/0108032].

\bibitem{belle}
BELLE Collaboration, H. Tajima,
`Belle B physics results', 
{\it Int. J. Mod. Phys.} {\bf A17} (2002) 2967
[arXiv:hep-ex/0111037].




%

\bibitem{micromegas} G. Belanger, F. Boudjema, A. Pukhov and
A. Semenov,
`micrOMEGAs: a program for calculating the relic density in the
MSSM',
{\it Comput. Phys. Commun.} {\bf 149} (2002) 103 [arXiv:hep-ph/0112278];

\noindent G. Belanger, F. Boudjema, A. Pukhov and A.G. Semenov, 
`MicrOMEGAs: Version 1.3',
[arXiv:hep-ph/0405253];



\noindent 
See also the web page http://wwwlapp.in2p3.fr/lapth/micromegas



\bibitem{bmumuexp} CDF Collaboration, 
D.~Acosta et al., `Search for $B_s^0\to\mu^+\mu^-$ and $B_d^0\to\mu^+\mu^-$ 
decays in $p\bar p$ collisions at $\sqrt{s}=1.96$ TeV',
{\it Phys.\ Rev.\ Lett.}  {\bf 93} (2004) 032001
[arXiv:hep-ex/0403032];

\noindent 
D0 Collaboration, V.~M.~Abazov et al.,
`A search for the flavor-changing neutral current decay 
$B_s^0\to\mu^+\mu^-$ in $p\bar p$ collisions at $\sqrt{s}=1.96$ TeV
with the D0 detector',
{\it Phys. Rev. Lett.} {\bf 94} (2005) 071802 [arXiv:hep-ex/0410039].
  %%CITATION = HEP-EX 0410039;%%


 \bibitem{modi}
 I. Busching, M. Pohl and R. Schlickeiser,
 `Excess GeV radiation and cosmic ray origin',
 {\it Astron. \& Astrophys.} {\bf 377} (2001) 1056
 [arXiv:astro-ph/0108321];

 \noindent  A.~D. Erlykin and A.~W. Wolfendale, 
 `Supernova remnants and the origin of cosmic radiation: evidence from
 low-energy gamma-rays',
 {\it J. Phys.} {\bf G28} (2002) 2329;

 \noindent 
 F.~A. Aharonian and A.~M. Atoyan,
 `Broad-band diffuse gamma-ray emission of the galactic disk',
 {\it Astron. \& Astrophys.} {\bf 362} (2000) 937
 [arXiv:astro-ph/0009009];

 \noindent 
 A.~W. Strong, I.~V. Moskalenko and O. Reimer,
 `Diffuse galactic continuum gamma rays. A model compatible
 with EGRET data and cosmic-ray measurements',
{\it Astrophys. J.} {\bf 613} (2004) 962
 [arXiv:astro-ph/0406254].

\noindent T. Kamae, T. Abe and T. Koi,
`Diffractive interaction and scaling violation in $pp\to \pi^0 $
interaction and GeV excess in galactic diffuse gamma-ray spectrum of
EGRET',
arXiv:astro-ph/0410617.


\bibitem{modi2} 
W. de Boer,
`Indirect evidence for WIMP annihilation from diffuse galactic gamma
rays',
arXiv:astro-ph/0412620.




%===================================  CANGAROO  =========================================

% \bibitem{Cangaroo2}

% A. Kawachi $et$ $al.$, Astroparticle Physics, {\bf 14} (2001) 261 

%============================================================================


\bibitem{HooperSilk}
D.~Hooper, I.~de la Calle Perez, J.~Silk, F.~Ferrer and S.~Sarkar,
`Have atmospheric Cherenkov telescopes observed dark matter?',
{\it J. Cosm. Astrop. Phys.} {\bf 0409} (2004) 002
[arXiv:astro-ph/0404205].



\bibitem{Belangernonouniv}
G.~Belanger, F.~Boudjema, A.~Cottrant, A.~Pukhov and A.~Semenov,
`WMAP constraints on SUGRA models with non-universal gaugino masses
and prospects for direct detection',
%prospects for direct detection,''
{\it Nucl.\ Phys.\ }  {\bf B706} (2005) 411
[arXiv:hep-ph/0407218].
%%CITATION = HEP-PH 0407218;%%



\bibitem{Fornengo}
 N.~ Fornengo, L.~ Pieri and S.~ Scopel, 
`Neutralino annihilation into gamma-rays in the Milky Way and in external galaxies',
{\it Phys. Rev.} {\bf D70} (2004) 103529
[arXiv:hep-ph/0407342].


 \bibitem{pythia}
 T. Sjostrand et al., 
 `High-Energy-Physics Event Generation with PYTHIA 6.1',
 {\it  Comput. Phys. Commun.}  {\bf 135} (2001) 238
 [arXiv:hep-ph/0010017].
% %CITATION = ASTRO-PH 0203488;%%



\end{thebibliography}
\bibliographystyle{unsrt}

\end{document}